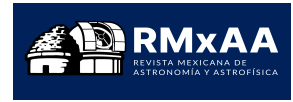

# The first 10 years of the HAWC Gamma-Ray Observatory: science results

A. Albert [19], R. Alfaro [1], C. Alvarez [2], A. Andrés [3], E. Anita-Rangel [3], M. Araya [4], J. C. Arteaga-Velázquez [5], K. P Arunbabu [34], D. Avila Rojas [3], H. A. Ayala Solares [6], R. Babu [7], P. Bangale [8], E. Belmont-Moreno [1], S. BenZvi [41], A. Bernal [3], C. Brisbois [16], K. S. Caballero-Mora [2], J.I. Cabrera Martínez [46, 48], T. Capistrán [3], A. Carramiñana [9], F. Carreón [3], S. Casanova [10], P. Colín-Farias [22], A.L. Colmenero-Cesar [5], U. Cotti [5], J. Cotzomi [11], S. Coutiño de León [35], E. De la Fuente [13], C. de León [5], R. DeLay [16], D. Depaoli [14], P. Desiati [12], N. Di Lalla [15], R. Diaz Hernandez [9], B. L. Dingus [19], M. A. DuVernois [12], J. C. Díaz-Vélez [12], R.W. Ellsworth [16], K. Engel [16], T. Ergin [7], C. Espinoza [1], K. L. Fan [16], K. Fang [12], H. Fleischhack [33], N. Fraija [3], S. Fraija [3], J. A. García-González [17], F. Garfias [3], N. Ghosh [18], H. Goksu [14], A. Gonzalez Muñoz [1], M. M. González [3], J. A. González [5], J. A. Goodman [16], S. Groetsch [18], J. Gyeong [30], J. P. Harding [19], S. Hernandez Cadena [31], I. Herzog [7], D. Huang [16], F. Hueyotl-Zahuantitla [2], P. Hüntemeyer [18], A. Iriarte [3], S. Kaufmann [20], N. Kelley-Hoskins [18], D. Kieda [21], A. Lara [22], R. Lauer [27], K. Leavitt [18], W. H. Lee [3], J. Lee [23], H. León Vargas [1], T. Lewis [18], J. T. Linnemann [7], A. L. Longinotti [3], G. Luis-Raya [20], K. Malone [19], A. Marinelli [43,44,45], S. Marinelli [7], O. Martinez [11], I. Martinez-Castellanos [24], J. Martínez-Castro [25], H. Martínez-Huerta [26], J. A. Matthews [27], J. McEnery [24], P. Miranda-Romagnoli [28], P. E. Mirón-Enriquez [3], T. Montaruli [42], J. A. Morales-Soto [5], E. Moreno [11], M. Mostafá [8], M. Najafi [18], A. Nayerhoda [10], L. Nellen [29], M. U. Nisa [7], R. Noriega-Papaqui [28], N. Omodei [15], M. Osorio-Archila [3], E. Ponce [11], Y. Pérez Araujo [1], E. G. Pérez-Pérez [20], J. Pretz [6], C. D. Rho [30], A. Rodriguez Parra [5], D. Rosa-González [9], M. Roth [19], E. Ruiz-Velasco [14], J. Ryan [49], H. Salazar [11], D. Salazar-Gallegos [7], F. Salesa Greus [10,35], A. Sandoval [1], M. Schneider [16], G. Schwefer [14], J. Serna-Franco [1], A. J. Smith [16], Y. Son [23], R. W. Springer [21], I. Taboada [37], O. Tibolla [20], K. Tollefson [7], I. Torres [9], R. Torres-Escobedo [31], R. Turner [18], F. Ureña-Mena [9], E. Varela [11], M. Vargas-Magaña [1], O. W. Vázquez Estrada [38], G. Vianello [36], L. Villaseñor [11], X. Wang [47], Z. Wang [47], I. J. Watson [23], I. Wisher [12], J. Wood [32], H. Wu [12], T. Yapici [41], S. Yu [6], S. Yun-Cárcamo [16], D. Zaborov [40] and H. Zhou [23]

[1]*Instituto de Física, Universidad Nacional Autónoma de México, Ciudad de México, México. The remaining affiliations are listed at the end of the paper.*

**Abstract**

The High-Altitude Water Cherenkov (HAWC) Observatory, located on the slopes of the Sierra Negra volcano in Mexico, began operations in March 2015. Over the past decade, HAWC has enabled the exploration of a broad range of topics in high-energy astrophysics and particle physics, resulting in more than 90 peer-reviewed publications. These studies have significantly advanced our understanding of several previously unexplored and poorly understood phenomena in the TeV energy regime. The present work provides an overview of the key scientific contributions of HAWC during its first ten years of operation.

**Resumen**

El Observatorio Cherenkov de Agua de Gran Altitud (HAWC), ubicado en las laderas del volcán Sierra Negra en México, inició operaciones en marzo de 2015. Durante la última década, HAWC ha permitido la exploración de una amplia gama de temas en astrofísica de altas energías y física de partículas, lo que ha dado lugar a más de 90 publicaciones arbitradas. Estos estudios han mejorado significativamente nuestra comprensión de varios fenómenos previamente inexplorados o poco comprendidos en el régimen de energía TeV. Este trabajo ofrece un resumen de las principales contribuciones científicas de HAWC durante sus primeros diez años de funcionamiento

## 1. INTRODUCTION

TeV gamma-rays are among the most energetic photons in the sky. They originate from some of the most extreme environments in the universe, including supernova explosions, binary stars, pulsars, active galactic nuclei, and gamma-ray bursts. The study of TeV gamma-rays is tightly coupled to that of cosmic rays, because the acceleration of these charged particles occurs in highly energetic astrophysical sources. Indeed, the interaction of accelerated cosmic rays with their astrophysical environment results in the emission of gamma-ray photons. Because these photons are neutral, they are not deflected by magnetic fields along their path to Earth (as in cosmic rays), where they can be intercepted and detected by our gamma-ray telescopes.

The last two decades have been a truly exciting epoch for the exploration of the Very-High-Energy (VHE) sky with the development and operation of many ground-based observing facilities: the Tibet Air Shower Gamma (AS$\gamma$) Experiment (operating since 1990; Amenomori et al., 1992), the Milagro Cherenkov Array (2000–2008; Atkins et al., 2000), the Astrophysical Radiation with Ground-based Observatory at YangBaJing (ARGO-YBJ) Experiment (operating since 2001; Montini, 2016), the Major Atmospheric Gamma Imaging Cherenkov (MAGIC) Telescopes (operating since 2004; Cortina et al., 2009), the High-Energy Stereoscopic System (H.E.S.S.) Cherenkov Telescopes (operating since 2004; Vincent, 2005), the Very Energetic Radiation Imaging Telescope Array System (VERITAS) Observatory (operating since 2007; Holder et al., 2006), and the Large High Altitude Air Shower Observatory (LHAASO, operating since 2019; Addazi et al., 2022). Among these, the High-Altitude Water Cherenkov Gamma-Ray Observatory (HAWC) is the longest-operating all-sky monitoring facility. HAWC started operations in March 2015 and was designed to detect gamma-ray photons and cosmic rays with energies ranging from $\sim 10^2$ GeV to $\sim 10^2$ TeV.

HAWC is located on the slope of the Sierra Negra volcano in Puebla, Mexico, at an altitude of 4,100 meters above sea level. With an area of 22,000 m$^2$, the completed main detector array is composed of 300 water tanks, each equipped with four upward-facing photomultiplier tubes (PMTs) that detect Cherenkov light arising from air showers. A more detailed description of HAWC's design and operation is available in Abeysekara & HAWC Collaboration (2023). With an instantaneous field of view (FoV) of 2 sr, HAWC observes 2/3 of the sky in the Northern Hemisphere with an active duty cycle of 95%.

This article presents a concise yet comprehensive review of the remarkable scientific output generated by the HAWC Observatory during its first ten years of operation. The topics were grouped according to the following main areas: the HAWC catalog of gamma-ray sources in §2, sources in the Galactic Plane in §3, compact sources (microquasars and active galaxies) in §4, transient events (gamma-ray bursts) in §5, cosmic rays in §6, and contributions to particle physics in §7.

## 2. HAWC Catalogs of Very-High-Energy Gamma-Ray Sources

### 2.1. The Early HAWC Catalogs of Very-High-Energy Gamma-Ray Sources

HAWC started collecting data during its construction phase with the array in a partial configuration. The first survey of the inner Galaxy region, reported in Abeysekara et al. (2016), was conducted using one-third of the total HAWC array and yielded the detection of ten TeV sources.

In its third year of operations, HAWC produced the first catalog of TeV gamma-ray sources (2HAWC), detected using a total of 507 days of data, this time collected by the complete HAWC detector array (Abeysekara et al., 2017a).

Remarkably, at that time, the 2HAWC catalog was the most sensitive TeV all-sky survey with 39 VHE detections, of which 19 were not associated with previously known TeV sources. With the continued accumulation of observed exposure and improved data reconstruction, the outcome of 2HAWC was superseded by the subsequent two catalogs, particularly by the third one, 3HWC, which was built by applying the same detection methodology as 2HAWC.

The two most recent major HAWC catalogs (3HAWC and 4HWC) are described in detail below, as well as a catalog of sources emitting above 56 TeV.

### 2.2. 3HWC: The Third HAWC Catalog of Very-High-Energy Gamma-Ray Sources

#### 2.2.1. 3HWC Overview

3HWC, the third HAWC catalog of VHE gamma-ray sources (Albert & HAWC Collaboration, 2020a), is the product of an all-sky search for point sources and extended sources with extensions up to 2°, assuming power-law energy spectra. It was based on 1,523 days of exposure and was collected between November 2014 and June 2019. The catalog analysis was sensitive to approximately two-thirds of the sky, covering a declination range from −26° to +64°. The flux sensitivity ranges from approximately $3 \times 10^{-15}$ TeV$^{-1}$ cm$^{-2}$ s$^{-1}$ (for hard spectrum sources at the center of HAWC's sensitive declination range) to $3 \times 10^{-14}$ TeV$^{-1}$ cm$^{-2}$ s$^{-1}$ (for soft spectrum sources at the edges of HAWC's sensitive declination range). These values were obtained assuming a power-law energy spectrum with a free spectral index and normalization given at a pivot energy of 7 TeV at a significance threshold of 5$\sigma$.

The 3HWC catalog comprises 65 sources, of which 17 are considered secondary sources because they are not well separated from neighboring sources. The expected number of false detections due to random background fluctuations was estimated to be 0.75. The source list and all-sky significance maps may be viewed interactively and downloaded in machine-readable formats on the HAWC collaboration's website[1].

Of the 65 sources in the 3HWC catalog, 37 have counterparts from the previous 2HWC catalog (Abeysekara et al., 2017a). This includes five sources clustering near the location of the Geminga pulsar, which are believed to be part of the extended Geminga TeV halo (see §3.4). At the time of the study, eight of the remaining 28 3HWC sources had potential associations with known TeV sources, and another 14 sources had potential associations with known GeV sources, as seen by the *Fermi*-LAT. Two sources were co-located with gamma-ray-quiet radio pulsars from the Australia Telescope National Facility (ATNF) catalog. Four sources did not have any immediate potential counterparts.

*2.2.2. Source Types within the 3HWC Catalog*

Most HAWC sources were clustered around the Galactic Plane, indicating a probable origin in the Milky Way. However, the catalog contains five sources with Galactic latitudes $|b| > 10°$. Only two of these are firmly identified as extragalactic sources: 3HWC J1104+381 (Mrk 421) and 3HWC J1654+397 (Mrk 501). Both are blazars known to be active at TeV energies and are detected by HAWC with high significance, allowing the study of their temporal variations. These sources are further discussed in §4.3. Searches for further known TeV-emitting Active Galactic Nuclei (AGNs) are discussed in §4.2 and §4.4.

The remaining three high-latitude sources are likely of galactic origin. 3HWC J0621+382 ($b = 10.97°$) was detected in the 0.5° extended source search. Although it lies near the GeV source 4FGL J0620.3+3804, classified as a blazar of unknown type (Abdollahi et al., 2020), it also has another possible counterpart in the ATNF pulsar PSR J0622+3749 (Manchester et al., 2005). Its extension makes a galactic origin more likely. This hypothesis has been strengthened by the detection of the likely LHAASO counterpart, LHAASO J0621+3755 (Aharonian & LHAASO-Collaboration, 2021) / 1LHAASO J0622+3754 (Cao & LHAASO-Collaboration, 2024), which is also extended and has been described as a potential TeV halo.

3HWC J1739+099 ($b = 20.34°$) was detected in the point-source search without a TeV counterpart at the time of the catalog study. It is co-located with the gamma-ray pulsar PSR J1740+1000 (Manchester et al., 2005) / 4FGL J1740.5+1005 (Abdollahi et al., 2020) and the center-filled supernova remnant (SNR) SNR G034.0+20.3 (Ferrand & Safi-Harb, 2012). This has since been detected by LHAASO as 1LHAASO J1740+0948u (Cao & LHAASO-Collaboration, 2024). It is likely of galactic origin and connected to the pulsar and/or SNR. 3HWC J1743+149 is further discussed in §2.2.3.

The 60 3HWC sources located within 10° of the Galactic Plane are likely of galactic origin, covering a large variety of source classes, which will be covered in more detail later in this paper. HAWC detected emissions from pulsar wind nebulae (cf., §3.1 and §3.3), supernova remnants (§3.2), pulsar halos (§3.4), star-forming regions (§3.5.3), and microquasars (§4.1).

*2.2.3. Unassociated Sources within the 3HWC Catalog*

Of the four 3HWC sources with no immediate potential counterparts, three are likely spurious point-source detections of extended TeV source components, hinting at their complex morphologies.

The remaining source, 3HWC J1743+149 ($b = 21.68°$), presents an intriguing puzzle. Its distance of 1.3° from the GeV-detected millisecond pulsar (MSP) PSR J1741+1351 (Manchester et al., 2005; Hooper & Linden, 2022) points towards a connection, making 3HWC J1743+149 a potential TeV halo. On the other hand, with a test statistic of 25.9, it is the second-lowest significance source in the catalog, raising the possibility of this being a false detection due to an unlikely but possible background fluctuation. A search for TeV emissions from the tail of PSR J1741+1351 with the VERITAS telescopes yielded no detection (Benbow et al., 2021). The upper limits (ULs) derived from the VERITAS observations are marginally compatible with the HAWC measurements, assuming that the source is slightly extended.

*2.2.4. Ultra-High-Energy View of the Sky*

Searches for emissions above ~100 TeV have long been of interest to astrophysicists; however, until HAWC came online, there were no catalogs of emissions above this energy threshold. The original motivation for studying this energy threshold, generally known as "ultra-high energy" (UHE), was its potential connection with the sources of galactic cosmic rays. There is a feature at approximately 1 PeV in the cosmic-ray spectrum known as the knee, with cosmic rays expected to be galactic in origin up to approximately this point. Cosmic rays with an energy of 1 PeV are associated with gamma-rays that are approximately one order of magnitude less energetic. These gamma-rays are hadronic in origin.

HAWC developed two independent energy estimation methods, called the "ground parameter" and "neural network". The ground parameter method relies on fitting a modified Nishimura-Kamata-Greisen (NKG) function to the lateral distribution function of the air shower and using that fit to extrapolate the energy. The neural network uses several inputs, including the fraction of charge in different annuli around the air shower core. The spectral fits obtained with the two estimators are identical within statistical errors, and the creation of these estimators has extended the energy range of the experiment and allowed for surveys of the UHE sky. These estimators were first applied to the Crab Nebula and resulted in the detection of this source above 100 TeV (see detailed description in section 3 of Abeysekara & HAWC Collaboration, 2019).

This source is a firmly identified leptonic accelerator (see §3.1), showing that HAWC had the sensitivity to detect UHE pulsar wind nebulae (PWNe) at these energies, despite

---
[1] data.hawc-observatory.org/datasets/3hwc-survey/index.php

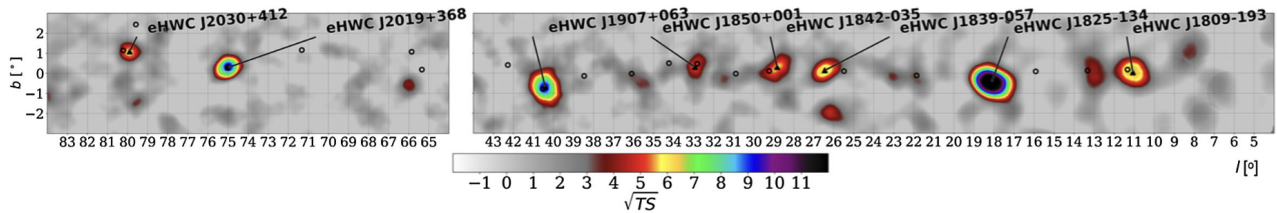

**Figure 1.** The inner Galactic Plane above 56 TeV, as of 2020. A disk with a radius of 0.5° is assumed as the source morphology. New observations, which are currently in preparation, have greatly increased the number of sources above this energy threshold. ©AAS. Reproduced with permission from Abeysekara et al. (2020).

the fact that the emission is suppressed in the Klein-Nishina regime.

In 2020, the HAWC Collaboration published a world-first catalog: the first collection of sources, using the 3HWC catalog method, listing all sources emitting above 56 TeV and sources emitting above 100 TeV (Abeysekara et al., 2020). At that time, this was the highest-energy astrophysical source catalog ever produced. This catalog consists of nine sources emitting significantly ($>5\sigma$) above 56 TeV, with three of the sources continuing above 100 TeV. All but one of these sources (the Crab Nebula) are morphologically extended in nature, and almost all are nearby extremely powerful pulsars with high-spindown power ($>10^{36}$ erg/s). Figure 1 shows the Galactic Plane above 56 TeV.

With more statistics, the last few years have brought about even more UHE detections. More than 20 sources are now observed above 56 TeV, with a new UHE catalog in preparation as of this writing. Several studies have focused on the different sources and regions of the sky where UHE emissions are prevalent. For example, it has been shown that the UHE emission is a generic feature near powerful pulsars (Albert & HAWC Collaboration, 2021a). Ultra-high-energy emission has also been observed near diverse sources, such as superbubbles and microquasars. This is discussed in the following sections.

### 2.3. 4HWC: The Fourth HAWC Catalog of Very-High-Energy Gamma-Ray Sources

4HWC, the fourth HAWC catalog of VHE gamma-ray sources, is a step forward from the previous 2HWC and 3HWC in terms of both data and modeling quality (Albert et al., 2023a). 4HWC is the first HAWC catalog to use the Pass 5 HAWC dataset, which provides significant improvements in gamma/hadron separation and angular resolution. In addition to greater data quality, the Pass5 dataset includes nearly 1,000 more days of data, with much greater sensitivity at larger zenith angles between 37 and 45° (Albert & HAWC Collaboration, 2024). To build upon the improved data quality, a new source-finding method was developed to significantly expand the modeling flexibility and power of the catalog. This method, called the Automated Likelihood Pipeline Search (ALPS) method, uses iterative multi-source fitting to explore a much larger and more continuous range of source morphologies and spectra than previous HAWC catalogs. The method is based on current techniques employed in dedicated HAWC analyses and the *Fermi*-LAT search for extended sources (Ackermann et al., 2012). To begin the method, point-like sources are added with a simple power-law assumption until a minimum source significance is no longer reached. Subsequently, all point-like sources were tested for evidence of larger extended emissions and for any curvature in their spectra. Any sources that drop below the significance threshold are dropped from the list of source models, and any nearby source models are refitted. In the final step, after all the sources are tested, the finalized source models are refitted again.

An all-sky search yields more than 80 sources, the majority of which exhibit spectral curvature (either a log parabola or a power law with an exponential cutoff) and a significant spatial extension. A publication containing the full catalog is currently under review. The alternate approach to fitting the extension and spectral curvature of each source illustrates a major change in the proportion of sources that are extended compared with previous HAWC catalogs. Figure 2 shows 4HWC sources detected in a portion of the Galactic Plane.

In addition to providing improved modeling of the morphology and spectral shape of gamma-ray sources, the ALPS method, in combination with the improved data quality of Pass 5, is able to discover several new possible gamma-ray sources. In total, of the ~80 sources identified by the 4HWC catalog, 12% do not have a counterpart in any TeV catalog and represent possible new TeV gamma-ray sources. As for sources with known associations, three are associated with AGNs, ~30 with pulsars, ~20 with SNRs, and the rest with high-mass X-ray binaries and low-mass X-ray binaries. Of the pulsar associations, a subcategory of possible TeV Halo candidates was defined as a pulsar 100–400kyr in age and with a spindown power of at least 1% of the Geminga pulsar. Of the ~30 pulsars, half are marked as very likely Halo Candidates, representing 40% of the sources.

## 3. The HAWC View of the Galactic Plane

### 3.1. The Crab Nebula

The Crab Nebula (M1), the remnant of the supernova witnessed all over the world in 1054 AD when it was visible even during the daytime for about three weeks, is a unique object in many aspects. Powered by high-energy electrons accelerated by the rapidly rotating neutron star left by the explosion, the nebula is observable throughout the electromagnetic spectrum, particularly in TeV gamma rays, which are frequently used as a standard candle. However, calibration with the Crab Nebula must consider the flaring activity reported in the low GeV range by the *AGILE* and *Fermi*-LAT collaborations (Tavani et al., 2011; Abdo et al., 2011).

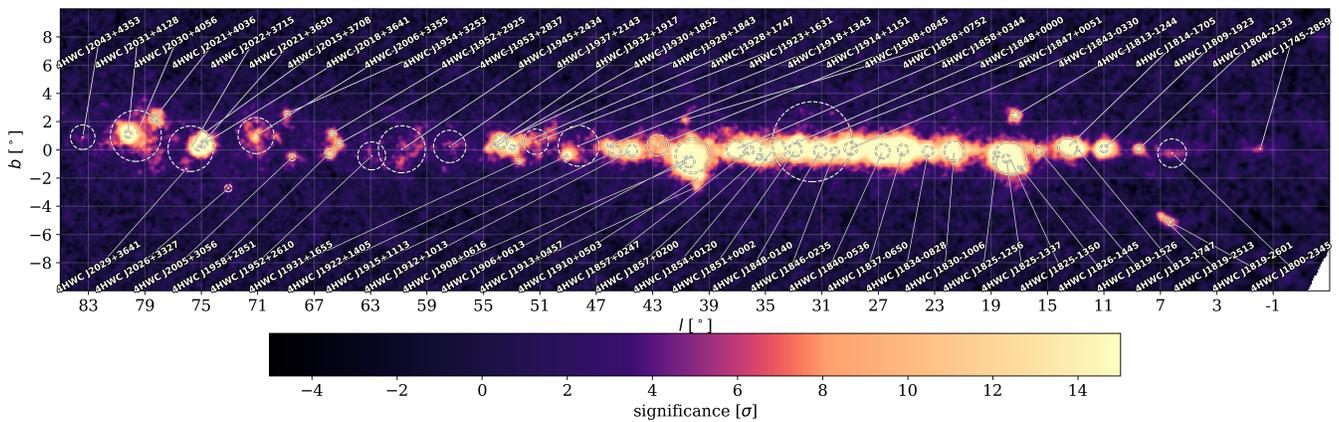

**Figure 2.** The inner Galactic Plane from the 4HWC catalog (paper to be submitted). The source locations and sizes are labeled. Dashed circle size corresponds to the $1\sigma$ width of the symmetric Gaussian model, where applicable.

The HAWC site provides a privileged view of the Crab, as it culminates within two degrees of zenith at Sierra Negra. The HAWC Collaboration has presented three papers dedicated to the Crab Nebula (2017, 2019, 2024), together with daily monitoring of this source and the active galaxies Mrk 421 and Mrk 501 (Abeysekara et al., 2017b). The publication Abeysekara et al. (2017b) presented a Bayesian block analysis applied to the early HAWC data of the three sources, finding no variability at TeV energies for the Crab Nebula, in contrast to the two Markarian BL Lac objects.

The HAWC observations of the Crab Nebula reported in Abeysekara et al. (2017c) served to validate and diagnose the performance of the detector. Using 507 live days of data, HAWC detected it with a significance in excess of $100\sigma$, fitting a log-parabola spectrum in the 1–37 TeV energy range. These observations provided the first definition of fHit bins, that is, the energy bins corresponding to the fraction of triggered phototubes in the detector array that are used for spectral analyses. It also demonstrated the capability of properly reconstructing showers and separating photons from hadrons, allowing for an early characterization of systematic errors and the absolute pointing of the instrument.

Two years later, Abeysekara & HAWC Collaboration (2019) presented one of the first measurements of Crab Nebula beyond 100 TeV (~concurrently with Amenomori et al., 2019a). Remarkably, extending the HAWC energy range beyond 100 TeV also allowed the search for PeVatrons and tests of fundamental physics (see §3.5 and §7.3).

In the most recent Crab Nebula publications (Albert & HAWC Collaboration, 2024), the HAWC Collaboration quantified the gain in the performance of the detector owing to improved air-shower reconstruction algorithms, internally identified as Pass 5. These algorithms suppress noise through a multiplane-fitting approach, providing gains in angular resolution and background rejection. They have significantly improved the performance of HAWC, particularly at large zenith angles, allowing the detection of southern sources such as the Galactic Center (GC) and the microquasar V4641 Sgr (see Sections 3.5.2 and 4.1, respectively).

### 3.2. Supernova Remnants

Supernova remnants are the aftermath of catastrophic stellar explosions and are considered primary sources of Galactic cosmic rays. Depending on whether they are Type II supernovas, they can also be PeVatron candidates (§3.5), capable of accelerating particles to PeV energies (Vink, 2022). Supernova remnants provide ideal environments for particle acceleration mechanisms, such as diffusive shock acceleration, possessing the necessary energy budget to account for the observed local cosmic-ray energy density (Bell, 2004; Caprioli et al., 2010; Bell et al., 2013).

Once accelerated, these particles can interact with the interstellar medium (ISM) or radiation fields, producing gamma-rays. For example, SNR G106.3+2.7 has been observed to interact with nearby molecular clouds, resulting in gamma-ray emissions above 56 TeV, likely from pion decay. The emission extends beyond 100 TeV, supporting the classification of SNR G106.3+2.7 as a PeVatron (Albert & HAWC Collaboration, 2020b; Alfaro et al., 2024a).

Similarly, IC 443 has been extensively studied, with recent HAWC observations indicating gamma-ray emission extending up to 30 TeV. This emission is consistent with a hadronic origin, suggesting that IC 443 is a potential

PeVatron candidate (Alfaro et al., 2025). Both IC 443 and G106.3+2.7 are considered Type II SNRs and are among the most promising candidates for PeVatrons (Vink, 2022).

HAWC has also observed other SNRs where the origin of the gamma-ray emission is complex due to overlapping sources, such as pulsars and PWNe. For instance, in the case of HAWC J1907+063, the emission surrounds both pulsar PSR J1907+0602 and SNR G40.5-0.5. Various models have been tested, including leptonic (pulsar-dominated), hadronic (SNR-dominated), and composite scenarios. An energy-dependent gamma-ray morphology study favored the leptonic model (Acharyya et al., 2024).

Another complex region is HESS J1809-193, where two SNRs, G11.1+0.1 and G11.0-0.0, along with the pulsar PSR J1809-1917, overlap with the observed emission. Similarly, the TeV source HAWC J1844-034 includes both the pulsar PSR J1844-0346 and the supernova remnant SNR G28.6-0.1, each possessing sufficient energy budgets to account for the observed gamma-ray emission (Albert et al., 2023b).

In addition, HAWC has conducted a survey of TeV emissions from Galactic SNRs, providing valuable insights into their capabilities as cosmic-ray accelerators (Fleischhack, 2019a). These observations contribute to our understanding of the mechanisms underlying particle acceleration in SNRs and their roles in cosmic-ray production.

Furthermore, modeling of the non-thermal emission of the gamma Cygni SNR up to the highest energies has been performed to investigate the underlying particle populations responsible for the observed gamma-ray emission (Fleischhack, 2019b; Abeysekara et al., 2021). These studies enhance our understanding of the acceleration processes occurring in SNRs and their contributions to the galactic cosmic-ray spectrum.

### 3.3. Pulsar Wind Nebulae

Pulsar wind nebulae form when relativistic winds from pulsars interact with their environment (Gaensler & Slane, 2006). It is believed that the spin-down energy of the pulsar is converted into magnetic energy and relativistic electrons, although protons or heavier nuclei may also be involved. The evolution of PWNe can be divided into three phases: (1) young PWNe driven by termination shocks; (2) middle-aged PWNe interacting with their surrounding SNRs; and (3) relic PWNe (Gelfand et al., 2009; Giacinti et al., 2020). When particles escape from relic PWNe after about 100 kyr, they form extended TeV halos (for an extensive discussion on TeV Halos, please see §3.4).

Pulsar wind nebulae are typically visible at all wavelengths across the electromagnetic spectrum. High-energy observations in the X-ray wavebands show synchrotron radiation from relativistic electrons and positrons, and VHE to UHE gamma-ray emissions are related to the inverse Compton scattering (ICS) of cosmic microwave background (CMB) photons, ambient infrared (IR), or optical stellar radiation (Pope et al., 2024). Pulsar wind nebulae are the most numerous gamma-ray sources detected at VHE energies, representing approximately 40% of the known galactic sources (H.E.S.S. Collaboration et al., 2018). Composite SNRs are particularly intriguing because they represent regions of ongoing PWN-SNR interaction and may accelerate particles to PeV energies (Ohira et al., 2018). Several middle-aged PWNe were detected at UHE gamma-ray energies using HAWC and LHAASO (Abeysekara et al., 2020; Cao & LHAASO-Collaboration, 2021).

One of the well-known PWNe observable in the Northern Sky is the Boomerang PWN, which was detected within the composite G106.3+2.7 system, has distinct head and tail regions, and was first observed in the radio wavelength range (Joncas & Higgs, 1990; Pineault & Joncas, 2000). The head consists of the radio PSR J2229+6114 and its compact boomerang-shaped nebula, and the tail contains the supernova ejecta from the SNR. In X-rays, *Chandra* observations detected a pulsation at 51.6 ms from the pulsar (Halpern et al., 2001), while XMM and Suzaku detected extended diffuse X-rays from both the head and tail regions (Fujita et al., 2021; Ge et al., 2021). The measured magnetic field of the Boomerang PWN was 140 $\mu$G (Liang et al., 2022). The *Chandra* ACIS image of the PWN fitted by a 3D torus model (Ng & Romani, 2004) shows that the brighter west side of the torus is caused by Doppler boosting of mildly relativistic magnetohydrodynamic outflow from the termination shock. The torus model also predicts that the pulsar should move along the spin axis such that the PWN morphology depends on both the proper motion of the pulsar and the density gradient of the region (Kolb et al., 2017). The *Fermi*-LAT detected GeV emission overlapping with PSR J2229+6114, which was also associated with the EGRET source 3EG J2227+6122 (Abdo et al., 2009a), where gamma-ray pulsations were observed above 0.1 GeV (Abdo et al., 2009b). In addition, Xin et al. (2019) reported extended gamma-ray emissions between 3 and 500 GeV, coinciding with the tail region. In the VHE–UHE gamma-ray range, the G106.3+2.7 system was observed by MAGIC, VERITAS, HAWC, Tibet AS$\gamma$ and LHAASO. VERITAS detected VHE emission from the tail region, and the TeV source was reported to overlap with dense molecular clouds (Acciari, 2009), whereas MAGIC detected TeV gamma-rays from the head region (The MAGIC Collaboration et al., 2022). Gamma rays with energies higher than 100 TeV were detected using HAWC (Albert & HAWC Collaboration, 2020b), Tibet AS$\gamma$ (Tibet AS$\gamma$ Collaboration et al., 2021), and LHAASO (Cao & LHAASO-Collaboration, 2021). The UHE source coincides with the VERITAS and *Fermi* sources in the tail region, as well as with PSR J2229+6114. UHE detection identified the Boomerang region as a PeVatron candidate (more details in §3.5), but its origin is still debated between leptonic and hadronic cases associated with the Boomerang PWN and SNR interaction with molecular clouds, respectively (Alfaro et al., 2024a).

Other extended TeV gamma-ray sources detected by HAWC that are potentially PWNe include 2HWC

J2006+341 (Albert et al., 2020) and HAWC J2031+415 (Alfaro & HAWC Collaboration, 2024a). 2HWC J2006+341 was discovered by HAWC (Abeysekara et al., 2017a) in the Cygnus Region of the Galaxy. PSR J2004+3429, located 0.4° away from the position of 2HWC J2006+341, has a characteristic age of ∼18 kyr, and the estimated distance and spin-down power of this pulsar are ∼11 kpc and $5.8 \times 10^{35}$ erg s$^{-1}$, respectively, (Barr et al., 2013). No other potential TeV sources were detected within 2HWC J2006+341.

In addition, the source HAWC J2031+415 lies in the Cygnus region and shows extended TeV emission modeled as a symmetric Gaussian (see Figure 5 and §3.5.3). It has a spectral shape of a power law with an exponential cutoff energy around 32 TeV, and is observable up to 151 TeV (Aliu et al., 2014; Abeysekara, A. U. et al., 2018). HAWC J2031+415 is predicted to be the UHE extension of TeV J2032+4130. The most likely power source for the PWN is PSR J2032+4127, an old pulsar with an estimated characteristic age of ∼200 kyr and an estimated spin-down luminosity of $1.5 \times 10^{35}$ erg s$^{-1}$, located at a distance of $1.33 \pm 0.06$ kpc (Manchester et al., 2005). Unlike leptonic models based on ICS, hadronic models are physically incompatible with HAWC J2031+415. Although the leptonic models fit to HAWC data revealed a very low magnetic field of 1.48 $\mu$G, this could indicate that the X-ray counterpart of the PWN may be fading. More X-ray and radio data will help constrain the leptonic model and confirm this hypothesis.

### 3.4. TeV Halos around the Geminga and Monogem Pulsars

As explained in §3.3, the diffusion of charged particles after ∼$10^5$ yr from relic PWNe results in the formation of an extended region of TeV emission— the so-called "TeV Halos." The discovery of this new class of TeV sources represents one of the greatest achievements of the HAWC observatory, as described below.

During the development of the second HAWC source catalog (2HWC; Abeysekara et al., 2017a), two extended sources associated with the Geminga and B0656 +14 pulsars (also known as "Monogem") were identified. Located at a distance of ∼250 pc, the Geminga pulsar is a well-known object that has been detected at various wavelengths, such as radio, optical, X-rays, and gamma-rays. It is a mature pulsar with a characteristic age of $3 \times 10^5$ yr and a spin-down luminosity of $3.26 \times 10^{34}$ erg s$^{-1}$ (Abdo et al., 2010). Regarding its VHE emission, Milagro previously reported the detection of a high-energy source associated with the PWN of Geminga with an angular size of ∼2.8° (Abdo et al., 2007).

Using a dataset of 507 days with enhanced resolution and sensitivity, HAWC was able to more accurately determine the extension of both sources (Figure 3), obtaining ∼5.5° for Geminga and ∼4.8° for Monogem (Abeysekara et al., 2017). Both Geminga and Monogem, as detected by HAWC, revealed characteristics that differed from those of "typical" PWNe. For instance, while Geminga's PWN observed in the X-ray band has a physical size of 0.2–0.3 pc, the emission detected at TeV energies extends to an approximate size of 24 pc. Furthermore, observations at other wavelengths show no correlation between matter (gas and dust) and high-energy emission, suggesting that the emission is produced by the ICS of CMB photons by energetic electrons and positrons. These findings indicate that these sources are not PWNe but rather a new subclass of Galactic gamma-ray sources, referred to as "pulsar halos," "TeV halos," or "inverse Compton halos."

The HAWC detection of Geminga and Monogem is particularly significant in light of earlier results from *PAMELA* (*Payload for Antimatter Matter Exploration and Light-nuclei Astrophysics*; Adriani et al. (2009a)), which were later confirmed by *Fermi*-LAT and AMS (Alpha Magnetic Spectrometer; Ackermann et al., 2012; Accardo et al., 2014), revealing an excess of positrons at energies > 10 GeV. Because of the energy losses experienced by the highest-energy positrons in the interstellar magnetic and radiation fields, this excess implies that its origin lies in cosmic-ray accelerators within a few hundred parsecs from Earth. Indeed, Geminga and Monogem are the closest pulsars to Earth, at distances of 250 and 280 parsecs, respectively, which makes them candidates for producing positron excess.

A diffusion model was applied to the HAWC data of Geminga and Monogem (Abeysekara et al., 2017). According to the model, the emission region arises from electrons and positrons diffusing away from the pulsar, which then up-scatter the CMB photons via the ICS. The angular sizes correspond to physical extents on the order of tens of parsecs, indicating that these particles escape the confinement of the PWN and travel away from their source (see Figure 3). Based on the brightness of the sources, the fluxes of electrons and positrons were estimated to be 40 and 4% of the spin-down power of Geminga and Monogem, respectively. The spatial and spectral morphologies of the gamma-ray flux are characterized as a function of several factors, including the diffusion coefficient $D_E$, where $E$ denotes the energy of the electrons. From HAWC observations, the diffusion coefficient for particles with energies of 100 TeV, $D_{100}$ was determined to be $4.5 \pm 1.2 \times 10^{27}$ cm$^2$ s$^{-1}$, which is approximately 100 times smaller than the average value reported for the Galaxy. This low value of $D_{100}$ could potentially arise from the additional effects of turbulent scattering. However, given this "slow" (or inhibited) diffusion, under the assumption of isotropic and homogeneous diffusion, the HAWC study concluded that Geminga and Monogem are unlikely to be the primary sources of the observed positron excess. Instead, this excess may be attributed to other sources, such as microquasars, SNRs, or the annihilation or decay of dark matter particles. Therefore, this work holds great importance for both high-energy gamma-ray astrophysics and pulsar astrophysics.

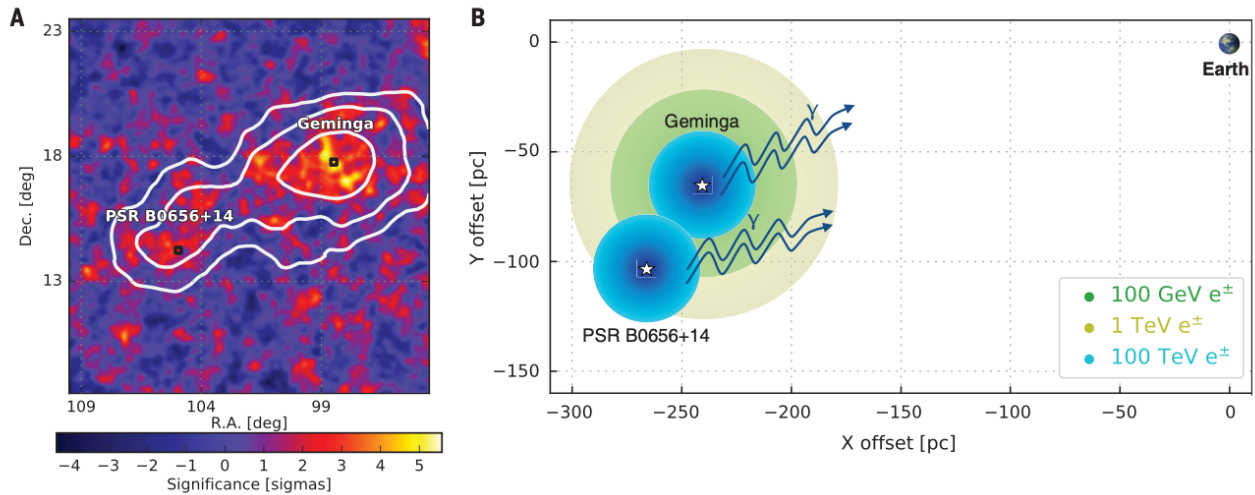

**Figure 3. Panel A:** HAWC (1–50 TeV) significance map of the Geminga and Monogem region. White contours trace the diffusion model fit at 5, 7, and 10σ. **Panel B:** Sketch of the two-pulsar region and the Earth projected onto the Galactic Plane. Different colors mark the spatial scale of diffusion of leptons at the corresponding energies on the left side of the figure. From the paper "Extended gamma-ray sources around pulsars constrain the origin of the positron flux at Earth", Abeysekara et al. (2017), *Science* 358, 911-914 (2017). [DOI:10.1126/science.aan4880]. Reprinted with permission from AAAS.

More recently, a follow-up study of the sky region containing halos around Geminga and Monogem was conducted (Albert et al., 2024a) using 2,398 days of data, which significantly increased the exposure of the previous observations. In addition to the larger dataset, the key difference compared with the 2017 analysis lies in the diffusion model used to fit observations. In 2024, a two-zone slow-diffusion model based on the properties of the pulsar was employed, whereas an earlier study used an approximate solution for the electron distribution, leading to an overestimation of the diffusion coefficient. Additionally, the efficiency was recalculated, yielding values of ∼6.1% for Geminga and ∼5.1% for Monogem, which are significantly different from those reported previously (40% and 4%). These efficiency values are consistent with earlier HAWC observations and with the detection of the Geminga halo in the high-energy band using *Fermi*–LAT data (Di Mauro et al., 2019). Furthermore, the diffusion coefficients around the halos were constrained to values between 4.8 and $7 \times 10^{27}$ cm$^2$/s for an energy of 100 TeV, thereby confirming the presence of a region with inhibited diffusion.

### 3.4.1. HAWC Search for TeV Halos around Other Pulsars

In 2023 (Albert et al., 2023), a new TeV Halo candidate around the pulsar PSR J0359+5414 was reported using 2,321 days of HAWC data. This pulsar was first identified as a radio-quiet pulsar in the third catalog of the *Fermi*-LAT (3FGL). A 95% UL on its extension was determined to be 0.41°, corresponding to a physical size of $R_{ul} = 25(d/3.45 \text{ kpc})$pc. Similar to Geminga and Monogem, the diffusion coefficient derived from HAWC observations was much lower than the average diffusion coefficient in the ISM. These results strongly indicate that the emission originates from a halo, thus classifying this HAWC source as a *candidate halo*. This is the first evidence of a halo around a radio-quiet pulsar, suggesting that halo formation is not dependent on the geometry of the gamma-ray and radio beams. Lastly, given its estimated age of 97 kyr, this pulsar is younger than both Geminga and Monogem pulsars. This is likely in a transitional stage between the relic and halo phases of PWNe, with the observed halo features indicating that high-energy particles are beginning to escape into the ISM.

As mentioned earlier (§3.4), the observed excess diffuse emission can be attributed to unresolved sources. One class of such sources, pulsar halos, may explain the excess at energies of a few TeV. Therefore, understanding the halos associated with MSPs is crucial for explaining the origin of GC excess. At GeV energies, an excess originating from the GC has been identified (e.g. Ackermann et al., 2017), which can be explained by dark matter annihilation or astrophysical processes, particularly those involving unresolved MSPs. Recent studies suggest that if MSPs produce TeV halos with an efficiency comparable to that of Geminga, then the MSPs contributing to the GeV excess would also generate TeV halos through ICS.

In 2025, results from a search for gamma-ray excesses associated with potential pulsar halos around MSPs were published (Abeysekara et al., 2025). A sample of 57 sources within HAWC's FoV was selected from catalogs such as the ATNF Pulsar Catalogue, the *Fermi*-LAT Pulsar Catalog (3PC), the West Virginia University Catalog (WVU), and the LOFAR Tied-Array All-Sky Survey (LOTAAS). No significant gamma-ray emissions associated with halos produced by MSPs were found, suggesting that these pulsars are not as efficient at producing halos as isolated pulsars. The absence of TeV halo detections around MSPs could be due to a lower efficiency in accelerating VHE electrons or to a lack of effective particle confinement in their vicinity. Although MSPs are known to produce intense GeV gamma-ray emissions, the HAWC results indicate that their contribution to Galactic diffuse emission at TeV energies is negligible. The absence of detectable TeV halos around MSPs suggests that their efficiency in producing very high-energy electrons is significantly lower than that of isolated middle-aged pulsars. Another search for extended gamma-ray emission was conducted on a sample of 36 isolated middle-aged pulsars (based on radio and gamma-ray selection) using 2,321 days of HAWC data (Albert & HAWC Collaboration, 2025). The resulting stacked gamma-ray emission yielded a $5.10\sigma$ significance against the background, indicating that TeV halos may be a common property of middle-aged pulsars.

### 3.5. PeVatron Sources

The study of UHE ($E \gtrsim 100$ TeV) gamma-ray sources has given rise to the emerging field of PeV astronomy (e.g., Cao & LHAASO-Collaboration, 2021). A PeVatron is commonly defined as a source capable of accelerating hadrons to energies of $\gtrsim 1$ PeV (e.g., Angüner, 2023). These sources play a central role in understanding the origin of cosmic rays and answering a fundamental question in high-energy astrophysics: What is the maximum energy to which nature can accelerate particles?

The term PeVatron began appearing in high-energy astrophysics literature in the early 2010s, particularly in connection with the H.E.S.S. collaboration. The detection of gamma-rays at energies greater than 100 TeV from an astronomical source implies that the parent particles, typically protons or nuclei, were likely accelerated to energies $\gtrsim 1$ PeV, which is analogous to the beam energies in terrestrial particle accelerators.

Ultra-high energy sources may originate from either leptonic mechanisms, such as ICS, or hadronic interactions, primarily via neutral pion decay following collisions of protons or nuclei with the surrounding matter (for example Abeysekara et al., 2020; Amenomori et al., 2019b; Cao et al., 2024). Indeed, one of the key open questions in studying UHE sources is determining the nature of their emission, whether it is hadronic or leptonic. Therefore, PeV astronomy provides a framework for investigating the origin of cosmic rays and related spectral features, such as the *knee* around ~3 PeV. An energy scale of 1 PeV serves as a threshold to distinguish PeVatron candidates from less extreme accelerators. However, the possible co-acceleration of leptons alongside hadrons complicates this classification and cannot be ruled out. Because gamma-ray observations alone are often insufficient to differentiate between hadronic and leptonic emission mechanisms, multiwavelength observations are essential. For instance, radio observations can help trace hadronic emission by probing the density of ambient nucleons (e.g., de la Fuente et al., 2023a,b), whereas X-ray data are critical for constraining leptonic contributions (e.g., Suzuki et al., 2025).

#### 3.5.1. PeVatron Observations

PeV astronomy emerged from decades of cosmic-ray research in the 20th century. However, following the Galactic Centre's H.E.S.S. study in 2016, the field has matured rapidly. The Galactic Center may represent the first observed PeVatron, as suggested by H.E.S.S. observations conducted between 2004 and 2013, indicating the presence of a proton accelerator (Aharonian et al., 2006; H.E.S.S. Collaboration et al., 2016), up to at least 1PeV. However, the term *PeVatron* was not used in 2006.

The year 2021 is generally regarded as a breakthrough, as HAWC, Tibet-AS($\gamma$), and the LHAASO observatory reported groundbreaking detection of gamma-rays with energies above 100 TeV. Thanks to the contributions of these observatories, the Northern Hemisphere sky has now been well studied at ultra-high energy (UHE). In this context, the HAWC collaboration reported a power-law gamma-ray spectrum extending to at least 200 TeV from the source HAWC J1825–134, which is considered a potential PeVatron candidate (Albert & HAWC Collaboration, 2021b). For more information on the GC region, see Section 3.5.2.

In 2021, the LHAASO collaboration reported the first 12 Galactic gamma-ray sources with photon energies up to 1.4 PeV (Cao & LHAASO-Collaboration, 2021), marking a significant milestone in the search for Galactic PeVatrons. Some of these coincide with already known TeV sources discovered by HAWC and listed in the 2HWC catalog (Abeysekara et al., 2017a), including 2HWC J1825–134, J1837–065, J1841–055, J1844–032, J1849+001, J1908+063, J1930+188, J1955+285 and J2031+415, which underlines HAWC's pioneering role in this field.

HAWC J2031+415, which is associated with the Cygnus OB2 region, provided the first observational evidence that massive star clusters, such as Cygnus OB2, can act as PeVatrons. Similarly, HAWC identified SNR G106.3+2.7 as a PeVatron candidate based on gamma-ray emissions exceeding 100 TeV. This discovery was later independently confirmed by Tibet-AS$\gamma$ (Tibet AS$\gamma$ Collaboration et al., 2021), illustrating the synergy between observatories.

Another important contribution of HAWC is the detection of extended TeV gamma-ray emission from the region around the Dragonfly Nebula (2HWC J2019+367), which is associated with the pulsar PSR J2021+3651. Multi-wavelength analyses, including X-ray and radio data, indicate that a large and possibly ancient pulsar wind nebula (PWN G75.2+0.1) is the origin of this emission. Although the gamma-ray signal is consistent with leptonic processes, modeling suggests that the electron energies may reach the PeV scale (Woo et al., 2023), making this source one of the strongest candidates for a leptonic PeV accelerator. HAWC detection was crucial for identifying this nebula as a potential PeVatron; however, its characterization was further refined by LHAASO observations. The latter underlines the fruitful synergy between wide-field gamma-ray observatories.

Subsequent studies investigated other candidates, including the Galactic Center (see §3.5.2) and LHAASO J2108+5157 (articles to be published in 2025). The first comprehensive LHAASO source catalogue that reaches the PeV range was published in 2024 (Cao & LHAASO-Collaboration, 2024). These results mark a turning point and build on the fundamental discoveries of previous instruments, particularly HAWC, whose ongoing observations are expected to further increase the population of known PeVatron candidates.

By the conventional definition, a PeVatron is a hadronic accelerator, that is, a source embedded in a dense medium that exhibits a hard gamma-ray spectrum without a cutoff above 100 TeV. In this context, HAWC has contributed to the study of particle acceleration mechanisms in microquasar jets. As discussed in §4.1, HAWC detected an extended, very energetic gamma-ray emission spatially associated with the microquasar SS 433, in which leptonic processes appear to dominate. In addition, a marginally significant extended gamma-ray signal was detected near V4641 Sgr, but its physical connection to the microquasar is uncertain, and its emission mechanism remains under investigation. These observations suggest that microquasars may contribute to the acceleration of cosmic rays in the Galaxy, complementing better-known accelerators such as supernova remnants. The role of microquasars remains an active and promising area of research in high-energy astrophysics, representing a remarkable discovery by HAWC.

According to the HAWC results, galactic PeVatron candidates can be roughly divided into five categories:

1. **Galactic center sources:** Extended gamma-ray emission above 100 TeV suggests PeV-scale acceleration within the central molecular zone.
2. **Binary systems and microquasars:** HAWC detected TeV emission from SS 433 (leptonic origin) and a marginal signal near V4641 Sgr, indicating possible particle acceleration in dynamical environments.
3. **Supernova remnants (SNRs):** Hard spectrum sources such as 2HWC J1930+188 (associated with G54.1+0.3) and G106.3+2.7 (confirmed by Tibet-AS$\gamma$) are strong PeVatron candidates.
4. **TeV halos:** Extended hard-spectrum sources such as 2HWC J1908+063 (PSR J1907+0602) and 2HWC J2019+367 (Dragonfly Nebula, PWN G75.2+0.1) can accelerate electrons to PeV energies (Woo et al., 2023).
5. **Massive star cluster:** 2HWC J2031+415, which coincides with Cygnus OB2 and the Cygnus Cocoon, provides early evidence that star clusters can act as PeVatrons, which was later confirmed by LHAASO observations (Cao & LHAASO-Collaboration, 2021).

Several of these sources — 2HWC J1825–134, J1837–065, J1841–055, J1844–032, J1849+001, J1908+063, J1930+188, J1955+285, and J2031+415 — were first catalogued by HAWC and later refined or confirmed by LHAASO (Cao & LHAASO-Collaboration, 2024), underlining the synergistic role of wide-field gamma-ray observatories in the identification of Galactic PeVatrons.

### 3.5.2. The Galactic Center PeVatron

As detailed in many sections of this review, within the Milky Way, several astrophysical objects have been proposed as potential PeVatrons, including SNRs (§3.2), young massive clusters in SFRs (§3.5.3), PWNe (§3.3), microquasars (§4.1), and GC.

The GC region was the first confirmed observational PeVatron candidate for the Milky Way. Located at the core of the Galaxy, the GC region hosts a complex and energetic environment, including the supermassive black hole (SMBH) Sagittarius $A^*$ (Sgr $A^*$), three massive star clusters within the Central Molecular Zone (CMZ): the Arches, the Quintuplet, and the Nuclear Cluster, and dense molecular gas structures (Aharonian et al., 2019; H. E. S. S. Collaboration et al., 2018). Several pointing gamma-ray instruments have detected emissions from the GC region up to ∼20 TeV, with particular interest in two key sources: Sgr $A^*$ (HESS J1745–290) and the unidentified source HESS J1746–285, which is spatially coincident with the Galactic radio arc (H.E.S.S. Collaboration et al., 2016).

Using seven years of data, HAWC has detected UHE gamma rays from this region for the first time, identifying nearly 100 gamma-ray events with energies exceeding 100 TeV (Figure 4; Albert et al. (2018a)). To date, this detection provides the strongest observational evidence that the GC is indeed a PeVatron, confirming the presence of extreme particle acceleration in the central region of the Milky Way. After subtracting the contributions from the two H.E.S.S. sources, the remaining emission detected by HAWC was best described by a single point source with a power-law energy spectrum that exhibited no indication of a spectral cutoff up to at least 114 TeV. A model-dependent analysis favors hadronic acceleration mechanisms and quasi-continuous cosmic-ray injection scenarios as the most plausible interpretations. Furthermore, the estimated gamma-ray luminosity of the PeVatron indicates a cosmic-ray energy density exceeding the Galactic average, strongly suggesting the presence of freshly accelerated protons in the 0.1–1 PeV range within the GC region. Finally, the total energy budget of protons above 100 TeV inferred from HAWC data was found to be consistent with previous measurements reported by H.E.S.S..

### 3.5.3. Star-Forming Regions

SFRs are considered areas of galaxies where gas and dust are cold enough to allow gravitational collapse of molecular clouds, resulting in the birth of new stars, which will eventually contribute to the global galaxy star formation rate. However, SFRs are also believed to be the sites where particle acceleration occurs (see Bykov et al. (2020) for review). Indeed, being rich in massive stars, SFRs are characterized by higher rates of stellar winds and supernova explosions, which are responsible for turbulence and shock waves in the ISM. This dynamic environment favors particle acceleration up to relativistic speeds, resulting in the production of extremely high-energy cosmic rays with energies well above the PeV threshold, making SFRs a very important target for VHE observatories, such as HAWC.

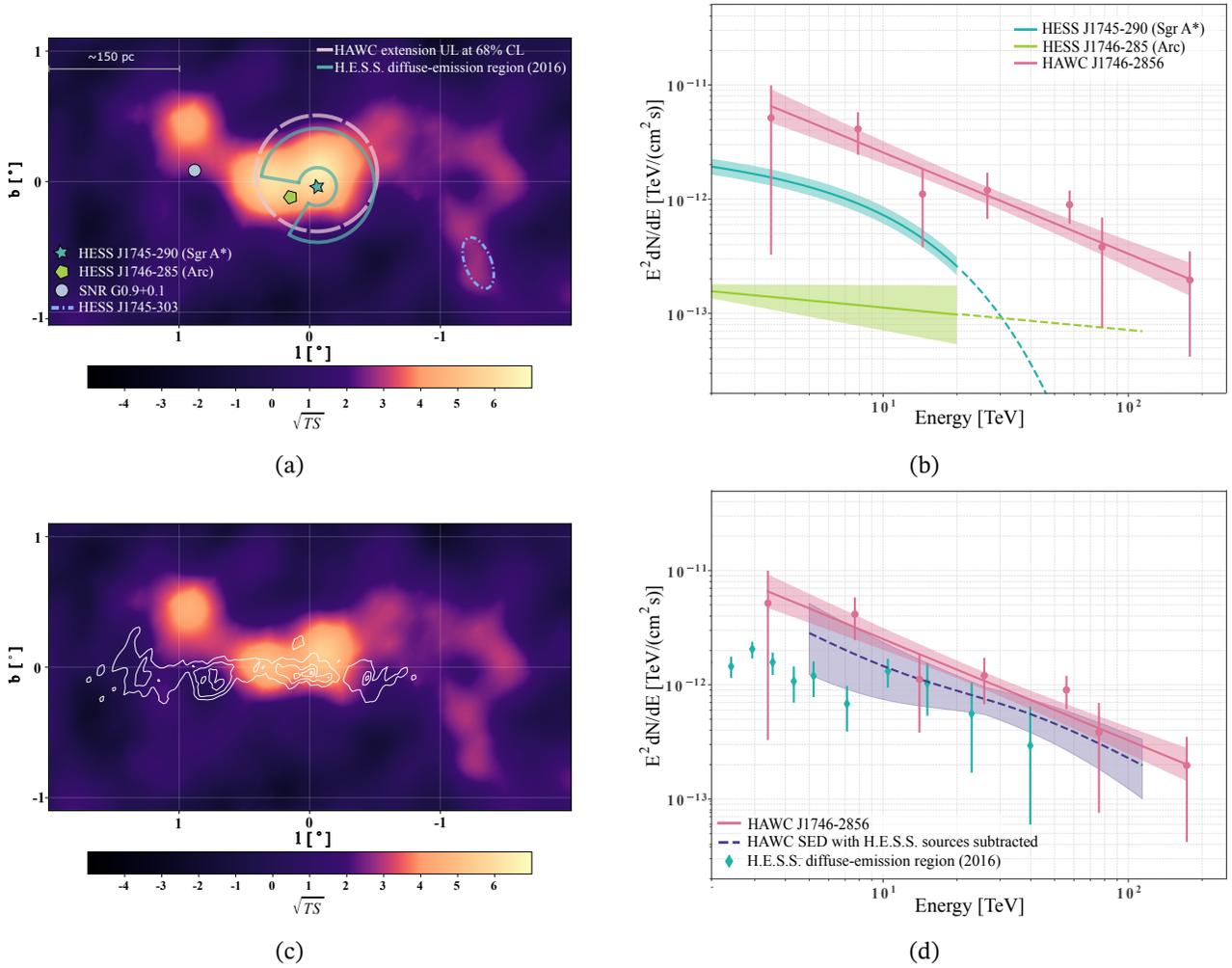

**Figure 4.** HAWC Galactic Center analysis results. ©AAS. All figures were reproduced with permission from Albert et al. (2018a). **(a)** Significance map obtained using the HAWC neural network energy estimator (on- and off-array events, as described in Abeysekara & HAWC Collaboration, 2019). The dashed circle outlines the extension upper limit of the HAWC point source at 68% CL. **(b)** Spectral energy distribution (SED) of two H.E.S.S. sources and the best-fit spectrum of HAWC J1746-2856. The dashed lines for the H.E.S.S. sources show the extrapolation of their best fit to the HAWC energy range. **(c)** HAWC emission after subtracting the two H.E.S.S. point sources. White contours are the density distribution contours of the ambient gas as traced by carbon monosulfide (CS; J1-0) line emission (Tsuboi et al., 1999). **(d)** Original best-fit of the HAWC spectral energy distribution (SED) and the result after subtracting the two H.E.S.S. point-source spectra.

Observational evidence for particle acceleration across the electromagnetic spectrum within SFRs include synchrotron radiation (radio band), associated with relativistic electrons, high ionization levels in molecular clouds, revealing the effect of local highly accelerated cosmic rays below ∼1 GeV, and, more relevant to this review, gamma-ray photons detected in the cavities carved by stellar winds and supernova explosions, known as "superbubbles."

The "Cygnus Cocoon" is one such superbubble, surrounding the Cygnus OB2 association of massive star clusters that is embedded within the larger (200 pc size) SFR Cygnus X. At a distance of 1.4 kpc from the Earth, the Cygnus Cocoon was discovered in the high-energy regime as an extended gamma-ray emission by the *Fermi* satellite (Ackermann et al., 2011). In Ackermann et al. (2011), it was proposed that the 1–100 GeV emission, with a gamma-ray luminosity of ∼ 9× $10^{34}$ erg $s^{-1}$ originating from a 50 pc-wide region, traces the accelerated particles in a superbubble.

The Cygnus Cocoon is coincident with one of the brightest regions in the Galactic Plane observed by HAWC (see Figure 2), where many TeV sources are present: the well-known PWN TeV J2032+4130; 2HWCJ2031+415,

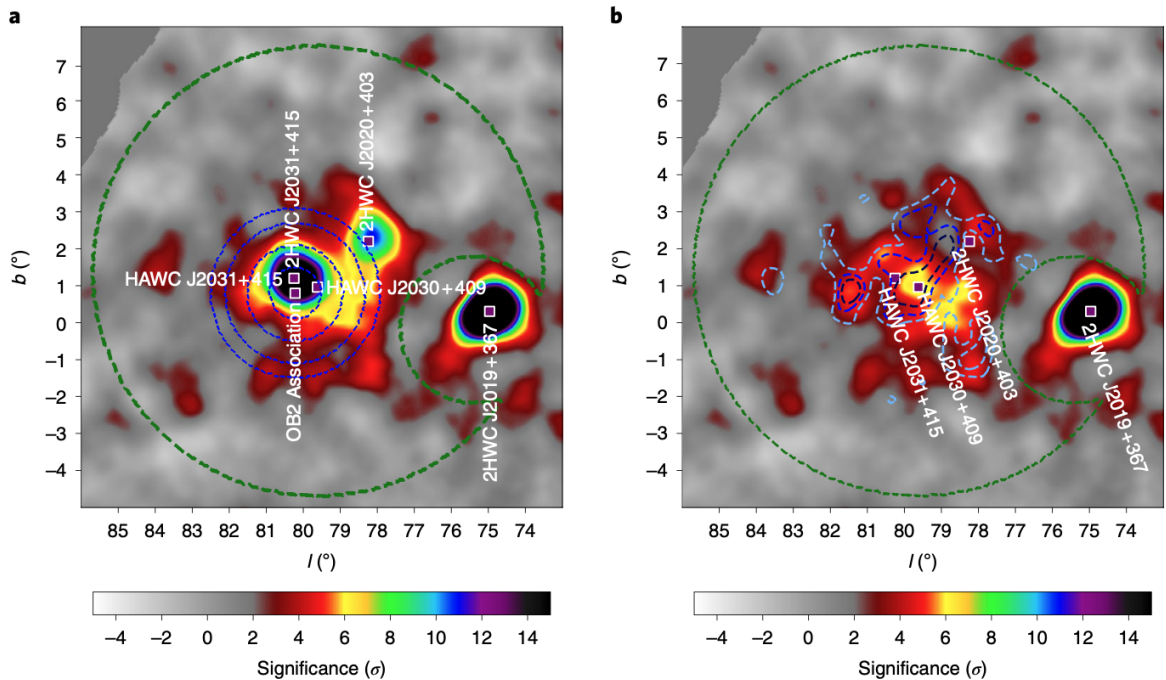

**Figure 5.** HAWC significance map of the Cocoon region adapted from Abeysekara et al. (2021) Nature Astronomy, vol. 5, p. 465-471, 2021; Springer Nature. **Left Panel:** Green contours show the region of interest masking the bright source 2HWC J2019+367 and the blue contours are centered on the OB2 star association and mark the ring-shaped regions where the cosmic-ray density was estimated (see Figure 6). **Right Panel:** Same map after subtraction of HAWC J2031+415 (PWN) and 2HWC J2020+403 (SNR). The light-shaded blue contours mark the *Fermi*-LAT emission for different flux levels (see Abeysekara et al., 2021, for further details)

which is coincident with the cocoon and recently confirmed by HAWC to be a PWN emitting at TeV energies (Alfaro & HAWC Collaboration, 2024a), and 2.36° away, 2HWC J2020+403, associated with the γCygni SNR.

Figure 5 shows the HAWC significance map of the Cocoon region obtained using 1,343 days of accumulated HAWC exposure (Abeysekara et al., 2021). Careful subtraction of the brightest sources was carried out to analyze the extended superbubble emission at TeV energies. The extension and emissivity of the HAWC source HAWC J2030+409 (see the right panel of Figure 5) were in good agreement with the *Fermi*-LAT measurements, indicating a contribution of approximately 90% of the total flux in the region of interest. HAWC sensitivity allowed the TeV spectral energy distribution (SED) of the Cocoon to be measured well above the 10 TeV limit previously known from ARGO observations (see the left panel of Figure 6).

The solid detection of such a high-energy spectrum disfavored a purely leptonic origin for the GeV and TeV emission in the Cygnus Cocoon, indicating that protons with $E > 100$ TeV must be present. Furthermore, the cosmic-ray energy densities of $10^{-2}$ eV cm$^{-3}$, calculated in concentric rings up to 55 pc from the Cygnus OB2 association (right panel in Figure 6), provided confirmation of a much higher value in this region when compared to the local cosmic ray energy density of $10^{-3}$ eV cm$^{-3}$. This study provided confirmation of particle acceleration in the Cygnus Cocoon in the 1–100 TeV energy regime, suggesting that powerful SFRs may act as PeVatrons (Abeysekara et al., 2021).

### 3.6. Galactic Diffuse Emission

Galactic diffuse gamma-ray emission originates from the interaction of cosmic rays with the ISM and radiation fields in our Galaxy. At TeV energies, this emission is primarily produced through two mechanisms: hadronic interactions, where cosmic-ray protons collide with interstellar gas, producing neutral pions that decay into gamma-rays, and leptonic processes, where high-energy electrons upscatter ambient photons through the ICS (Nayerhoda et al., 2019). The spatial and spectral characteristics of this emission provide crucial information about cosmic-ray transport and distribution throughout the Galaxy and the structure of the ISM, as well as the larger-than-expected emission of gamma-rays in the GC (De La Torre Luque et al., 2023).

Initial HAWC studies focused on developing analysis strategies for detecting and measuring diffuse emission (Hüntemeyer & Ayala Solares, 2013), expanding upon earlier Milagro observations that have reported enhanced emissions compared to theoretical predictions (Abdo et al., 2008a). Subsequent HAWC analyses refined the techniques for measuring both the spectrum and angular distribution of this emission along the Galactic Plane (Nayerhoda et al., 2019). These analyses also reported a measured gamma-ray emission along the Galactic Plane higher than the one found by the DRAGON (Diffusion of cosmic RAys in

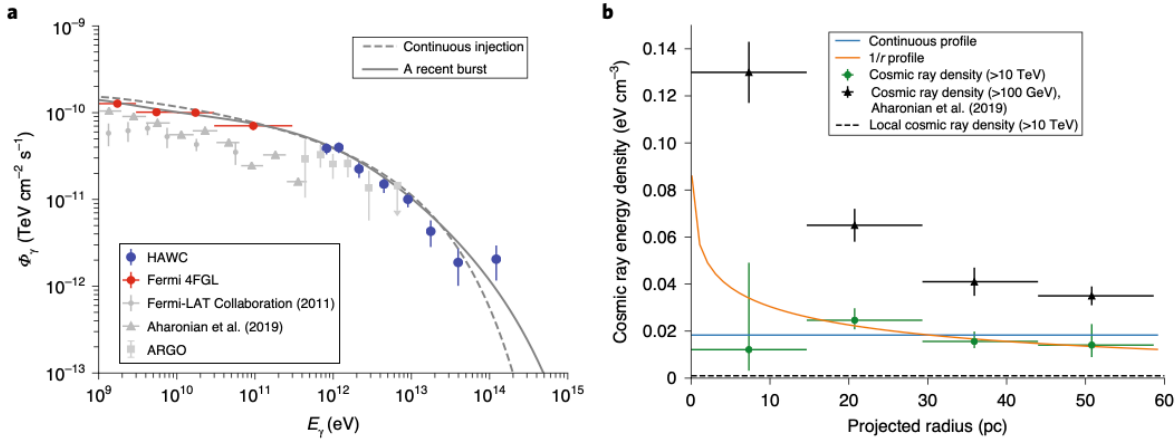

**Figure 6.** Gamma-ray spectrum and cosmic-ray density in the Cocoon region adapted from Abeysekara et al. (2021) Nature Astronomy, vol. 5, p. 465-471, 2021; Springer Nature. **Left Panel:** high-energy SED from different experiments; the HAWC data are marked in blue. The grey solid and dashed lines correspond to hadronic models applied to the data in this region. **Right Panel:** Cosmic-ray energy density profile derived for four annuli centered at the OB2 association; the green points corresponds to the HAWC estimate above 10 TeV.

Galaxy modelizatiON) base model (Evoli et al., 2008, 2017) predictions by a factor of two (Nayerhoda et al., 2019), with potential contributions from unresolved sources and mis-modeled known sources.

The latest HAWC analysis (Alfaro & HAWC Collaboration, 2024b) provides measurements of the Galactic diffuse emission between longitudes 43° and 73°, finding a spectral index of $-2.61 \pm 0.03$. The measured emission exceeded the predictions from the DRAGON cosmic-ray transport model by a factor of ∼2. This excess can be explained by contributions from unresolved sources and TeV halos around PWNe (De La Torre Luque et al., 2023).

Importantly, the fitted index is statistically indistinguishable from the slope of the local cosmic-ray spectrum above ∼300 GeV spectral hardening, as reported by *PAMELA* (Ahn et al., 2010) and AMS–02 (Aguilar et al., 2015). Detecting the same harder index after averaging over a $\Delta \ell \simeq 30°$ swath of the inner Galaxy implies that the 300 GeV break is a Galactic-scale feature rather than a peculiarity of the solar neighborhood. These new measurements require transport models to (i) invoke additional TeV-energy emitters (e.g., unresolved pulsar halos) and (ii) include a spatially extended hard component in the Galactic cosmic-ray population.

## 4. Compact Objects seen by HAWC

Accreting compact objects constitutes an important source of highly energetic radiation in the Universe. This is true in the case of SMBHs ($M_{BH} = 10^{6-9} M_\odot$) powering AGNs and in the case of stellar-mass compact objects, which are generally associated with Galactic X-ray and radio-emitting binaries, called microquasars, because of their similarities with their more massive extragalactic counterparts. Both types of sources (AGNs and microquasars) are powered by accretion processes and share typical observational properties, such as strong X-ray emission, flux variability, and multi-wavelength spectral energy distribution (SED), but only a subset of them show radio jets that are often associated with the production of gamma-ray emission (Mirabel & Rodríguez, 1999; Blandford et al., 2019). In the following section, we review the contributions of the HAWC Observatory for accreting compact objects.

### 4.1. Microquasars and Binaries

The well-known microquasar SS 433 is a binary system composed of a supergiant star whose Roche lobe is accreted by a compact object (either a neutron star or a black hole). This microquasar lies very close to the SNR W50 (less than 2° separation), with an estimated distance of approximately 40 pc between the SNR lobes and the central source of SS 433. Several prior multi-wavelength studies of SS 433 have established the presence of sub-relativistic jets of ionized matter outflowing at v = 0.26c, radio and X-ray emission, likely produced by the synchrotron mechanism of electrons accelerated in the lobe-jet interaction. Such campaigns also established super-Eddington accretion in the central object, but only the upper limits for emission at energies higher than 100 GeV were measured. The HAWC Observatory provided the very first detection of the W50-SS 433 system at TeV energies (Abeysekara & HAWC Collaboration, 2018), presenting an exquisite view of the spatially resolved lobes up to an energy of 25 TeV (left panel of Figure 7). The discovery of VHE emission in the lobes of this system implies that the gamma-rays detected by HAWC are produced by particles of significantly higher energy, opening the door for subsequent observations from other gamma-ray facilities. Detailed modeling of the HAWC emission (more in Abeysekara & HAWC Collaboration

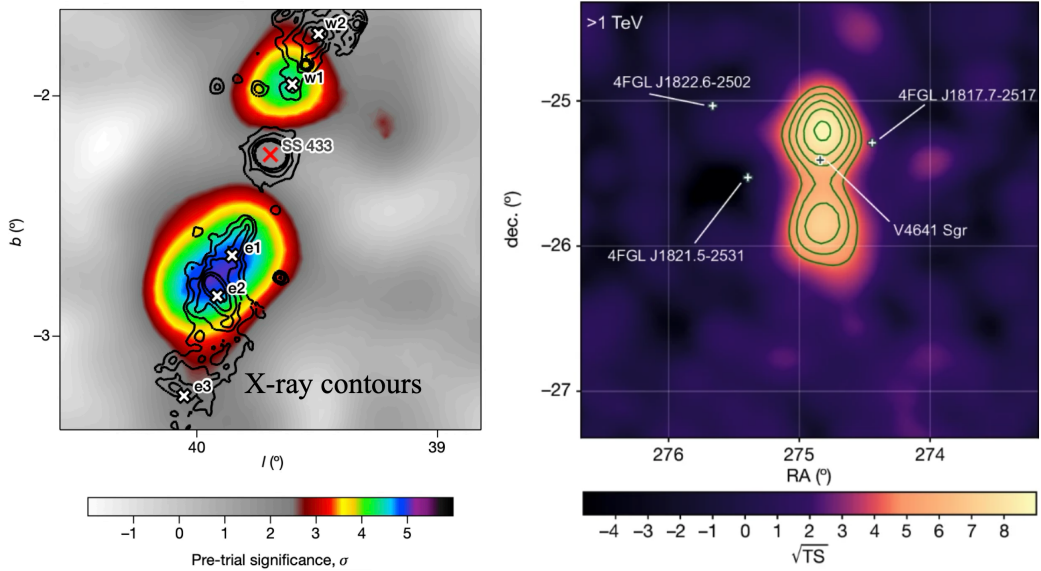

**Figure 7. Left Panel:** The HAWC TeV image of the SS 433/W50 region in Galactic coordinates, reproduced from Abeysekara & HAWC Collaboration (2018) Nature, Volume 562, Issue 7725, p.82–85, 2018, Springer Nature. The statistical significance in the color scale is estimated before accounting for statistical trials. The gamma-ray excess in SS 433 was measured after fitting and subtracting the contribution of the extended source MGRO J1908 +06. The labels e1, e2, e3, w1, and w2 mark jet termination regions observed in the X-ray, whereas the X label marks the location of the central binary. Black contours trace *Chandra* X-ray emission. **Right Panel:** HAWC significance map of the microquasar V4641 Sgr, reproduced from Alfaro & HAWC Collaboration (2024c) Nature, Volume 634, Issue 8034, pp. 557–560, 2024, Springer Nature. White crosses mark the positions of the other gamma-ray sources in the field of view.

(2018)) exhibited a clear preference for leptonic acceleration mechanisms.

That study was based on a dataset collected between November 2014 and December 2017 (1,017 days). Since then, HAWC has continued to increase in exposure and analysis methods. A longer dataset (2,565 days) with an improved reconstruction (Alfaro & HAWC Collaboration, 2024c) has allowed us to localize the gamma-ray emission in each jet lobe and identify that the best-fit positions of the gamma-ray emission are close to the X-ray emission hotspots "e1″ and "w1″ shown in Figure 7. The gamma-ray SEDs in each lobe were modeled using a power law with no cutoff and reached maximum energies of 56 and 123 TeV for the east and west regions, respectively, providing a strong indication of an efficient electron acceleration in the extended jet.

A few years after the first discovery of VHE gamma-ray emission from SS 433, HAWC detected extremely highly energetic photons in another microquasar, V4641 Sgr (Alfaro & HAWC Collaboration, 2024c), which is a binary system located slightly off the Galactic Plane and composed of black hole accreting matter from a type B main-sequence star. One of its peculiarities is its radio jet, which exhibits the fastest superluminal motion in the Milky Way. Despite this extreme behavior, gamma-ray emission from V4641 Sgr was not detected prior to HAWC observations. The data collected over a period of ∼7 years (2015–2022) yielded significant gamma-ray emissions at the position of the microquasar with a significance of $8.8\sigma$ above 1 TeV and $5.2\sigma$ above 100 TeV.

The HAWC map could either be fitted with one single extended source or with two point-like sources lying at an approximate distance of tens of pc from the central binary (see the right panel of Figure 7). Interestingly, as reported by Alfaro & HAWC Collaboration (2024c), the two gamma-ray sources show a very similar spectral shape, reaching energies higher than 200 TeV without any indication of cutoff in the spectrum, leading to a likely association of the gamma-ray emission with the microquasar lobes, which is consistent with the HAWC findings for SS 433. In this case, HAWC provided the first observational information on large-scale jets in V4641 Sgr. The dominant acceleration mechanism responsible for the persistent and high-energy photons observed in the HAWC spectrum of V4641 Sgr could not be firmly determined without additional multiwavelength information, but the recent LHAASO detection of 800 TeV photons from this source (LHAASO-Collaboration, 2024) corroborated the suggested interpretation of this microquasar as a hadronic PeVatron (see discussion in §3.5.1).

Based on the findings described above, HAWC has confirmed that large-scale jets in microquasars constitute an important source of Galactic cosmic rays. It also supports the important analogy of microquasars with their more massive extragalactic counterparts, that is, radio-loud active galaxies where powerful radio jets show strong interaction with the ambient gas, resulting in extended bubbles of radio-/X-ray-emitting gas that are not detectable in the majority of cases.

HAWC has also set limits on the production of gamma-ray emission from the central binary in microquasars (different from the emission in the extended lobes), which is expected when the accreting object has a high-mass companion star whose mass is transferred through stellar winds. Indeed, some of these so-called high-mass microquasars (HMMQ) show emissions up to GeV energies and sometimes even higher (Dubus, 2006; Zdziarski et al., 2017). Based on 1,523 days of observations, HAWC's search for TeV emission from binary objects in a sample of four HMMQ (LS 5039, Cyg X-1, Cyg X-3, and SS 433) returned no evidence for such radiation (Albert et al., 2021a). By placing a limit on the efficiency of emission above 1 TeV, the HAWC study concluded that the non-thermal mechanism postulated to explain hard X-ray/MeV emission from HMMQs is less plausible, favoring the alternative scenario of thermal Comptonization in the binary accretion disk.

Furthermore, in 2025, new HAWC observations of the gamma-ray binary LS 5039 showed not only that its spectrum extends up to 200 TeV but also that the spectral emission up to 118 TeV is modulated by the orbital motion of the binary system (Alfaro & HAWC-COLLABORATION, 2025). This novel result is consistent with gamma-ray photons produced close to the stellar system, where orbital effects are imprinted, as observed by HAWC.

### 4.2. HAWC Active Galactic Nuclei Survey

At the center of any active galaxy lies a very energetic core (AGN), whose emission often dominates over the integrated light of the entire galaxy. As mentioned at the beginning of §4, these regions exhibit intense emission over the entire electromagnetic spectrum (broadband SED), rapid variability, and, in 10% of the cases, the presence of radio outflow known as jets (Padovani et al., 2017). The mechanism behind the nuclear emission is accretion onto the SMBH, forming an accretion disk. Active galactic nuclei with strong magnetic fields and/or rotating black holes exhibit relativistic jets and radio lobes extending to the Mpc scale. Jets are sometimes also detected in the optical and X-ray bands and are generally expelled perpendicularly with respect to the accretion-disk plane. The two AGN types that have been observed emitting up to TeV are blazars and radio galaxies, although the latter are not as numerous. On the other hand, blazars dominate the extragalactic gamma-ray (Ajello et al., 2017) and the TeV sky.

Among AGNs, both blazars and radio galaxies have jets pointing at relatively small angles from the observer's line of sight, with radio galaxies typically showing larger angles than blazars. Indeed, the jet in blazars points directly at or at angles $\theta < 10°$ from the observer's line of sight. Whereas for radio galaxies, the angles are $\theta > 10°$. TeV AGNs present radiative emission across a broad spectrum, from the radio band up to gamma-rays in GeV and TeV ranges, demonstrating a characteristic double-peaked SED. Among blazars, sources whose emission reaches the TeV energy range, some are designated as "high-synchrotron-peak (HSP)" blazars. For those sources, their synchrotron emission peaks in the X-ray band and their second-component peaks are found at TeV energies. This makes AGNs, especially blazars, the main scientific interest for surveying and monitoring with HAWC. With its high duty cycle of approximately 95% and wide FoV (∼2 sr), HAWC is particularly well-suited for studying these sources over a long time.

Albert et al. (2021b) reported the first survey of AGNs transiting within HAWC's FoV over a period of 4.5 years. The survey was designed to carry out a systematic search for TeV emission in 138 AGNs selected from the third catalog of hard *Fermi*-LAT sources (3FHL) with redshifts ≤ 0.3 and at declinations within 19°± 40°. A search for the TeV counterparts of the 138 3FHL sources was carried out by performing a maximum-likelihood test and assuming a point source at the 3FHL position. The convolution with HAWC's response functions was performed using a power-law model and considering the extragalactic background light (EBL) absorption (with the three most common EBL models; see details in Albert et al. (2021b)).

The HAWC detections resulting from that work are dominated by the highly significant blazars Mrk 421 and Mrk 501 ($65\sigma$ and $17\sigma$, respectively). A few more AGNs showed marginal significance: the radio galaxy M87 and the blazars VER J0521+211 and 1ES 1215+303 were detected slightly above $3\sigma$, whereas the collective emission, excluding Mrk 421 and Mkr 501, showed a significance of $4.2\sigma$. The HAWC AGN survey provided the first information on persistent TeV emission from a large sample of gamma-ray-selected AGN. As HAWC's data collection continues to increase, more AGNs have been detected thanks to deeper exposure and improved data reconstruction, as reported in Ureña-Mena et al. (2023), who replicated the AGN survey described above using a data set of 5.7 years. In addition to solid detections for Mrk 421, Mrk 501, M87, and 1ES 1215+303, they found eight AGNs (seven blazars and one radio galaxy) with marginal detections, confirming the potential of HAWC as a survey observatory for quiescent VHE emissions from AGNs.

More detailed studies on longer temporal and spectral monitoring of the brightest AGNs are described in the following sections.

### 4.3. Long-Term Monitoring of Blazars: Spectra and Light Curves of Mrk 421 and Mrk 501

Markarian 421 and Markarian 501 are two of the most luminous blazars in the extragalactic sky and are located at redshifts of 0.031 and 0.034, respectively. These sources have been thoroughly studied across multiple energy bands and are generally described by a Synchrotron Self-Compton (SSC) leptonic model, which accounts for most of their emission in a majority of cases. However, certain observations, such as orphan flares and outliers in previous multiwavelength comparisons, indicate the presence of additional emission mechanisms contributing to the VHE emission (Błażejowski et al., 2005; Acciari et al., 2009).

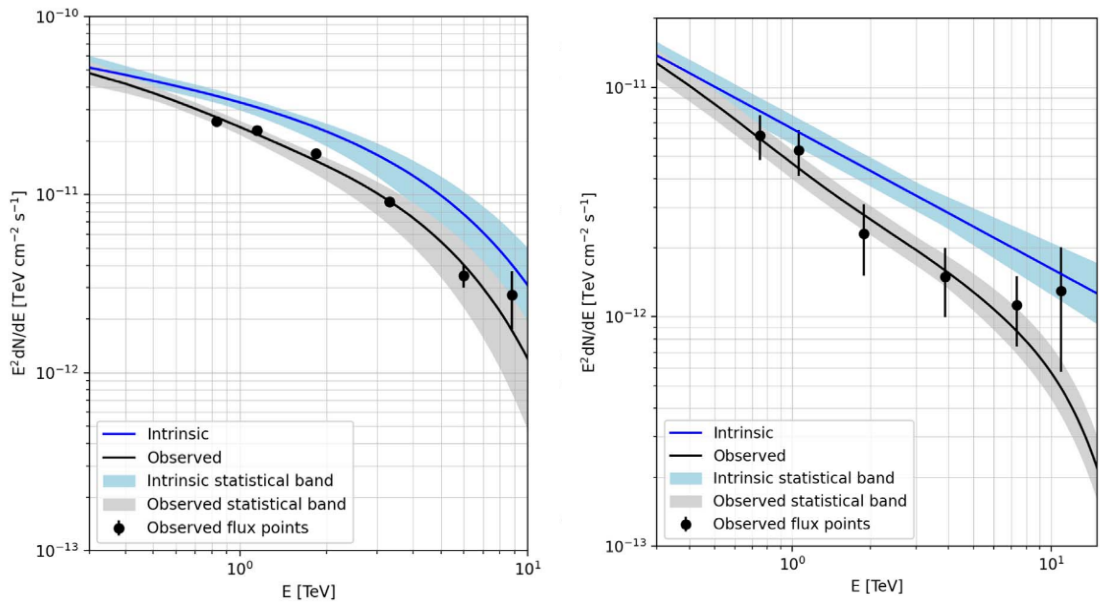

**Figure 8.** HAWC spectra of the blazars Mrk 421 (**left**) and Mrk 501 (**right**), ©AAS. Reproduced with permission from Albert et al. (2022a). Black circles indicate the observed flux points. The blue and black lines mark, respectively, the intrinsic and observed spectra with their statistical uncertainties in corresponding colors.

HAWC performed a time-averaged analysis of the TeV SED for both Mrk 421 and Mrk 501 using a data set comprising 1,038 days (Albert et al., 2022a). This study revealed that the intrinsic spectrum of Mrk 421 can be represented by a power law with an exponential cutoff (see the left side of Figure 8). Additionally, this analysis concluded that the maximum energy at which the source was detected is 9 TeV. In the case of Mrk 501, its intrinsic spectra can be represented by a simple power law, and the maximum energy at which the source was detected is 12 TeV (see the right side of Figure 8).

Both Mrk 421 and Mrk 501 are constantly monitored by HAWC; this continuous and unbiased monitoring can be used in multi-wavelength studies. The first HAWC analysis of these sources utilized 511days of data and explored the possibility of an X-ray/gamma-ray correlation using *Swift*-BAT data. There is no conclusive evidence for such a correlation (Abeysekara et al., 2017b).

Later, Alfaro et al. (2025) presented a multi-wavelength study using HAWC and *Swift*-XRT contemporary observations covering the period from 2015 to 2020. HAWC data, in an energy range from 300 GeV to 100 TeV, detected both low and high emission states, particularly for Mrk 421, as shown in the light curve in Figure 9. These light curves were compared with the corresponding light curves from *Swift*-XRT in the 0.3 to 10 keV energy range. Determining the correlation between the two observed bands can provide further understanding of the emission mechanisms at very high energies detected by HAWC.

Indeed, under the prescription of the SSC model, an X-ray/gamma-ray correlation is expected. For Mrk 421, a linear dependence was found in Alfaro et al. (2025), suggesting that a multiple-zone SSC model may be needed to explain the X-ray and gamma-ray emissions. This work also effectively measured the harder-when-brighter behavior of gamma-ray emission.

### 4.4. The HAWC View of Radio galaxies: M87 and More

M87 is a radio galaxy located at the center of the Virgo cluster at a distance of approximately 16 Mpc (Mei et al., 2007). Its closeness and the presumed larger angle between the jet and the observer's line of sight compared to blazars have enabled various studies. These include the observation of the shadow of its SMBH, M87*, as well as measurements of its mass, spin, and magnetic field (Collaboration et al., 2019; Event Horizon Telescope Collaboration et al., 2021). In addition, optical observations of HST-1 indicate a complex jet structure (Raue et al., 2012).

M87 was the first radio galaxy observed at energies >750 GeV in 1999 by HEGRA (Aharonian et al., 2003). Since then, imaging atmospheric Cherenkov telescopes (IACTs) such as MAGIC, H.E.S.S., and VERITAS (MAGIC Collaboration et al., 2020; H. E. S. S. Collaboration et al., 2024; Beilicke & VERITAS Collaboration, 2012), as well as water Cherenkov observatories, such as HAWC, have been monitoring its activity. Throughout the years, M87 has exhibited both low and flaring states, with rapid variability, implying compact emission regions (Beilicke et al., 2007). However, the exact location of the emission region and emission mechanism at these energies are not completely known.

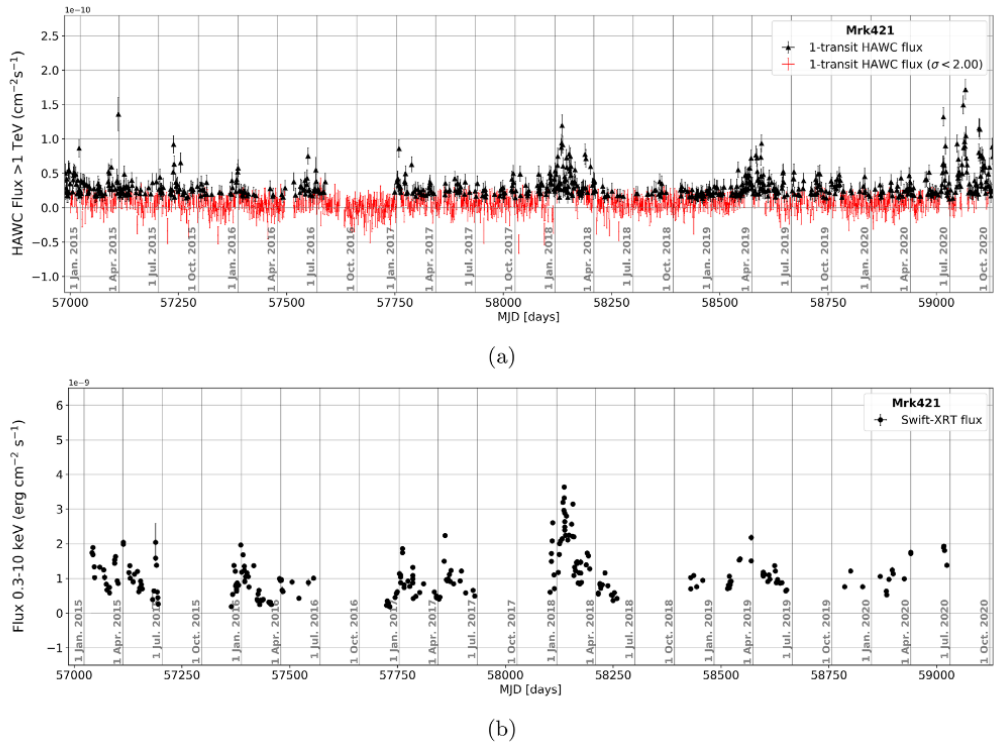

**Figure 9.** HAWC (E > 1 TeV, **top**) and *Swift*-XRT (E = 0.3–10 keV, **bottom**) long term monitoring of Mrk 421 showing correlated flaring periods over ∼6 years, ©AAS. Reproduced with permission from Alfaro et al. (2025). The red points mark HAWC transitions with flux >1 TeV below the 2$\sigma$ threshold.

M87 is the closest radio galaxy in HAWC's FoV and therefore it has been continuously monitored since the beginning of HAWC's operation in 2014. Alfaro & HAWC-Collaboration (2022) presents the results of this monitoring from 2014 to 2019, reporting a marginal evidence of VHE emission. HAWC monitoring provided the first view of M87 at energies above 1 TeV integrated over ∼5 years, showing that this average emission is comparable to the measurements obtained by the observatories MAGIC, H.E.S.S., and VERITAS during low states of activity. Alfaro & HAWC-Collaboration (2022) also reported the modeling of the broadband SED, which can be considered a good description of the quiescent activity of this radio galaxy. A comparison of the HAWC limits with the various available VHE observations from other facilities included in that study showed that the SSC leptonic model alone is insufficient to describe the multi-wavelength SED; hence, a lepto-hadronic model was explored to explain the overall SED. The leptonic model yielded a good fit of the low-energy bands, while the hadronic model explained the VHE emission in the quiescent state observed by HAWC, MAGIC, and H.E.S.S., but also in an H.E.S.S. orphan flare observed in 2005. Continued monitoring of M87 using HAWC will improve the significance of detection, as shown in Capistrán et al. (2021). Because the significance increases with exposure time, better constraints on the nature of the source VHE emission are obtained.

In fact, (Alfaro et al., 2025), reports the continuous TeV monitoring of four radio galaxies over 7.5 years with HAWC detection thresholds for M87 (>5$\sigma$), 3C 264 (slightly above 3$\sigma$) and IC 310 and NGC 1275 (both below 2$\sigma$). The light curves and cumulative significance reported in this work show that the long-term behavior of these four radio galaxies is similar to the variability pattern of the blazar Mrk 421, possibly implying common emission mechanisms in their jets.

## 5. Transient Events with HAWC

### 5.1. HAWC Limits on Gamma-Ray Bursts

Gamma-Ray Bursts (GRBs) are among the most energetic phenomena observed in the Universe. They are typically categorized into two distinct classes, based on their duration: short GRBs ($T_{90} < 2$ s) and long GRBs ($T_{90} > 2$ s) (Kouveliotou et al., 1993). Short GRBs are generally associated with mergers of compact objects, such as neutron star-neutron stars (NS-NS) or neutron star-black hole (NS-BH) systems (Abbott et al., 2017a; Berger, 2014), whereas long GRBs are thought to result from the core collapse of massive stars, commonly associated with Type Ic supernovae (Woosley & Bloom, 2006).

The high-energy spectra of GRBs typically exhibit two dominant emission components. The first is attributed to synchrotron radiation, which is emitted by relativistic electrons that are accelerated within the jet structure. This mechanism can account for photon energies extending into the GeV regime and is a widely accepted model for prompt and afterglow emission phases (Zhang, 2011). The second major component involves the SSC process, in which synchrotron photons are upscattered to higher energies by the same population of electrons responsible for their production (Sari & Esin, 2001; Fan & Piran, 2008). This inverse Compton-scattering process can extend the spectral range to TeV energies.

Until recently, GRBs have not been detected in the TeV energy range. However, in the last decade, some GRBs—GRB 180720B (Abdalla et al., 2019), GRB 190114C (MAGIC Collaboration et al., 2019), GRB 190829A (H.E.S.S. Collaboration et al., 2021), and GRB 221009A (Kennea et al., 2022; Lesage et al., 2022; Huang et al., 2022)—have been observed at TeV energies. Notably, GRB 221009A, dubbed the Brightest of All Time (BOAT), has challenged existing models of GRB emission and underscored the importance of searching for VHE emission in these events; see, e.g., Li & Ma (2024); González et al. (2023); Alves Batista (2022).

The HAWC Observatory performs real-time all-sky monitoring for GRBs using a blind search designed to detect the electromagnetic counterparts of GRBs at TeV energies. In addition, HAWC features an alert system that enables the follow-up of GRBs from their earliest moments, providing valuable insights into the physics of GRBs at early times and in the VHE regime.

HAWC analysis of GRBs has provided valuable insights into the physical models of VHE emission. For example, Abeysekara et al. (2015) analyzed GRB 130427A using data from an early phase of the observatory when only 10% of the final detector array was operational. At that time, GRB 130427A was the most powerful GRB ever detected, and Abeysekara et al. (2015) derived an upper limit for its VHE emission. Another important study focused on a sample of GRBs detected by the *Fermi* Gamma-Ray Space Telescope and the Neil Gehrel *Swift* Observatory, which were within HAWC's FoV at their reported trigger times. Alfaro & HAWC Collaboration (2017) analyzed a set of 64 GRBs that met this criterion and derived flux ULs for them. In particular, the flux UL for GRB 170206A, one of the brightest GRBs detected by *Fermi*, ruled out the presence of a strong SSC component, highlighting the crucial role of the upper limits in constraining physical models. More recently, the HAWC Collaboration performed an analysis focused exclusively on short GRBs. Using a sample of 20 short GRBs—two detected by *Fermi* and 18 by *Swift*, all within HAWC's FoV at their reported trigger times—Albert et al. (2022b) derived flux ULs and placed constraints on the microphysical parameters of the SSC forward-shock model. Assuming a redshift of $z = 0.3$ for the GRBs with the highest fluence at keV energies, they obtained an interstellar density as low as $10^{-2}$ cm$^{-3}$.

In the case of GRB 221009A, the GRB occurred outside of the HAWC FoV. Eight hours after the prompt emission, the position of the GRB started to transit the HAWC detector, and after it finished, we performed a search and calculated a limit on the flux (Ayala & HAWC Collaboration, 2022). Interestingly, during the HAWC observations, the *Fermi* telescope observed a 400 GeV photon in the direction of the GRB (Axelsson et al., 2025). Our observations constrain a possible re-flare of the GRB and are consistent with a smooth decay in the GRB emission.

### 5.2. HAWC contribution to Gravitational Waves

Multi-messenger astronomy was initiated by the identification of gravitational waves (GWs), which encompass a broad spectrum of astrophysical entities including electromagnetic signals, neutrino observations, and GW signatures. The LIGO interferometers, European interferometer Virgo, and KAGRA interferometer are the foundations of the burgeoning field of GW astronomy. These interferometers have carried out four observation runs since the beginning: observation runs 1, 2, 3, and 4 (O1, O2, O3, and O4, respectively). The historic detection of GW170817 (Abbott et al., 2017b), the merger of binary neutron stars (BNS) that fell within Run O2. Gravitational-wave multi-messenger astronomy, as a novel and highly prospective field of astrophysics, was unequivocally confirmed by the observations of GRB 170817A and the subsequent kilonova emission across the electromagnetic spectrum. The GW170917/GRB 170817A event was also within the FoV of the HAWC Observatory. Observations with the HAWC Observatory began on August 17$^{\text{th}}$, 2017 at 20:53 UTC and ended 2.03 h later (Abbott et al., 2017b). Although no significant excess was detected, upper limits were placed for energies larger than 40 TeV.

## 6. Cosmic Rays

Although HAWC was originally designed to study gamma-rays, over its ten years of operation it has proven to be an excellent detector of TeV cosmic rays. HAWC has contributed significantly to the measurement of the all-particle energy spectrum (Alfaro et al., 2017, 2025), the elemental spectrum of the light mass group (H+He) (Albert et al., 2022c), and to the characterization of both large- and small-scale anisotropies in the arrival directions of cosmic rays (Abeysekara et al., 2014; Abeysekara et al., 2018a, 2019). These studies are particularly valuable because they probe an energy range where few precise measurements exist and where direct and indirect detection techniques overlap (Mollerach & Roulet, 2018; Tomassetti, 2023). Moreover, some HAWC results revealed unexpected features in the cosmic-ray flux, potentially challenging the standard models of cosmic-ray production and propagation in our galaxy. These analyses have also demonstrated the capability of high-altitude water Cherenkov detectors (WCDs) to perform precise TeV cosmic ray measurements, a key advantage that will be further leveraged by the future Southern Wide-field Gamma-ray Observatory (SWGO) observatory (Albert et al., 2019; Abreu et al., 2025).

At the latitude at which HAWC is located, the Moon crosses the FoV of the experiment and obstructs the flux of cosmic rays that reach the Earth. This produces a shadow in the cosmic-ray skymaps in the direction of our satellite (Alfaro et al., 2024b), which provides independent information about the cosmic-ray energy and angular resolution of the instrument (Alfaro et al., 2017), and can even be exploited to limit the presence of antiprotons in the cosmic-ray flux at TeV energies (Abeysekara et al., 2018b).

A shadow in the flux of Galactic cosmic rays can also be observed due to the presence of the Sun. However, this shadow is modulated by the solar activity due to the time evolution of the magnetic field in the solar atmosphere. Using HAWC data on Galactic TeV cosmic rays, we can also investigate this dependence. This was performed as a function of the number of sunspots and the magnitude of the photospheric magnetic field at different latitudes during a complete solar cycle (Alfaro et al., 2024b). HAWC results imply that the Sun shadow can be used to explore the magnetic fields of the low solar atmosphere. The interactions of Galactic cosmic rays with the atmosphere of the Sun produce a TeV gamma-ray emission, which was discovered with HAWC in the 0.5–2.6 TeV range with a dedicated analysis published in Albert et al. (2023a). By studying the magnitude of this emission and its dependence on solar activity, clues can be obtained to improve our current understanding of the interactions of Galactic cosmic rays with the solar atmosphere, which we know is not complete, as hinted at by the differences found between the model predictions and the HAWC data in Albert et al. (2023a).

Collisions of cosmic rays with the atmosphere create air showers of secondary particles that travel down to the ground (Engel et al., 2011). Through this mechanism, the contribution of all cosmic-ray interactions in the atmosphere produces a steady flux of secondary particles on the Earth's surface that is sensitive to space weather (Miroshnichenko, 2015) and atmospheric effects (Myssowsky & Tuwim, 1926; Steinke, 1930; Dorman, 2004). In HAWC, analyses of the variations in the counting rate of these secondary particles have allowed us to study the activity of the Sun and the propagation of solar magnetic perturbations in the interplanetary medium (Akiyama et al., 2020; Alvarez et al., 2021; Alfaro et al., 2024b). Unexpectedly, HAWC was also able to detect the local atmospheric perturbations caused by the passage of the shock waves generated by the explosion of the Hunga Tonga-Hunga Ha'apai volcano at the beginning of 2022 (Alfaro & HAWC-Collaboration, 2024).

Muons are among the most penetrating components of the secondary flux of cosmic rays. Once produced in the atmosphere, they travel in straight lines to the ground, carrying information on hadronic interactions (Engel et al., 2011) and atmosphere (Blackett, 1938; Forro, 1947; Taricco et al., 2022). The relative abundance of the muon component in the flux of secondary cosmic rays increases with zenith angle due to the atmospheric absorption of the nuclear and electromagnetic components (Cazón et al., 2004; Collaboration, 2014). The tracks of horizontal-going muons can be detected in HAWC thanks to its modular and compact design, as well as its large lateral detector area. Using the rock of the nearby Pico de Orizaba volcano and horizontal muons, searches for Earth-skimming muon neutrinos have also been implemented at HAWC (Albert et al., 2022d). In the following subsections, we summarize some of the aforementioned air-shower and cosmic-ray studies, as well as their main results.

### 6.1. All-Particle Spectrum of Cosmic Rays

Cosmic rays with energies from 10 TeV to 1 PeV offer an interesting window for high-energy astrophysical phenomena in our Galaxy (Moskalenko, 2023). However, precise measurements in this energy regime are not straightforward, because of the loss of exposure to direct and indirect cosmic ray experiments in this energy interval (Mollerach & Roulet, 2018; Tomassetti, 2023). For direct/indirect detectors, this typically occurs above/below a few hundreds of TeV. For several years, this situation has hampered detailed studies of the energy spectrum and mass composition of cosmic rays at TeV energies and constrained progress in the investigation of the origin and physics of these energetic particles. However, recent progress in cosmic-ray detection has changed this situation (Tomassetti, 2023). In this regard, one of the relevant achievements of HAWC has been to provide precise measurements of the spectrum and composition of cosmic rays in the TeV energy regime.

In Alfaro et al. (2017), HAWC published its first measurement of the total spectrum of cosmic rays, which covered the energy range from 10 to 500 TeV. This is an important achievement for several reasons. First, because it provided the first high-precision data on the all-particle spectrum in the 10–100 TeV interval; second, because it was able to bridge the gap between direct and indirect cosmic-ray experiments using precision data; and finally, because it found the presence of a softening in the total spectrum of cosmic rays at a few tens of TeV. This feature, previously hinted at by data from the NUCLEON satellite (Panov et al., 2017), was able to be confirmed by HAWC. The analysis was carried out using 7.7 months of extensive-air-shower (EAS) data. The primary energy of the shower events was estimated with a maximum-likelihood procedure using Monte Carlo (MC) predictions for pure protons based on the FLUKA (Ferrari et al., 2005) and QGSJET-II-03 (Ostapchenko, 2006a,b) hadronic interaction models, as well as the lateral distributions of the effective charge recorded by the detectors. Detector resolution and composition effects were take into account by means of an unfolding procedure (D'Agostini, 1995). Although the primary energy estimation shows a dependence on the primary composition, a general agreement with direct data and EAS measurements was found between 10 TeV and 500 TeV.

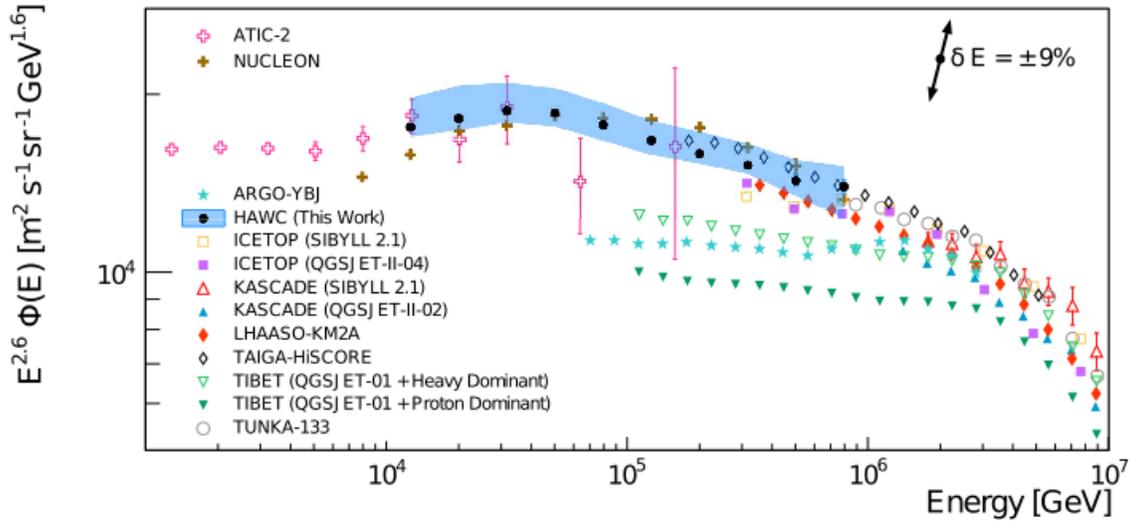

**Figure 10.** Updated total spectrum of cosmic rays measured with the HAWC observatory multiplied by an $E^{-2.6}$ energy scale (black circles). The spectrum was compared with various cosmic-ray measurements obtained from direct and indirect experiments. We included data from ATIC-2 (open pink crosses; Panov et al., 2009), NUCLEON (brown crosses; Grebenyuk et al., 2019), ARGO-YBJ (light blue stars; D'Amone et al., 2015), ICETOP (yellow open squares and violet squares; Aartsen et al., 2020), KASCADE (upward open and solid triangles), LHAASO-KM2A (red solid diamonds; Cao et al., 2024), TAIGA-HiScore (open black diamonds; Prosin et al., 2019), TIBET-III (downward open and solid triangles; Amenomori et al., 2008), and TUNKA-133 (open gray circles; Prosin et al., 2014). The error band represents the systematic error of HAWC data points. The error bars correspond to statistical errors. Figure from Alfaro et al. (2025).

An updated result for HAWC on the all-particle energy spectrum was recently released (Alfaro et al., 2025). The measured spectrum now extends up to 1 PeV, covering two decades of energy and reaching the domain of PeV air-shower experiments. The study was based on a large dataset collected during an effective time of 5.3 years. This analysis was performed using the modern QGSJET-II-04 high-energy hadronic interaction model (Ostapchenko, 2011). An improvement to this study, in comparison with the previous one (Alfaro et al., 2017), was to reduce the systematic errors caused by the modeling of the detector. The updated measurement of HAWC on the all-particle energy spectrum of cosmic rays is shown in Figure 10, where it is compared with the data from direct detectors and other air-shower experiments. From this figure, we observe that the spectrum of cosmic rays as measured by HAWC is in agreement with several direct detectors, in particular with ATIC-2 (between 10 and 200 TeV; Panov et al. (2009)) and NUCLEON (from 20 TeV to 1 PeV; Grebenyuk et al. (2019)). Above a few hundred TeV, the HAWC result is also consistent with different EAS experiments, such as TUNKA-133 (Prosin et al., 2014), TAIGA-HiScore (Prosin et al., 2019), ICETOP (Aartsen et al., 2020), and LHAASO-KM2A (Cao et al., 2024), but this is in disagreement with TIBET-III (Amenomori et al., 2008) and ARGO-YBJ (D'Amone et al., 2015). Softening reported by HAWC in 2017 at tens of TeV was also observed in the new analysis, as shown in Figure 10. Here, the break is observed at $40.2 \pm 0.1(\text{stat.}) \pm^{6.2}_{6.4}(\text{syst.})$ TeV. The spectral index before the cutoff was found to be $\gamma = -2.53 \pm 0.01(\text{stat.}) \pm^{0.04}_{0.05}(\text{syst.})$, and after, $\gamma = -2.71 \pm 0.01(\text{stat.}) \pm^{0.03}_{0.04}(\text{syst.})$. The feature was measured with statistical significance greater than $5\sigma$.

### 6.2. Cosmic-Ray Composition

The capability to study the mass composition of cosmic rays with HAWC was demonstrated in a study by Albert et al. (2022c), in which the energy spectrum of cosmic-ray light nuclei (H+He) was determined with high precision and statistics (see Figure 11). The spectrum was measured from 6 to 158 TeV and obtained from a subsample of EAS data, which was dominated by shower events produced by H and He nuclei. The subsample was selected by applying a cut on the lateral shower age as a function of the estimated primary energy derived from MC simulations. The relative abundance of light nuclei in the selected dataset is greater than 82%. Unfolding techniques were applied to correct the spectrum for a limited experimental resolution. The spectrum was also corrected for the contamination of events induced by heavy nuclei in the subsample. The main analysis was performed in the framework of the QGSJET-II-04 hadronic interaction model and showed the existence of softening in the spectrum of the H+He mass group of cosmic rays at $24^{+3.6}_{-3.1}$ TeV with a statistical significance of $4.1\sigma$. The cutoff was produced by a decrease in the spectral index from $\gamma = -2.51 \pm 0.02$ to $\gamma = -2.83 \pm 0.02$. Softening was also observed when using the EPOS-LHC high-energy hadronic interaction model (Pierog et al., 2015). Hints about the existence of this structure were first reported by ATIC-2 (Panov et al., 2009), CREAM (Yoon et al., 2017), and NUCLEON (Panov et al., 2017). Nowadays, DAMPE (Alemanno et al., 2024) also confirmed such a structure, which seems to be produced by individual cutoffs in the spectra of the H and He nuclei at tens of TeV, as pointed out by DAMPE (An et al., 2019; Alemanno et al., 2021), CALET (Adriani et al., 2022, 2023), and updated NUCLEON (Grebenyuk et al., 2019) data. As shown in Figure 11, the H+He elemental spectrum reported by HAWC is in full agreement with the results from DAMPE (Alemanno et al., 2024) within the 6–158 TeV interval, and with those from NUCLEON (Grebenyuk et al., 2019) between 20 and 126 TeV; however, it is higher than the JACEE spectrum (Takahashi, 1998). Within systematic uncertainties, the HAWC measurements are also in agreement with the results of ATIC-2 (Panov et al., 2009) and CREAM (Yoon et al., 2017) at tens of TeV. Close to 100 TeV, the HAWC spectrum is in agreement with the Tibet AS$\gamma$ (Amenomori et al., 2019b) and ARGO-YBJ (Bartoli et al., 2012) data, but it does not confirm the plain behavior reported by ARGO-YBJ from 10 to 100 TeV. The EAS-TOP data at approximately 80 TeV (Aglietta et al., 2004) are compatible with the HAWC results within systematic uncertainties.

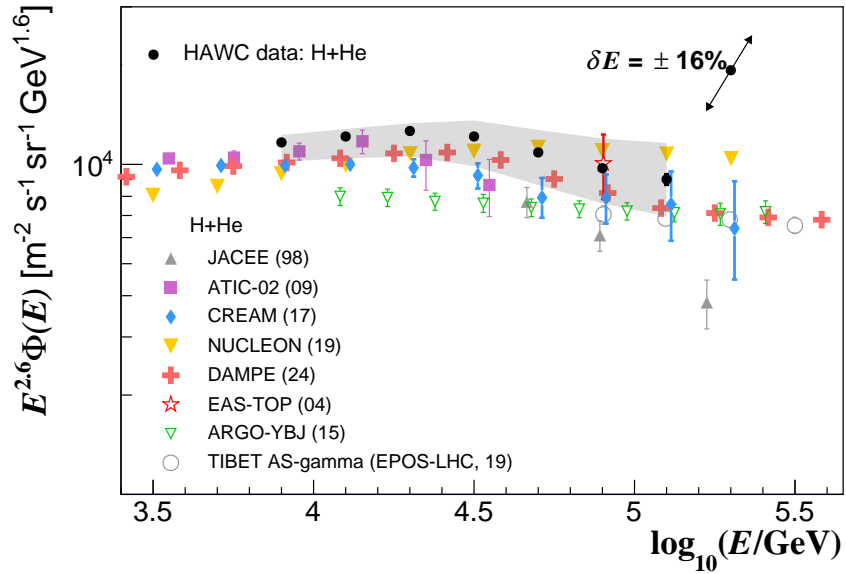

**Figure 11.** H+He spectrum of cosmic rays measured with the HAWC observatory multiplied by an $E^{-2.6}$ energy scale (black circles). The spectrum is compared with various cosmic-ray measurements from direct and indirect experiments: JACEE (gray upward triangles; Takahashi, 1998), ATIC-2 (purple squares; Panov et al., 2009), CREAM (blue diamonds; Yoon et al., 2017), NUCLEON (downward orange triangles; Grebenyuk et al., 2019), DAMPE (red crosses; Alemanno et al., 2024), EAS-TOP (open red stars; Aglietta et al., 2004), ARGO-YBJ (downward green open triangles; Bartoli et al., 2012), and Tibet AS$\gamma$ (open gray circles; Amenomori et al., 2019b). The error band represents the systematic error of HAWC measurements. Error bars indicate statistical errors. Figure updated with permission from Albert et al. (2022c) Physical Review D, Volume 105, Issue 6, article id.063021 Copyright (2022) by the American Physical Society.

### 6.3. Cosmic-Ray Anisotropies

Ground-based experiments in both the Northern and Southern Hemispheres have detected significant variations in the arrival direction distribution of TeV to PeV cosmic rays, across both large (≥60°) and medium angular scales (<60°) (Nagashima et al., 1998; Hall et al., 1999; Amenomori et al., 2005, 2006; Guillian et al., 2007; Abdo et al., 2008b, 2009; Aglietta et al., 2009; Munakata et al., 2010; Abbasi et al., 2010, 2011; De Jong, 2011; Abbasi et al., 2012; Aartsen et al., 2013; Bartoli et al., 2013; Abeysekara et al., 2014; Bartoli et al., 2015; Aartsen et al., 2016; Amenomori et al., 2017; Bartoli et al., 2018; Abeysekara et al., 2018c). The large-scale anisotropy structures have an observed amplitude of $\sim 10^{-3}$ (see, for example, Figure 12), while the small-scale structures have an amplitude $10^{-4}$. The HAWC Observatory has analyzed the cosmic ray sky in the TeV range, covering energies from 2.0 to 72.8 TeV (Abeysekara et al., 2018a, 2019).

#### 6.3.1. Energy-Dependent Anisotropy

Similar to other experiments in the Northern and Southern Hemispheres, HAWC observes an anisotropy whose arrival direction distribution depends on the cosmic-ray energy, as shown in Figure 13 (Abeysekara et al., 2018a). This anisotropy is dominated by a dipolar moment with a phase at RA $\alpha \sim 40°$, the amplitude of which gradually increases from $8 \times 10^{-4}$ at 2 TeV to $14 \times 10^{-4}$ at approximately 30 TeV before decreasing above this energy. A significant large-scale signal was also detected in the quadrupolar, octupolar, and smaller angular-scale moments, consistent with previous observations, but with better energy resolution and fitting precision.

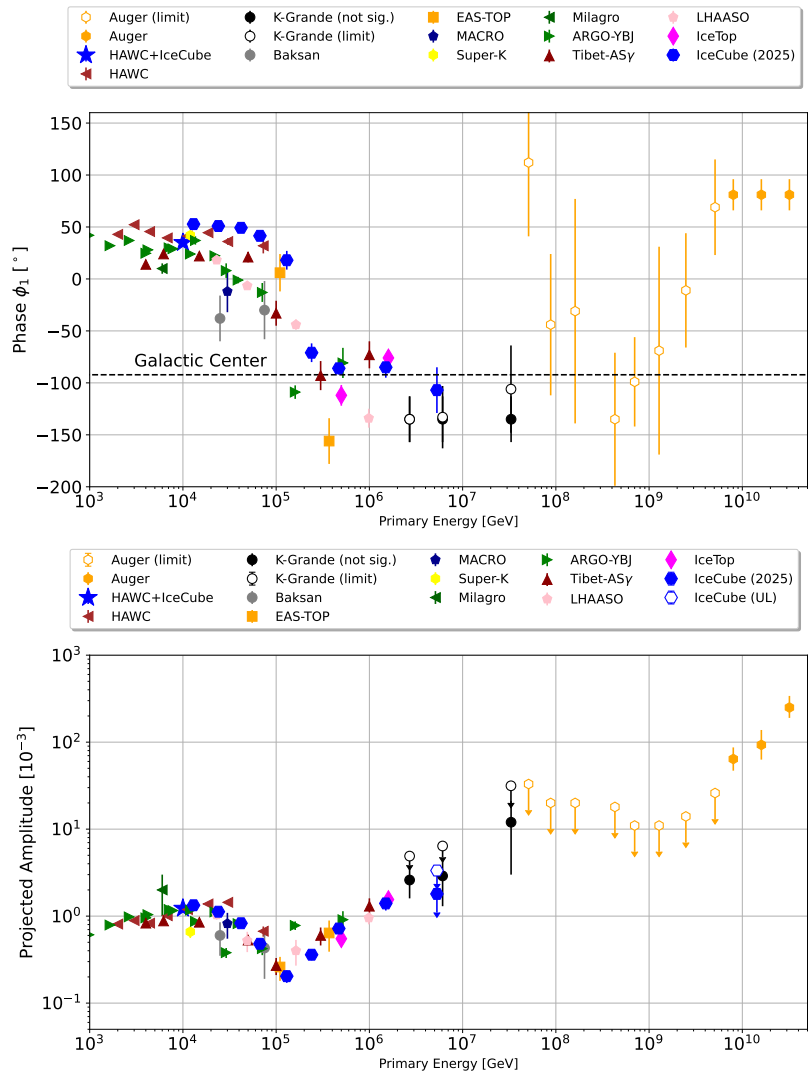

**Figure 12.** Reconstructed dipole component amplitude (**bottom panel**) and phase (**top panel**) from HAWC along previously published TeV–PeV results from other experiments (Abbasi et al., 2025). The results shown are from Abeysekara et al. (2018c); Chiavassa et al. (2015); Alekseenko et al. (2009); Aglietta et al. (2009); Ambrosio et al. (2003); Guillian et al. (2007); Abdo et al. (2009); Bartoli et al. (2015); Amenomori et al. (2005); and Aartsen et al. (2013, 2016).

#### 6.3.2. All-Sky Anisotropy

One limitation of this and previous measurements from other experiments is that their limited FoV makes it difficult to precisely characterize anisotropy in terms of spherical harmonic components. This complicates the quantitative study of its features, such as the dipolar or quadrupolar component, as correlations between the multipolar spherical harmonic terms $a_{\ell m}$ skew the interpretation of cosmic-ray distributions.

To overcome such limitations, HAWC conducted a joint analysis with the IceCube Neutrino Observatory, combining data from both experiments, and published the first analysis of the full celestial sphere with a median primary energy of 10 TeV (Abeysekara et al., 2019). The sky map, composed of data covering both hemispheres (cf. Figure 14) and its corresponding angular power spectrum (see Figure 15) helps mitigate the biases resulting from partial sky coverage, thus providing a crucial tool for investigating the propagation of cosmic rays through the local ISM and their interactions with interstellar and heliospheric magnetic fields. This analysis enabled the measurement of the horizontal dipolar components of the anisotropy, obtaining amplitudes of $\delta 0h = 9.16 \times 10^{-4}$ and $\delta 6h = 7.25 \times 10^{-4}$, respectively. Furthermore, the direction of the local interstellar magnetic field was inferred to be located at 229° ± 3.5° in right ascension and 11° ± 3° in declination. Based on these results, an estimate of the corresponding vertical dipolar component was obtained, with an amplitude of $3.97 \times 10^{-4}$.

The HAWC Collaboration is currently preparing an updated analysis of the cosmic-ray anisotropy in the Northern Hemisphere with eight years of data, as well as an update to the all-sky analysis with IceCube in order to extend the observations to energies of 500 TeV. Such maps are necessary to reduce observational biases in the determination of large-scale angular structures and to eliminate correlations between spherical harmonics. The new data will provide a better understanding of the anisotropy features shaped by propagation in the ISM and the heliosphere.

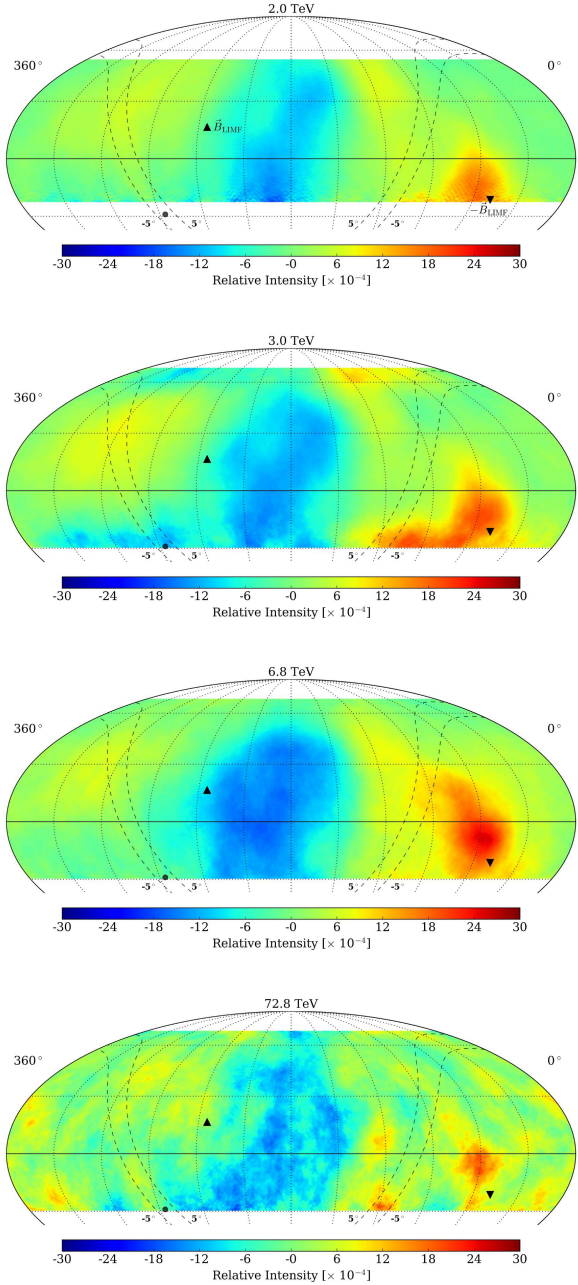

**Figure 13.** Anisotropy maps in statistical significance (smoothed 10°) for four energy bins separated by a likelihood-based reconstructed energy variable. ©AAS. Reproduced with permission from Abeysekara et al. (2018a).

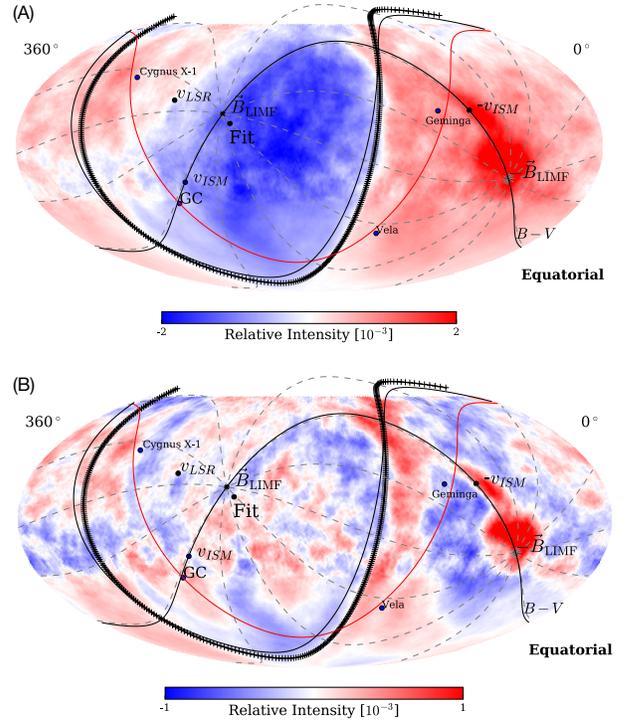

**Figure 14.** **(A)** Relative intensity of cosmic rays at 10 TeV median primary energy. **(B)** Small-scale anisotropy with dipole, quadrupole, and octupole spherical harmonics removed. The black curve composed of crosses corresponds to a fit to the boundary between the large-scale excess and deficit regions. The magnetic equator from Zirnstein et al. (2016) is represented by a black curve. Two nearby SNRs, Geminga and Vela, are shown for reference, along with the Galactic Plane. ©AAS. Reproduced with permission from Abeysekara et al. (2019).

### 6.4. Limits on the Rate of Antiprotons/Protons in the Cosmic-Ray Flux

Within standard models of cosmic-ray propagation, it is expected that antiprotons are produced when primary cosmic rays collide with the interstellar gas in our galaxy (Donato et al., 2001). In order to probe these models, it is important to measure the flux ratio between antiprotons and protons. This ratio has been measured in several experiments, such as the BESS (Mitchell et al., 2005), HEAT (Beach et al., 2001), CAPRICE (Boezio et al., 1997), *PAMELA* (Adriani et al., 2009b) and AMS-02 (Aguilar et al., 2016). In particular, direct measurements up to 450 GeV by AMS-02 showed an increase in the antiproton-to-proton ratio, although a decrease was expected from the models. The observed excess could be explained by unknown astrophysical sources or even by uncertainties in the interaction models (Gaisser & Schaefer, 1992; Auchettl & Balázs, 2012; Jin et al., 2015). Thus, measuring the $\bar{p}/p$ ratio at higher energies will help to form a better understanding of the cosmic-ray spectrum and propagation models. The HAWC Observatory, as a ground-based detector with a large effective area, can be used to investigate the $\bar{p}/p$ ratio at TeV energies.

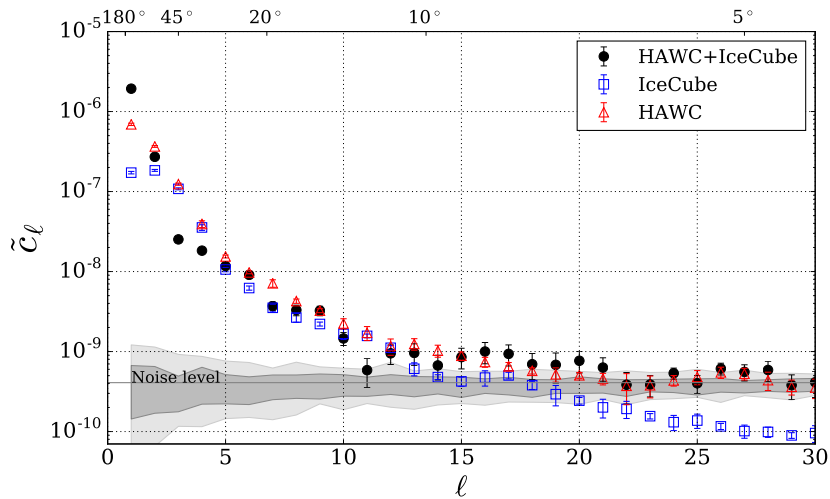

**Figure 15.** Angular power spectrum of the cosmic-ray anisotropy at 10 TeV. The gray band represents the 90% confidence level around the level of statistical fluctuations for isotropic sky maps. The noise level is dominated by limited statistics for the portion of the sky observed by HAWC. The IceCube dataset alone had a lower noise level and was sensitive to higher $\ell$ components. The dark and light gray bands represent the power spectra for isotropic sky maps at the 68% and 95% confidence levels, respectively. The errors do not include systematic uncertainties from partial sky coverage. ©AAS. Reproduced with permission from Abeysekara et al. (2019).

In 2018, HAWC set constraints on the $\bar{p}/p$ ratio at energies between 1 and 10 TeV by studying the deficit produced by the Moon (Moon shadow) in cosmic-ray flux with ~2.8 years of data (Abeysekara et al., 2018d). The Moon blocks the incoming cosmic-ray flux, resulting in a deficit displaced from the Moon's actual position due to deflection by the Earth's magnetic field. The angular deflection depends on the charge and energy of the cosmic rays, providing an estimate of the composition of the spectrum of $Z = 1.30 \pm 0.02$, meaning that $(70 \pm 2)\%$ of the detected flux originates from protons. In theory, another shadow in the opposite direction can be produced by antiprotons, with its relative intensity being proportional to the antiproton flux blocked by the Moon. With this in mind, assuming that HAWC data contain both proton and antiproton Moon shadows and fitting the shape to a two-dimensional Gaussian, 95% CL upper limits for the $\bar{p}/p$ ratio were calculated (as shown in Figure 16) (Abeysekara et al., 2018d). These limits set an experimental bound for the propagation models predicting an increase in the $\bar{p}/p$ ratio at very high energies, demonstrating HAWC's capability for setting strong constraints at energies that are not currently accessible for direct detection experiments.

### 6.5. Sun Shadow

High-energy Galactic cosmic rays can reach space closer to the Sun. In fact, some of them interact with the magnetic field and particles in the low-solar atmosphere (at coronal heights and even lower altitudes), causing a decrease in the intensity of Galactic cosmic rays, which can be observed by HAWC and is known as the Sun shadow. We studied the variation of the Sun shadow during the decrease, minimum, and increase phases of solar cycles 24 and 25 and found that the shape of the Sun shadow is correlated with the solar activity, being diffuse during higher activity phases when the Galactic cosmic rays are randomly deflected by the intense toroidal magnetic field of the active region belt (Alfaro et al., 2024b). However, the Sun shadow is well defined, similar to the Moon shadow, when solar activity reaches its minimum and the weak polar magnetic field barely deviates from the Galactic cosmic rays. This is shown in Figure 17 where the Sun shadow was integrated over a year from 2016, when activity was relatively high, up to 2021. Minimum activity occurred between 2019 and 2020, and the Sun shadow showed the largest deficit in relative intensity. Our study showed that Sun shadow may be used to explore low-coronal magnetic fields, and we found that galactic cosmic rays with energies in the range of ~12 to ~33 TeV interact the most with these magnetic fields (Alfaro et al., 2024b).

### 6.6. Low-Energy Galactic-Cosmic-Ray Anisotropies in the Solar Wind

The Heliosphere is dominated by the Sun and is full of plasma blown out by the star. This plasma carries the magnetic field generated in the solar interior, expelled at different space and time scales (Oughton & Engelbrecht, 2021). These magnetic fields interact with Galactic cosmic rays, and those interactions are dependent on the magnetic field strength and the Galactic-cosmic-ray energy. Low-energy Galactic cosmic rays (with energies of less than tens of GeV) are modulated by the solar wind accordingly with the 22 year solar magnetic cycle (Valdes-Galicia & Gonzalez, 2016). These modulations are well known and have been observed for decades by neutron monitors that are sensitive to relatively low-energy Galactic cosmic rays, with high statistical fluctuations due to the low collecting area of these detectors.

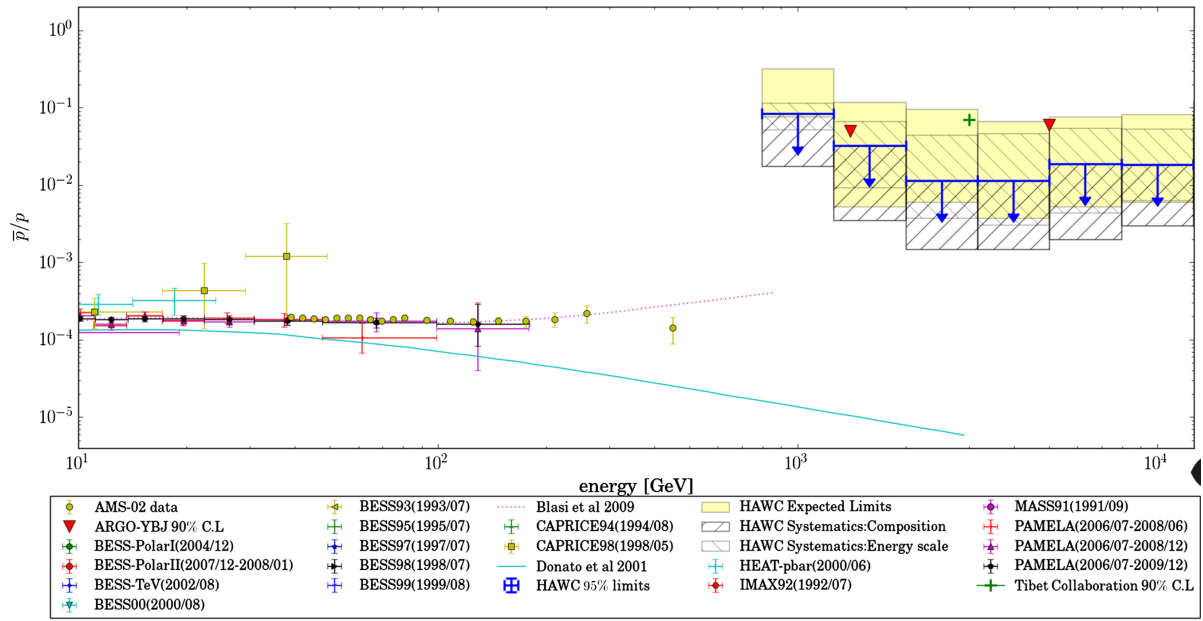

**Figure 16.** HAWC upper limits at TeV energies of the antiproton-to-proton ratio, along with direct measurements in the GeV range. The HAWC sensitivity and systematic uncertainties are shown as yellow and shaded bands, respectively. The expected ratio from purely secondary production of antiprotons is shown as a solid line (Donato et al., 2001). The ratio resulting from a primary antiproton production in supernovae is shown as a dotted line (Blasi & Serpico, 2009). 90% CL upper limit from ARGO-YBJ (Bartoli et al., 2012) and Tibet AS$\gamma$ (Tibet As$\gamma$ Collaboration et al., 2007) are compared with the HAWC 95% C.L. limits. Figure reproduced with permission from Abeysekara et al. (2018d) Physical Review D, Volume 97, Issue 10, id.102005 Copyright (2018) by the American Physical Society.

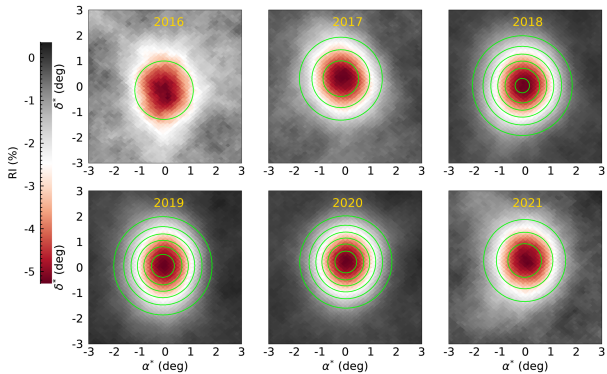

**Figure 17.** Relative intensity (color scale) and significance (green contour lines) of the deficit of Galactic cosmic rays arriving from the region of the sky where the Sun is located instantly. Each map was integrated for one year from 2016 to 2021 (marked in each map). ©AAS. Reproduced with permission from Alfaro et al. (2024b).

However, the large collecting area of HAWC makes it a well-suited instrument for studying the details of Galactic-cosmic-ray solar modulation (Alvarez et al., 2021). In particular, the high sensitivity of HAWC (with statistical fluctuations < 0.01%) allowed us to observe the funnel effect that the helicoidal magnetic field, transported by interplanetary coronal mass ejections, exerted on Galactic cosmic rays (Lara et al., 2024).

Galactic-cosmic-ray anisotropies have been observed for decades without clear explanation until recently, when HAWC observations showed that a relatively large Galactic-cosmic-ray anisotropy was caused by an interplanetary magnetic flux rope (MFR) observed near Earth at the same time (Akiyama et al., 2020). Figure 18 shows the Galactic-cosmic-ray enhancement as seen by the total HAWC scaler rate (black curve in the bottom panel) on October 14$^{th}$, 2016. This enhancement corresponds to the Galactic-cosmic-ray anisotropy generated by the rotation of the magnetic field, as shown in the middle panel, where the components of the magnetic field are plotted. This MFR was part of an interplanetary coronal mass ejection with moderate speed (< 400 km/s) and magnetic field (~25 nT), as shown in the top and second panels of Figure 18, respectively.

### 6.7. Gamma-rays from the Sun

Predicted by Seckel et al. (1991), observations with EGRET (Orlando & Strong, 2008), and *Fermi*-LAT (Abdo et al., 2011; Ng et al., 2016; Linden et al., 2018; Tang et al., 2018) identified the solar disk as a bright source of gamma-rays between 0.1 and 200 GeV, primarily resulting from $\pi^0$ decay due to cosmic-ray interactions in the Sun's atmosphere. However, inconsistencies between the observed gamma-ray flux and the theoretical predictions have prompted investigations at higher energies. TeV observations are critical because the solar-minimum gamma-ray spectrum differs significantly from the cosmic-ray spectrum, providing insights into cosmic-ray propagation through the magnetic field of the Sun.

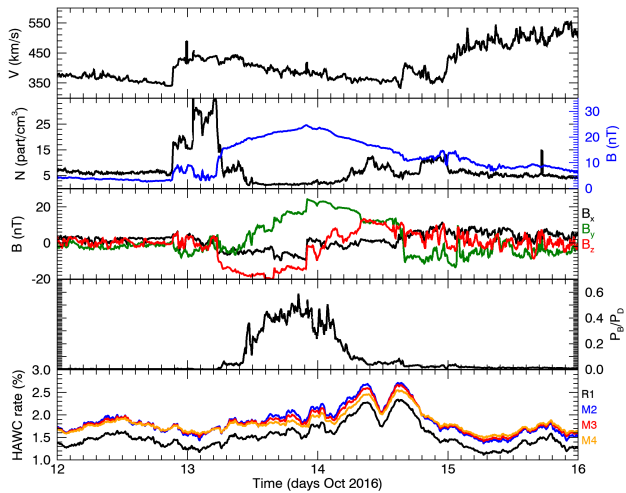

**Figure 18.** From top to bottom: The solar wind velocity; density (black) and total magnetic field (blue); components of the magnetic field (x, y, and z and in black, green, and red curves, respectively) showing the rotation of the field, i.e., a magnetic flux rope (MFR); the ratio between the magnetic and dynamic pressures; and the Galactic-cosmic-ray rate observed by HAWC single (black), double (blue), triple (red), and full (orange) PMT coincidences— the double-peak enhancement observed on Oct. 14, 2016 was caused by the MFR. Adapted from Akiyama et al. (2020).

The first detection of TeV gamma-rays from the Sun came from 6.1 years of data with HAWC (Albert et al., 2023a). The unique observational capabilities of HAWC allowed it to conduct a thorough analysis, leveraging an improved dataset, advanced reconstruction algorithms, and data-driven techniques to isolate gamma-ray signals from the background noise of cosmic rays. This improved analysis indicated a consistent gamma-ray signal with significance levels ranging from $4.2\sigma$ to $6.3\sigma$ for different energy bins. The data also revealed that the flux is notably higher during the Solar minimum compared to the Solar maximum, supporting findings from the *Fermi*-LAT regarding an anticorrelation with solar activity.

The spectral characteristics of the emitted gamma-rays suggest a steep index ($\gamma = 3.62 \pm 0.14$) compared to previously observed values at GeV energies, indicating a transition in emission mechanisms, possibly due to the interaction of cosmic rays with solar magnetic fields. A break in the gamma-ray spectrum is observed around 400 GeV, suggesting underlying complexities in cosmic-ray interactions that are yet to be fully understood. These VHE observations by HAWC further elucidate the role of solar magnetic fields in redirecting Galactic cosmic rays and enhancing the rate of hadronic interactions within the atmosphere of the Sun (Li et al., 2024). Figure 19 shows the gamma-ray spectrum of the Sun including 6.1 years of HAWC data, while Figure 20, the significance map for the corresponding gamma-ray emission at energies < 2 TeV. Both figures were obtained from Albert et al. (2023a).

### 6.8. The Hunga-Tonga Event

On January 15th, 2022, HAWC was able to detect a once-in-a-century phenomenon. Over 9,000 km away, the Hunga Tonga-Hunga Ha'apai volcano, located in the South Pacific, produced the largest volcanic explosion in modern times (Wright et al., 2022). This event produced shock waves in the atmosphere, which modified the medium in which the cosmic rays propagated.

In Alfaro & HAWC-Collaboration (2024), we reported the effect of the shock waves produced by volcanic activity on the secondary particle rate measured by HAWC. The observatory not only records the properties of the primary cosmic rays that enter the atmosphere but also keeps track of the number of secondary particles produced in the upper layers of the atmosphere, which continuously reach the HAWC WCDs. Figure 21 shows the effect of the shock waves on the particle rate measured by the observatory. As they propagated above HAWC, a small fraction of the secondary particles was absorbed, producing a deficit in the rate ($\Delta R$) shown in Figure 21 at approximately 13:00 UTC. We measured the propagation speed of the waves in two different directions around Earth. The speed at which the waves approached the HAWC location through the shorter arc of Earth's circumference was $\approx$ 316 m/s, whereas when the wave traveled through the longer arc, the speed was $\approx$ 312 m/s. The high-altitude location of HAWC makes these measurements unique in the study of this complex phenomenon.

The importance of this observation also comes from the fact that this was the first time that the propagation of this type of wave was characterized using cosmic rays. Other observatories looked for a signal without success (Abbrescia et al., 2022), but the large sensitive area of the HAWC detectors, as well as their sophisticated data-acquisition system (Abeysekara et al., 2018e), allowed HAWC to capture a very detailed view of this rare phenomenon. With this publication, we showed that due to the very high combined data rate acquired by all HAWC PMTs (tens of MHz), it is possible to observe atmospheric phenomena with very high time resolution, complementing the traditional methods.

### 6.9. Horizontal Muons and Neutrinos

Characterization of the muon flux is an important element in the validation of hadronic interaction models. In the vertical direction, the muon flux increases with atmospheric depth, reaching a maximum at a depth of $\sim$200 g/cm$^{-2}$ and then decreases towards the sea level (Tanabashi et al., 2018). However, at large zenith angles, the energy of the primary cosmic rays, which are the progenitors of the muons, is larger than that in the vertical case. This occurs because muons experience both energy losses and decay while propagating in the atmosphere, so only VHE muons are able to survive the approximately 40 times larger atmospheric depth in the horizontal direction compared to vertical muons.

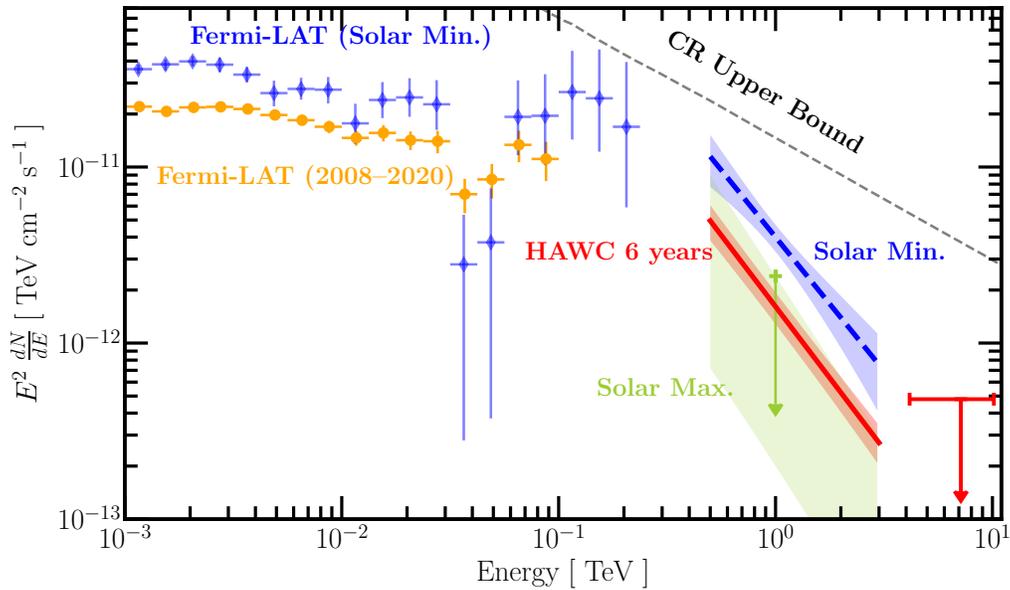

**Figure 19.** Gamma-ray spectrum of the solar disk. The 6.1-year spectrum by HAWC is shown by the solid red line. The 90% CL upper limit at 7 TeV is indicated by a red arrow. The spectrum at the Solar minimum is indicated by a dashed blue line. The shaded bands indicate statistical uncertainties. The Solar maximum flux at 1 TeV is shown as the 1σ UL. The *Fermi*–LAT spectra over the full Solar cycle (Linden et al., 2022) (orange) and at the Solar minimum (Tang et al., 2018) (blue) are also shown. The grey dashed line shows the theoretical maximum of the gamma-ray spectrum (Linden et al., 2018). Figure reproduced with permission from Albert et al. (2023a) Physical Review Letters, Volume 131, Issue 5, article id.051201 Copyright (2023) by the American Physical Society.

Muons can be produced by charged current weak interactions, for example, during the decay of hadrons or when a neutrino collides with matter. In the latter case, if the process occurs at high energies (∼TeV), then the charged lepton approximately follows the direction of the incoming neutrino, enabling the identification of astrophysical neutrino sources. This is the idea behind the use of Earth-skimming neutrinos as a method to identify the location of cosmic accelerators in the Universe (Feng et al., 2002). For this method to work, the neutrino interaction should take place in the outer layer of the Earth's crust, so that the charged lepton, or its decay products, can escape and be measured. The traditional setup for detecting neutrinos involves the use of underground facilities. However, the possibility of placing this type of detector on Earth's surface is of great interest and would allow a reduction in the costs of construction and operation of such observatories if successful. One of the main obstacles in pursuing such an approach is the ability to control the huge background caused by vertical muons (Spiering, 2012).

HAWC was designed to measure secondary particles from air showers reaching the detector from a 45° cone centered around the detector zenith. However, the modular design of the observatory also allows the use of each WCD of the array as part of a horizontal particle tracker (León Vargas et al., 2017). Using the same datasets utilized by other HAWC analyses, that is, data acquired by the same multiplicity trigger, we searched for events that contained sets of PMTs that measured Cherenkov light in a temporal sequence consistent with individual relativistic particles with horizontal propagation. The left panel of Figure 22 shows one of the 122 candidate signals that point back to the base of the Pico de Orizaba volcano from our publication Albert et al. (2022d). These 122 candidates are dominated by the background produced by non-horizontal muons that are scattered in that direction and thus seem to point back to the base of the volcano. This background can be controlled using a muon energy threshold (>100 GeV), where these scatterings are negligible.

The results of our study are significant because, as previously mentioned, it was not possible to control the very large background of secondary particles with vertical propagation that hit each WCD. Although we were not able to conclusively report a neutrino detection, HAWC has shown for the first time that it is possible to implement the Earth-skimming neutrino detection method with an above-ground air-shower detector. Moreover, the tools that we developed to implement our search for Earth-skimming neutrinos can be used to characterize the flux of high-energy horizontal muons. The right panel of Figure 22

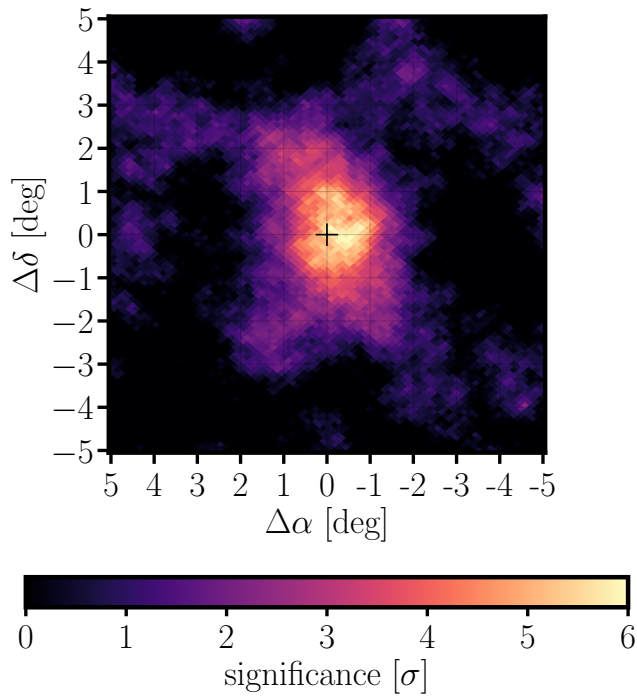

**Figure 20.** Significance map for the gamma-ray emission from the Sun for reconstructed energies <2 TeV. The map is centered at the Sun's location. This was done using 6.1 years of HAWC data collected from November 2014 to January 2021. Figure reproduced with permission from Albert et al. (2023a) Physical Review Letters, Volume 131, Issue 5, article id.051201 Copyright (2023) by the American Physical Society.

shows the track of a horizontal muon crossing the entire detector array.

## 7. HAWC Contributions to Particle Physics

Over the past decade, HAWC has provided valuable insights into high-energy particle physics, thereby complementing traditional astrophysical studies. With its ability to detect VHE gamma-rays and continuously monitor the sky, HAWC has provided valuable constraints on dark matter signatures, investigated the possible existence of dark matter candidates, such as Weakly Interacting Massive Particles (WIMPs) and Axion-Like Particles (ALPs), and plays a role in testing fundamental physics by placing some of the strongest limits on Lorentz Invariance Violation (LIV) and Primordial Black Hole (PBH) evaporation. By complementing other experiments and offering a unique perspective of extreme astrophysical environments, HAWC continues to refine our understanding of the universe at the highest energy scales. This section presents an overview of its key findings and their broader implications in particle physics.

### 7.1. Dark Matter Searches

Indirect search for dark matter using VHE gamma-rays is a key frontier in modern astrophysics. The HAWC

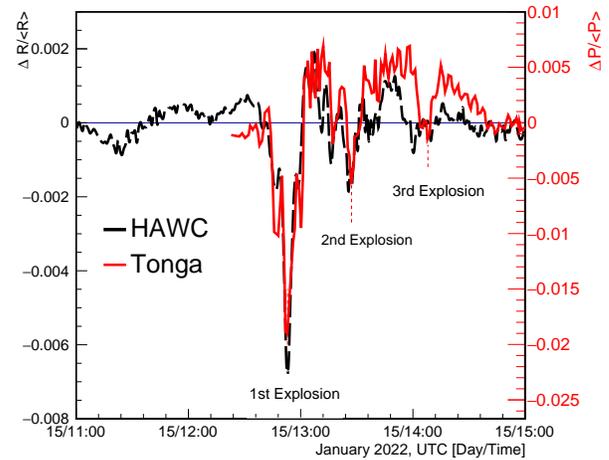

**Figure 21.** Comparison of the particle-rate change detected with HAWC and the pressure change measurement near the Tonga volcano (Wright et al., 2022). The measurement from HAWC coincided with the first passage of the shockwave above the observatory. The pressure measurement was shifted in time to show how the structures of both signals were very similar, allowing HAWC to observe the sequence of explosions that occurred 9,000 km away. The figure was reproduced from Alfaro & HAWC-Collaboration (2024) Advances in Space Research, Volume 73, Issue 1, p. 1083-1091, Copyright Elsevier 2024.

Observatory, with its continuous sky-monitoring capability and sensitivity to the multi-TeV regime, has played a crucial role in exploring potential dark matter signals from various astrophysical targets. This work summarizes HAWC's contributions over the past decade in probing two major dark matter candidates: WIMPs and ALPs. By analyzing a diverse set of sources, including dwarf galaxies, the Galactic Halo, and extragalactic structures, HAWC has set some of the most stringent constraints on dark matter interactions, refining theoretical models and expanding our understanding of the high-energy Universe.

The results obtained by HAWC in the indirect search for dark matter have been fundamental in expanding the frontiers of VHE gamma-ray astrophysics. Its unique capability to continuously monitor large regions of the sky has allowed for some of the most stringent constraints on WIMP and ALP annihilation and decay, complementing and strengthening the constraints obtained by other experiments. Beyond ruling out theoretical models, HAWC has opened new opportunities to explore astrophysical structures previously unconsidered in the search for dark matter, solidifying its role as a key observatory in the exploration of the extreme Universe and in understanding one of the most fundamental questions in modern physics.

The following sections present the key results obtained by HAWC in the search for these dark matter candidates in detail, highlighting their impact on gamma-ray astrophysics.

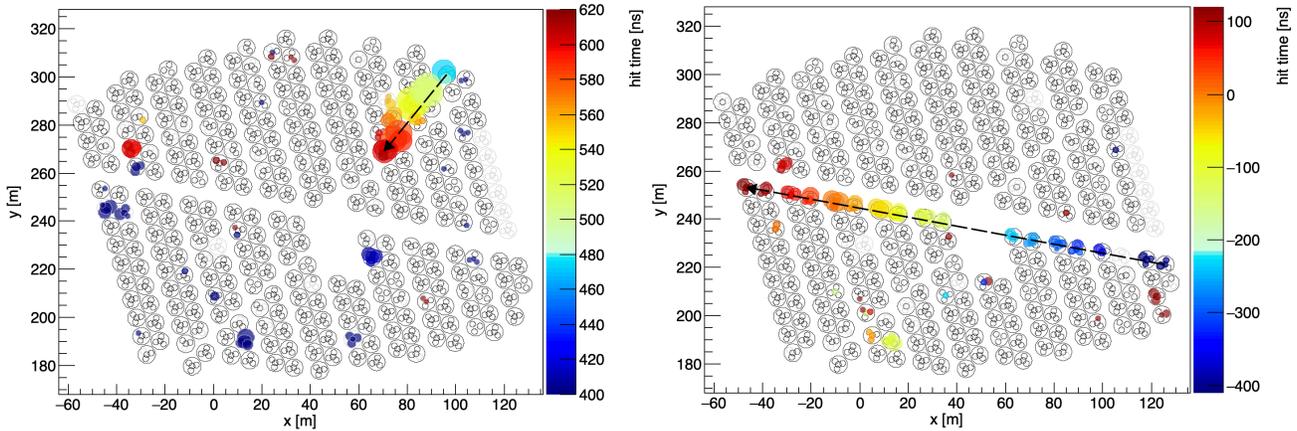

**Figure 22. Left Panel:** Event display of one of the muon candidate signals that point back to the base of the Pico de Orizaba volcano. The color code indicates the time at which each PMT is activated, and the size of each circle is proportional to the measured charge. The arrow shows the reconstructed propagation direction of the near horizontal muon as it traveled through the HAWC WCDs. **Right Panel:** Event display of one of the muon candidates that cross the full HAWC observatory.

### 7.1.1. Weakly Interacting Massive Particles

The indirect search for dark matter in the form of WIMPs has been a key research area for the HAWC Observatory. Given its ability to detect gamma-rays in the multi-TeV regime and its wide FoV, HAWC has enabled the exploration of various astrophysical sources, where a higher density of dark matter is expected. Among them are Dwarf Spheroidal (dSph) and Dwarf Irregular (dIrr) galaxies, the Galactic Halo and Center, and extragalactic structures such as the Virgo Cluster and the Andromeda Galaxy (M31).

Each of these studies provided crucial insights into the possible presence of WIMPs in different cosmic environments. Although no unambiguous signal of dark matter annihilation or decay has been detected, this has allowed for the establishment of unprecedented constraints on the interaction cross section and lifetime of these particles, significantly contributing to the exploration of their nature and constraining theoretical models.

**Dwarf Spheroidal and Dwarf Irregular Galaxies** Dwarf galaxies, particularly dSph and dIrr galaxies, have been fundamental targets in the indirect search for dark matter because they exhibit a high mass-to-light ratio and are therefore expected to contain large amounts of WIMPs with minimal astrophysical background. HAWC has conducted detailed observations of 15 dSph and 31 dIrr galaxies within its FoV, analyzing more than 2,141 days of data. Because no significant excess of gamma-ray emission attributable to WIMP annihilation or decay was found, constraints on the annihilation cross section were set at the level of $\sim 10^{-23}$ cm$^3$ s$^{-1}$ for WIMP masses around 40 TeV, depending on the annihilation channel, and decay lifetimes of $> 10^{27}$ s (approximately $3.17 \times 10^{19}$ years) for WIMP masses around 100 TeV (Albert et al., 2018b; Oakes et al., 2019; Albert et al., 2020a; Salazar-Gallegos et al., 2021; Armand et al., 2022; Alfaro et al., 2023).

**The Virgo Cluster and the Andromeda Galaxy** Galaxy clusters and massive galaxies represent key environments for the indirect search for dark matter due to their high mass content and relative proximity. HAWC has conducted detailed studies of the Virgo Cluster and the Andromeda Galaxy, leveraging its sensitivity in the energy range from 300 GeV to 100 TeV. Based on observations spanning more than five years, the possible WIMP annihilation signals in these systems were analyzed. Since no significant excess of gamma-ray emission was detected, the limits on the annihilation cross-section were set at $2 \times 10^{23}$ cm$^3$ s$^{-1}$ for WIMP masses near 50 TeV, and the lifetime constraints for dark matter decay in these regions reached values of $> 10^{27}$ s (approximately $3.17 \times 10^{19}$ years) for masses above 30 TeV (Albert et al., 2018c, 2024b).

**Galactic Halo and Galactic Center** The Galactic Halo and Center are priority regions in the search for dark matter because of the expected high density of WIMPs in these areas. HAWC has conducted detailed studies on these structures using well-motivated substructure models to evaluate the possible presence of WIMP annihilation or decay signals. With more than five years of observations, constraints on the annihilation cross-section and lifetime of dark matter particles in the multi-TeV regime have been obtained, with annihilation cross-section limits around $10^{-23}$ cm$^3$ s$^{-1}$ for WIMP masses of approximately 10–100 TeV and decay lifetime limits exceeding $10^{27}$ s (about $3.17 \times 10^{19}$ years) for masses above 20 TeV (Abeysekara et al., 2014; Albert et al., 2023b). New observations, currently in preparation, will place stringent constraints on WIMP annihilation and decay using HAWC observations of the GC.

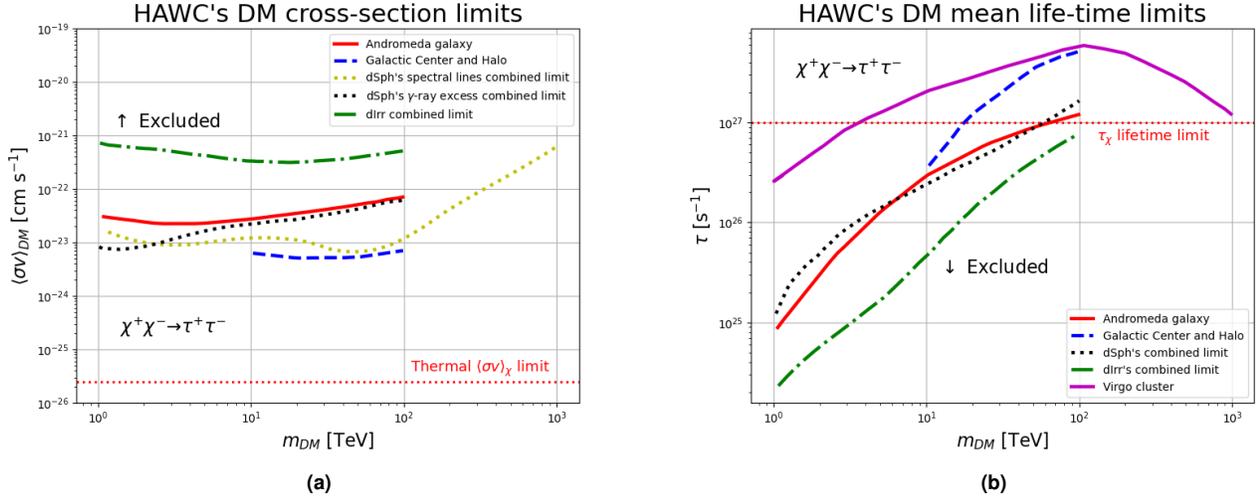

**Figure 23. (a):** Thermal cross-section limits assuming WIMP annihilation. The sources studied with HAWC are the Andromeda Galaxy (Albert et al., 2018c), the Galactic Center and Galactic Halo (Albert et al., 2023b), a dSph combined limit analyzing spectral lines (Albert et al., 2020a), a dSph combined limit searching for a gamma-ray excess (Albert et al., 2018b), and a dIrr combined limit (Alfaro et al., 2023). **(b):** Mean lifetime limits assuming WIMP decay. The sources studied with HAWC are the Andromeda Galaxy (Albert et al., 2018c), the Galactic Center and Galactic Halo (Albert et al., 2023b), a dSph combined limit searching for a gamma-ray excess (Albert et al., 2018b), a dIrr combined limit (Alfaro et al., 2023), and the Virgo Cluster (Albert et al., 2024b).

Figure 23 shows exclusion limits for the WIMP annihilation and decay obtained with HAWC, covering results from previous studies such as those conducted at the GC (Albert et al., 2023b), dSph galaxies (Albert et al., 2018b, 2020a), dIrr galaxies (Alfaro et al., 2023), the Virgo Cluster (Albert et al., 2024b), and the Andromeda Galaxy (Albert et al., 2018c); all limits are for the $\chi^+\chi^- \to \tau^+\tau^-$ channel. These limits illustrate how HAWC has significantly improved restrictions on dark matter from sources in various astrophysical environments.

### 7.1.2. Axion-Like Particles

The HAWC Observatory has explored the possible existence of ALPs, a dark matter candidate, by studying Galactic VHE sources. In particular, the source eHWC J1908+063, one of the brightests detected by HAWC, was analyzed using data from the Pass 5 data reconstruction. This study evaluates the conversion of photons into ALPs within a galactic magnetic field, a phenomenon that would modify the observed gamma-ray spectrum. Because no evidence of such conversion was found, constraints on ALP-photon coupling were established, improving the upper limits to $g_{a\gamma} < 10^{-12}$ GeV$^{-1}$ for ALP masses in the range of fractions of neV up to tens of $\mu$eV.

These results are highly relevant, as ALP searches often rely on extragalactic sources, where background radiation absorption complicates the analysis. In contrast, HAWC demonstrated that high-energy Galactic sources can effectively constrain ALP models, eliminating uncertainties related to extragalactic attenuation. By setting competitive limits on photon-ALP conversion in the multi-TeV regime, HAWC complements and improves previous constraints obtained with telescopes, such as the *Fermi*-LAT and H.E.S.S., strengthening the exploration of this type of dark matter candidate (HAWC Collaboration et al., 2024).

### 7.2. Primordial Black Holes

Currently, the only known mechanisms of black hole formation are through the gravitational collapse of dying massive stars or the merger or cannibalism of a massive star by its binary companion, possibly a neutron star. It follows, then, that only black holes with masses greater than the Chandrasekar limit (∼1.4 solar masses) can be produced in our era. On the other hand, the conditions in the early Universe were markedly different: shortly after the Big Bang, conditions in the Universe were characterized by high densities that would have resulted in associated mechanisms of significant compression. This would have been conducive to the formation of black holes spanning an enormous range of masses, from the Planck mass to SMBHs. These black holes are known as Primordial Black Holes (Page & Hawking, 1976; Carr et al., 2010).

In the contemporary Universe, PBHs in certain mass ranges may constitute a non-negligible fraction of dark matter (e.g., Carr et al. (2010, 2016)), making them a source class for which the community is enthusiastic about finding proof. For stellar-mass black holes and larger, the Hawking radiation (Page & Hawking, 1976) is nearly negligible. However, for lower-mass PBHs, this process is predicted to dominate their evolution— PBHs with initial masses of ∼$5 \times 10^{14}$ g are expected to be expiring today, producing short, few-seconds-long bursts of high-energy gamma radiation in the GeV–TeV energy range, which is ideal for observation with HAWC.

In Albert et al. (2020b), PBH bursts were searched. Adapting a technique used in GRB analysis, a continuous search for transients at energies above a few hundred GeV with sliding time windows of lengths 0.2, 1, and 10 s, as well as stepped, partially overlapping square spatial bins to pinpoint the location of any potential signal, was performed using HAWC data spanning from March 2015 through May 2018. Finding no statistically significant PBH signal in the data, the upper limits on the local burst rate density of PBHs were computed. HAWC was able to set an upper limit at the 99% CL of ∼3400 pc$^{-3}$ yr$^{-1}$ using a burst duration of 10 s. At the time, this was the most restrictive limit for nearby PBHs.

We stress that these types of limits on PBH burst events in dedicated searches yield important information about the early Universe, such as constraints on the cosmological density fluctuation spectrum on scales smaller than those constrained by the CMB.

### 7.3. Testing Lorentz Invariance

Lorentz invariance is a fundamental symmetry of nature that forms the backbone of Einstein's theory of Special Relativity and the Standard Model of particle physics. It dictates that the laws of physics remain the same regardless of the observer's frame of reference, implying that the speed of light in vacuum is constant for all inertial observers. However, many quantum gravity theories, including Loop Quantum Gravity, String Theory, and noncommutative geometry models, suggest that at Planck-scale energies (∼$1.22 \times 10^{19}$ GeV), Lorentz symmetry may be violated or deformed (Addazi et al., 2022; Kostelecky & Samuel, 1989; Colladay & Kostelecký, 1998). Such violations could lead to modifications in the dispersion relations of high-energy particles, resulting in observable effects such as energy-dependent time delays in GRBs, anomalous photon decay at ultra-high energies, vacuum Cherenkov radiation, and photon splitting (see Martínez-Huerta et al. (2020) and references therein). Because these effects are expected to be cumulative over long distances and increasingly significant at high energies, astroparticle observatories such as HAWC provide an optimal environment for testing potential departures from Lorentz invariance.

HAWC has provided some of the most stringent constraints on Lorentz invariance violations by analyzing gamma-rays above 100 TeV from astrophysical sources. Using observations of photons from the Crab Nebula and other extreme sources, HAWC has tested certain quantum gravity models that predict strong modifications to particle interactions below the Planck scale. Tests of Lorentz invariance violation (LIV) in the photon sector can be expressed as modifications to the photon dispersion relation, written in terms of the energy scale, $E_{\text{LIV}}^{(n)}$. Here, n denotes the order of the correction: n=1 corresponds to linear (first-order) terms and n=2 refers to quadratic (second-order) terms. The constraints imposed by HAWC on LIV-induced photon decay are particularly significant. The constraints imposed by HAWC on LIV-induced photon decay are particularly significant. If LIV were present at these energy scales, photons above a given threshold energy would decay into electron-positron pairs, leading to an abrupt cutoff in the gamma-ray spectrum (Martínez-Huerta & Pérez-Lorenzana, 2017). The absence of such a cutoff in the HAWC datasets upper bounds on the LIV energy scale of $E_{\text{LIV}}^{(1)} > 2.2_{1.75}^{2.69} \times 10^{31}$ eV for first-order modifications and $E_{\text{LIV}}^{(2)} > 0.8_{0.69}^{0.91} \times 10^{23}$ eV for second-order modifications (Albert et al., 2020). These limits exceed the previous constraints by up to two orders of magnitude and surpass the Planck energy scale by a factor of more than 1,800, making them among the most stringent LIV limits in the photon sector at the time of their publication.

In addition to photon-decay studies, HAWC also explored additional LIV-related phenomena, such as energy-dependent time delays in GRBs, which could indicate dispersion effects due to modifications in the speed of light at high energies (Nellen, 2016). Although experiments such as the *Fermi*-LAT and VERITAS have previously constrained such time delays, HAWC's continuous sky coverage and access to VHE gamma-rays have allowed for more systematic and long-term searches. Similarly, photon splitting ($\gamma \to 3\gamma$) is another possible consequence of LIV, which suppresses the highest-energy photons in the spectrum of distant astrophysical sources (Rubtsov et al., 2017; Astapov et al., 2019). The non-detection of such anomalous radiation in HAWC data further constrains the LIV-induced modifications to standard quantum electrodynamics. HAWC's detection of gamma-rays well above 100 TeV without any observed suppression effectively rules out this possibility at the measured energy scales, and sets the upper bound for second-order modifications $E_{\text{LIV}}^{(2)} > 12_{10.3}^{14.0} \times 10^{23}$ eV through the cutoff absence up to 244 TeV in HAWC J1825-134 (Albert et al., 2020).

The comparison of these constraints with those from other high-energy observatories further highlights the importance of HAWC contributions.

Before 2021, other experiments and analyses using *Fermi*-LAT, High-Energy Gamma-Ray Astronomy (HEGRA), and H.E.S.S. data placed limits of the order of $E_{\text{LIV}}^{(1)} \sim 10^{29}$ eV and $E_{\text{LIV}}^{(2)} \sim 10^{20}$ eV, respectively, corresponding to first– and second–order modifications (n=1,2) in the photon dispersion relation (see Martínez-Huerta et al. (2020) and references therein).

The HAWC limits surpass them significantly, restricting a broader class of LIV parameters and proving HAWC to be an effective instrument for investigating LIV in the highest-energy regime. Future Cherenkov observatories, such as the Cherenkov Telescope Array (CTA) or SWGO, are expected to improve upon these constraints.

The absence of any LIV signatures in HAWC data has profound implications for quantum gravity models. Many effective field theory (EFT) approaches that predict observable LIV effects in the TeV–PeV range have been constrained, requiring Planck-scale physics to either preserve Lorentz invariance or exhibit deviations only at significantly higher energies, such that the HAWC constraints were translated into the theoretical parameters (Kostelecky & Russell, 2011). Additionally, the lack of detected photon decay at very high energies suggests that the space-time fabric remains continuous at observable scales, challenging some models that predict space-time granularity at energies below the Planck scale. These results strongly favor scenarios in which quantum gravity effects do not introduce measurable violations of Lorentz symmetry in the photon sector at the currently accessible energy scales.

## 8. Summary

In the first decade of operation, HAWC has significantly advanced our understanding of VHE sky and cosmic ray science. The remarkable stream of publications summarized herein shows that a wide-field Cherenkov array monitoring observatory in the TeV band is fully capable of producing crucial and often unexpected contributions in several domains of High-Energy and Multi-Messenger Astrophysics. From the discovery of new types of TeV sources to insights on particle acceleration in pulsars and compact objects, from cosmic-ray properties to PeVatron sources, HAWC has cemented a truly impressive legacy that any future high-energy observatory will look at as a benchmark.

We acknowledge the support from: the US National Science Foundation (NSF); the US Department of Energy Office of High-Energy Physics; the Laboratory Directed Research and Development (LDRD) program of Los Alamos National Laboratory; Consejo Nacional de Ciencia y Tecnología (CONACyT), México, grants LNC-2023-117, 271051, 232656, 260378, 179588, 254964, 258865, 243290, 132197, A1-S-46288, A1-S-22784, CF-2023-I-645, cátedras 873, 1563, 341, 323, Red HAWC, México; DGAPA-UNAM grants IG101323, IN111716-3, IN111419, IA102019, IN106521, IN114924, IN110521, IN102223; VIEP-BUAP; PIFI 2012, 2013, PROFOCIE 2014, 2015; the University of Wisconsin Alumni Research Foundation; the Institute of Geophysics, Planetary Physics, and Signatures at Los Alamos National Laboratory; Polish Science Centre grant, 2024/53/B/ST9/02671; Coordinación de la Investigación Científica de la Universidad Michoacana; Royal Society - Newton Advanced Fellowship 180385; Gobierno de España and European Union-NextGenerationEU, grant CNS2023- 144099; The Program Management Unit for Human Resources & Institutional Development, Research and Innovation, NXPO (grant number B16F630069); Coordinación General Académica e Innovación (CGAI-UdeG), PRODEP-SEP UDG-CA-499; Institute of Cosmic Ray Research (ICRR), University of Tokyo. H.F. acknowledges support from NASA under award number 80GSFC21M0002. C.R. acknowledges support from National Research Foundation of Korea (RS-2023-00280210). HM acknowlegdes support from grant CBF 2023-2024-1630. We acknowledge the significant contributions of former members of the HAWC collaboration: R. Arceo, E. Almaraz, M. Alvarez, J. Álvarez Romero, J. R. Angeles Camacho, B. M. Baughman, A. Becerril, D. Berley, O. Blanco, J. Braun, M. Castillo Maldonado, T. DeYoung, M. A. Diaz Cruz, D. W. Fiorino, J. L. Flores, D. Garcia, G. Garcia-Torales, V. Grabski, Z. Hampel-Arias, E. Hernandez, A. Imran, J. Jablonski, M. Lamprea, J. Martínez, T. Niranjan Ukwatta, T. O. Oceguera-Becerra, C. Rivière, M. Rosenberg, K. Sparks Woodle, A. Tepe, P. Vanegas, X. Judith Vázquez, D. Warner, T. Weisgarber and P. W. Younk. We also acknowledge the significant contributions over many years of Stefan Westerhoff, Gaurang Yodh and Arnulfo Zepeda Domínguez, all deceased members of the HAWC collaboration. Thanks to Scott Delay, Luciano Díaz, Eduardo Murrieta and Janina Nava for technical support.

## ■ References

Aartsen, M. G., Abbasi, R., Abdou, Y., et al. 2013, ApJ, 765, 55, doi: 10.1088/0004-637X/765/1/55

Aartsen, M. G., Abbasi, R., Ackermann, M., et al. 2020, PhRvD, 102, 122001, doi: 10.1103/PhysRevD.102.122001

Aartsen, M. G., Abraham, K., Ackermann, M., et al. 2016, ApJ, 826, 220, doi: 10.3847/0004-637X/826/2/220

Abbasi, R., Abdou, Y., Abu-Zayyad, T., et al. 2010, ApJL, 718, L194, doi: 10.1088/2041-8205/718/2/L194

—. 2011, ApJ, 740, 16, doi: 10.1088/0004-637X/740/1/16

—. 2012, ApJ, 746, 33, doi: 10.1088/0004-637X/746/1/33

Abbasi, R., et al. 2025, ApJ, 981, 182, doi: 10.3847/1538-4357/adb1de

Abbott, B. P., Abbott, R., Abbott, T. D., et al. 2017a, PhRvL, 119, 161101, doi: 10.1103/PhysRevLett.119.161101

—. 2017b, ApJ, 848, L12, doi: 10.3847/2041-8213/aa91c9

Abbrescia, M., Avanzini, C., Baldini, L., et al. 2022, NatSR, 12, 19978, doi: 10.1038/s41598-022-23984-2

Abdalla, H., Adam, R., Aharonian, F., et al. 2019, Natur, 575, 464, doi: 10.1038/s41586-019-1743-9

Abdo, A. A., Allen, B., Berley, D., et al. 2007, ApJL, 664, L91, doi: 10.1086/520717

Abdo, A. A., Allen, B., Aune, T., et al. 2008a, ApJ, 688, 1078, doi: 10.1086/592213

Abdo, A. A., et al. 2008b, PhRvL, 101, 221101, doi: 10.1103/PhysRevLett.101.221101

Abdo, A. A., Allen, B. T., Aune, T., et al. 2009a, ApJL, 700, L127, doi: 10.1088/0004-637X/700/2/L127

Abdo, A. A., Ackermann, M., Ajello, M., et al. 2009b, ApJ, 706, 1331, doi: 10.1088/0004-637X/706/2/1331

Abdo, A. A., et al. 2009, ApJ, 698, 2121. http://stacks.iop.org/0004-637X/698/i=2/a=2121

—. 2010, ApJ, 720, 272, doi: 10.1088/0004-637X/720/1/272

Abdo, A. A., Ackermann, M., Ajello, M., et al. 2011, Sci, 331, 739, doi: 10.1126/science.1199705

Abdo, A. A., Ackermann, M., Ajello, M., et al. 2011, ApJ, 734, 116, doi: 10.1088/0004-637X/734/2/116

Abdollahi, S., Acero, F., Ackermann, M., et al. 2020, ApJS, 247, 33, doi: 10.3847/1538-4365/ab6bcb

Abeysekara, A. U., Albert, A., Alfaro, R., & Collaboration, H. 2017, Sci, 358, 911, doi: 10.1126/science.aan4880

Abeysekara, A. U., & HAWC Collaboration. 2018, Natur, 562, 82, doi: 10.1038/s41586-018-0565-5

—. 2019, ApJ, 881, 134, doi: 10.3847/1538-4357/ab2f7d

—. 2023, NIMPA, 1052, 168253, doi: 10.1016/j.nima.2023.168253

Abeysekara, A. U., Alfaro, R., Alvarez, C., et al. 2014, ApJ, 796, 108, doi: 10.1088/0004-637X/796/2/108

Abeysekara, A. U., et al. 2014, ApJ, 796, 108, doi: 10.1088/0004-637X/796/2/108

Abeysekara, A. U., Alfaro, R., Alvarez, C., et al. 2014, Phys. Rev. D, 90, 122002, doi: 10.1103/PhysRevD.90.122002

Abeysekara, A. U., et al. 2015, APh, 62, 125, doi: 10.1016/j.astropartphys.2014.08.004

—. 2016, ApJ, 817, 3, doi: 10.3847/0004-637X/817/1/3

—. 2017a, ApJ, 843, 40, doi: 10.3847/1538-4357/aa7556

—. 2017b, ApJ, 841, 100, doi: 10.3847/1538-4357/aa729e

—. 2017c, ApJ, 843, 39, doi: 10.3847/1538-4357/aa7555

—. 2018a, ApJ, 865, 57, doi: 10.3847/1538-4357/aad90c

—. 2018b, PhRvD, 97, 102005, doi: 10.1103/PhysRevD.97.102005

—. 2018c, ApJ, 865, 57. http://stacks.iop.org/0004-637X/865/i=1/a=57

—. 2018d, PhRvD, 97, 102005, doi: 10.1103/PhysRevD.97.102005

—. 2018e, NIMPA, 888, 138, doi: 10.1016/j.nima.2018.01.051

—. 2019, ApJ, 871, 96, doi: 10.3847/1538-4357/aaf5cc

—. 2020, PhRvL, 124, 021102, doi: 10.1103/PhysRevLett.124.021102

Abeysekara, A. U., Albert, A., Alfaro, R., et al. 2021, Nature Astronomy, 5, 465, doi: 10.1038/s41550-021-01318-y

Abeysekara, A. U., et al. 2025, PhRvD, 111, 043014, doi: 10.1103/PhysRevD.111.043014

Abeysekara, A. U., et al. 2018, ApJ, 861, 134, doi: 10.3847/1538-4357/aac4a2

Abreu, P., et al. 2025. https://arxiv.org/abs/2506.01786

Accardo, L., Aguilar, M., Aisa, D., et al. 2014, PhRvL, 113, 121101, doi: 10.1103/PhysRevLett.113.121101

Acciari, e. a. 2009, ApJL, 703, L6, doi: 10.1088/0004-637X/703/1/L6

Acciari, V. A., Aliu, E., Aune, T., et al. 2009, ApJ, 703, 169, doi: 10.1088/0004-637X/703/1/169

Acharyya, A., Adams, C. B., Bangale, P., et al. 2024, ApJ, 974, 61, doi: 10.3847/1538-4357/ad698d

Ackermann, M., Ajello, M., Allafort, A., et al. 2011, Sci, 334, 1103, doi: 10.1126/science.1210311

Ackermann, M., Ajello, M., Allafort, A., et al. 2012, PhRvL, 108, 011103, doi: 10.1103/PhysRevLett.108.011103

Ackermann, M., Ajello, M., Albert, A., et al. 2017, ApJ, 840, 43, doi: 10.3847/1538-4357/aa6cab

Addazi, A., Belotsky, K., & Vitaly, B. 2022, Chin. Phys. C, 46, 035001, doi: 10.48550/arXiv.1905.02773

Addazi, A., Alvarez-Muniz, J., Alves Batista, R., et al. 2022, PrPNP, 125, 103948, doi: https://doi.org/10.1016/j.ppnp.2022.103948

Adriani, O., Barbarino, G. C., Bazilevskaya, G. A., et al. 2009a, Natur, 458, 607, doi: 10.1038/nature07942

—. 2009b, PhRvL, 102, 051101, doi: 10.1103/PhysRevLett.102.051101

Adriani, O., et al. 2022, PhRvL, 129, 101102, doi: doi:10.1103/PhysRevLett.129.101102

—. 2023, PhRvL, 130, 171002, doi: doi:10.1103/PhysRevLett.130.171002

Aglietta, M., et al. 2004, APh, 21, 223, doi: doi:10.1016/j.astropartphys.2004.01.005

—. 2009, ApJ, 692, L130. http://stacks.iop.org/1538-4357/692/i=2/a=L130

Aguilar, M., Aisa, D., Alpat, B., et al. 2015, PhRvL, 114, 171103, doi: 10.1103/PhysRevLett.114.171103

Aguilar, M., Ali Cavasonza, L., Alpat, B., et al. 2016, PhRvL, 117, 091103, doi: 10.1103/PhysRevLett.117.091103

Aharonian, F., & LHAASO-Collaboration. 2021, PhRvL, 126, 241103, doi: 10.1103/PhysRevLett.126.241103

Aharonian, F., Yang, R., & de Oña Wilhelmi, E. 2019, NatAs, 3, 561, doi: 10.1038/s41550-019-0724-0

Aharonian, F., Akhperjanian, A., Beilicke, M., et al. 2003, A&A, 403, L1, doi: 10.1051/0004-6361:20030372

Aharonian, F., Akhperjanian, A. G., Bazer-Bachi, A. R., et al. 2006, Natur, 439, 695, doi: 10.1038/nature04467

Ahn, H. S., Allison, P., Bagliesi, M. G., et al. 2010, ApJL, 714, L89, doi: 10.1088/2041-8205/714/1/L89

Ajello, M., Atwood, W. B., Baldini, L., et al. 2017, ApJS, 232, 18, doi: 10.3847/1538-4365/aa8221

Akiyama, S., et al. 2020, ApJ, 905, 73, doi: 10.3847/1538-4357/abc344

Albert, A., Alfaro, R., Alvarez, C., et al. 2020, ApJL, 903, L14, doi: 10.3847/2041-8213/abbfae

Albert, A., & HAWC Collaboration. 2020a, ApJ, 905, 76, doi: 10.3847/1538-4357/abc2d8

—. 2020b, ApJL, 896, L29, doi: 10.3847/2041-8213/ab96cc

—. 2021a, ApJL, 911, L27, doi: 10.3847/2041-8213/abf4dc

—. 2021b, ApJL, 907, L30, doi: 10.3847/2041-8213/abd77b

—. 2024, ApJ, 972, 144, doi: 10.3847/1538-4357/ad5f2d

—. 2025, PhRvL, 134, 171005, doi: 10.1103/PhysRevLett.134.171005

Albert, A., Alfaro, R., Alvarez, C., et al. 2018a, ApJ, 853, 154, doi: 10.3847/1538-4357/aaa6d8

—. 2018b, ApJ, 853, 154, doi: 10.3847/1538-4357/aaa6d8

—. 2018c, JCAP, 2018, 043, doi: 10.1088/1475-7516/2018/06/043

Albert, A., et al. 2019, Science Case for a Wide Field-of-View Very-High-Energy Gamma-Ray Observatory in the Southern Hemisphere. https://arxiv.org/abs/1902.08429

—. 2020a, PhRvD, 101, 103001, doi: 10.1103/PhysRevD.101.103001

Albert, A., Alfaro, R., Alvarez, C., et al. 2020b, JCAP, 2020, 026, doi: 10.1088/1475-7516/2020/04/026

Albert, A., Alfaro, R., Alvarez, C., et al. 2020, PhRvL, 124, 131101, doi: 10.1103/PhysRevLett.124.131101

Albert, A., Alfaro, R., Alvarez, C., et al. 2021a, ApJL, 912, L4, doi: 10.3847/2041-8213/abf35a

Albert, A., Alvarez, C., Camacho, J. R. A., et al. 2021b, ApJ, 907, 67, doi: 10.3847/1538-4357/abca9a

Albert, A., Alfaro, R., Alvarez, C., et al. 2022a, ApJ, 929, 125, doi: 10.3847/1538-4357/ac58f6

Albert, A., et al. 2022b, ApJ, 936, 126, doi: 10.3847/1538-4357/ac880e

—. 2022c, PhRvD, 105, 063021, doi: 10.1103/PhysRevD.105.063021

—. 2022d, APh, 137, 102670, doi: 10.1016/j.astropartphys.2021.102670

Albert, A., Alfaro, R., Alvarez, C., et al. 2023a, in Proceedings of 38th International Cosmic Ray Conference — PoS(ICRC2023), Vol. 444, 759, doi: 10.22323/1.444.0759

—. 2023b, ApJ, 954, 205, doi: 10.3847/1538-4357/ace967

Albert, A., Alfaro, R., Arteaga-Velázquez, J. C., et al. 2023, ApJL, 944, L29, doi: 10.3847/2041-8213/acb5ee

Albert, A., et al. 2023a, PhRvL, 131, 051201, doi: 10.1103/PhysRevLett.131.051201

—. 2023b, JCAP, 12, 038, doi: 10.1088/1475-7516/2023/12/038

Albert, A., Alfaro, R., Alvarez, C., et al. 2024a, ApJ, 974, 246, doi: 10.3847/1538-4357/ad738e

Albert, A., et al. 2024b, PhRvD, 109, 043034, doi: 10.1103/PhysRevD.109.043034

Alekseenko, V. V., et al. 2009, NuPhS, 196, 179, doi: 10.1016/j.nuclphysbps.2009.09.032

Alemanno, F., et al. 2021, PhRvL, 126, 201102, doi: 10.1103/PhysRevLett.126.201102

—. 2024, PhRvD, 109, L121101, doi: 10.1103/PhysRevD.109.L121101

Alfaro, R., Alvarez, C., Araya, M., et al. 2025, arXiv e-prints, arXiv:2501.12613, doi: 10.48550/arXiv.2501.12613

Alfaro, R., & HAWC Collaboration. 2017, ApJ, 843, 88, doi: 10.3847/1538-4357/aa756f

Alfaro, R., & HAWC-Collaboration. 2022, ApJ, 934, 158, doi: 10.3847/1538-4357/ac7b78

Alfaro, R., & HAWC Collaboration. 2024a, ApJ, 975, 198, doi: 10.3847/1538-4357/ad782a

—. 2024b, ApJ, 961, 104, doi: 10.3847/1538-4357/ad00b6

—. 2024c, Natur, 634, 557, doi: 10.1038/s41586-024-07995-9

Alfaro, R., & HAWC-Collaboration. 2024, AdSpR, 73, 1083, doi: 10.1016/j.asr.2023.09.049

Alfaro, R., & HAWC-COLLABORATION. 2025, Orbital Modulation of Gamma-Rays Beyond 100 TeV from LS 5039. https://arxiv.org/abs/2503.20947

Alfaro, R., et al. 2017, PhRvD, 96, 122001, doi: 10.1103/PhysRevD.96.122001

—. 2023, ApJ, 945, 25, doi: 10.3847/1538-4357/acb5f1

—. 2024a, A&A, 691, A89, doi: 10.1051/0004-6361/202451514

—. 2024b, ApJ, 966, 67, doi: 10.3847/1538-4357/ad3208

Alfaro, R., Alvarez, C., Andrés, A., et al. 2025, ApJ, 980, 88, doi: 10.3847/1538-4357/ad9d3d

Alfaro, R., Alvarez, C., Anita-Rangel, E., et al. 2025, arXiv e-prints, arXiv:2506.16031. https://arxiv.org/abs/2506.16031

Alfaro, R., Alvarez, C., Arteaga-Velázquez, J., et al. 2025, APh, 167, 103077, doi: https://doi.org/10.1016/j.astropartphys.2024.103077

Aliu, E., Aune, T., Behera, B., et al. 2014, ApJ, 783, 16, doi: 10.1088/0004-637X/783/1/16

Alvarez, C., Angeles Camacho, J. R., Arteaga-Velázquez, J. C., et al. 2021, SoPh, 296, 89, doi: 10.1007/s11207-021-01827-z

Alves Batista, R. 2022, arXiv, doi: 10.48550/arXiv.2210.12855

Ambrosio, M., Antolini, R., Baldini, A., et al. 2003, PhRvD, 67, 042002, doi: 10.1103/PhysRevD.67.042002

Amenomori, M., Cao, Z., Ding, L. K., et al. 1992, PhRvL, 69, 2468, doi: 10.1103/PhysRevLett.69.2468

Amenomori, M., et al. 2005, ApJL, 626, L29, doi: 10.1086/431582

—. 2006, Sci, 314, 439, doi: 10.1126/SCIENCE.1131702

—. 2008, ApJ, 678, 1165, doi: 10.1086/529514

—. 2017, ApJ, 836, 153, doi: 10.3847/1538-4357/836/2/153

—. 2019a, PhRvL, 123, 051101, doi: 10.1103/PhysRevLett.123.051101

—. 2019b, EPJ Web Conf., 208, 03001, doi: doi:10.1051/epjconf/201920803001

An, Q., et al. 2019, SciA, 5, eaax3793, doi: doi:10.1126/sciadv.aax3793

Angüner, E. O. 2023, Turkish Journal of Physics, 47, 40, doi: 10.55730/1300-0101.2738

Armand, C., Moulin, E., Poireau, V., et al. 2022, in 37th International Cosmic Ray Conference, 528, doi: 10.22323/1.395.0528

Astapov, K., Kirpichnikov, D., & Satunin, P. 2019, JCAP, 1904, 054, doi: 10.1088/1475-7516/2019/04/054

Atkins, R., Benbow, W., Berley, D., et al. 2000, NIMPA, 449, 478, doi: https://doi.org/10.1016/S0168-9002(00)00146-7

Auchettl, K., & Balázs, C. 2012, in JPhCS, Vol. 384, JPhCS (IOP), 012016, doi: 10.1088/1742-6596/384/1/012016

Axelsson, M., Ajello, M., Arimoto, M., et al. 2025, ApJS, 277, 24, doi: 10.3847/1538-4365/ada272

Ayala, H., & HAWC Collaboration. 2022, GRB Coordinates Network, 32683, 1

Barr, E. D., Champion, D. J., Kramer, M., et al. 2013, MNRAS, 435, 2234, doi: 10.1093/mnras/stt1440

Bartoli, B., et al. 2012, PhRvD, 85, 022002, doi: 10.1103/PhysRevD.85.022002

Bartoli, B., Bernardini, P., Bi, X. J., et al. 2013, PhRvD, 88, 082001, doi: 10.1103/PhysRevD.88.082001

Bartoli, B., Bernardini, P., Bi, X. J., et al. 2015, PhRvD, 91, 112017, doi: 10.1103/PhysRevD.91.112017

Bartoli, B., Bernardini, P., Bi, X. J., et al. 2015, ApJ, 809, 90, doi: 10.1088/0004-637X/809/1/90

—. 2018, ApJ, 861, 93, doi: 10.3847/1538-4357/aac6cc

Beach, A. S., Beatty, J. J., Bhattacharyya, A., et al. 2001, PhRvL, 87, 271101, doi: 10.1103/PhysRevLett.87.271101

Beilicke, M., & VERITAS Collaboration. 2012, in American Institute of Physics Conference Series, Vol. 1505, High Energy Gamma-Ray Astronomy: 5th International Meeting on High Energy Gamma-Ray Astronomy (AIP), 586–589, doi: 10.1063/1.4772328

Beilicke, M., Aharonian, F., Benbow, W., et al. 2007, in American Institute of Physics Conference Series, Vol. 921, The First GLAST Symposium (AIP), 147–149, doi: 10.1063/1.2757288

Bell, A. R. 2004, MNRAS, 353, 550, doi: 10.1111/j.1365-2966.2004.08097.x

Bell, A. R., Schure, K. M., Reville, B., & Giacinti, G. 2013, MNRAS, 431, 415, doi: 10.1093/mnras/stt179

Benbow, W., Brill, A., Buckley, J. H., et al. 2021, ApJ, 916, 117, doi: 10.3847/1538-4357/ac05b9

Berger, E. 2014, ARA&A, 52, 43, doi: 10.1146/annurev-astro-081913-035926

Blackett, P. 1938, PhysRev, 54, 973, doi: 10.1103/PhysRev.54.973

Blandford, R., Meier, D., & Readhead, A. 2019, ARA&A, 57, 467, doi: 10.1146/annurev-astro-081817-051948

Blasi, P., & Serpico, P. D. 2009, PhRvL, 103, 081103, doi: 10.1103/PhysRevLett.103.081103

Błażejowski, M., Blaylock, G., Bond, I. H., et al. 2005, ApJ, 630, 130, doi: 10.1086/431925

Boezio, M., Carlson, P., Francke, T., et al. 1997, ApJ, 487, 415, doi: 10.1086/304593

Bykov, A. M., Marcowith, A., Amato, E., et al. 2020, Space Sci. Rev., 216, 42, doi: 10.1007/s11214-020-00663-0

Cao, Z., Aharonian, F., Axikegu, et al. 2024, PhRvL, 132, 131002, doi: Caprioli

Cao, Z., & LHAASO-Collaboration. 2021, Natur, 594, 33, doi: 10.1038/s41586-021-03498-z

—. 2024, ApJS, 271, 25, doi: 10.3847/1538-4365/acfd29

Capistrán, T., Avila Rojas, D. O., González, M. M., Fraija, N. I., & HAWC-Collaboration. 2021, in Proceedings of 37th International Cosmic Ray Conference — PoS(ICRC2021), Vol. 395, 839, doi: 10.22323/1.395.0839

Caprioli, D., Amato, E., & Blasi, P. 2010, APh, 33, 307, doi: https://doi.org/10.1016/j.astropartphys.2010.03.001

Carr, B., Kuhnel, F., & Sandstad, M. 2016, PhRvD, D94, 083504, doi: 10.1103/PhysRevD.94.083504

Carr, B. J., Kohri, K., Sendouda, Y., & Yokoyama, J. 2010, PhRvD, D81, 104019, doi: 10.1103/PhysRevD.81.104019

Cazón, L., Vázquez, R. A., Watson, A. A., & Zas, E. 2004, APh, 21, 71, doi: 10.1016/j.astropartphys.2003.12.009

Chiavassa, A., Apel, W. D., Arteaga-Velázquez, J. C., et al. 2015, in International Cosmic Ray Conference, Vol. 34, 34th International Cosmic Ray Conference (ICRC2015), 281, doi: 10.22323/1.236.0281

Collaboration, T. E. H. T., Akiyama, K., Alberdi, A., et al. 2019, The Astrophysical Journal Letters, 875, L1, doi: 10.3847/2041-8213/ab0ec7

Collaboration, T. P. A. 2014, JCAP, 2014, 019, doi: 10.1088/1475-7516/2014/08/019

Colladay, D., & Kostelecký, V. A. 1998, PhRvD, D58, 116002, doi: 10.1103/PhysRevD.58.116002

Cortina, J., Goebel, F., & Schweizer, T. 2009, arXiv e-prints, arXiv:0907.1211, doi: 10.48550/arXiv.0907.1211

D'Agostini, G. 1995, NIMPA, 362, 487, doi: 10.1016/0168-9002(95)00274-X

D'Amone, A., et al. 2015, PoS(EPS-HEP2015), 402, doi: doi:10.22323/1.234.0402

De Jong, J. 2011, ICRC2011, 4, 46, doi: 10.7529/ICRC2011/V04/1185

de la Fuente, E., Toledano-Juárez, I., Kawata, K., et al. 2023a, A&A, 675, L5, doi: 10.1051/0004-6361/202346681

de la Fuente, E., Toledano-Juarez, I., Kawata, K., et al. 2023b, PASJ, 75, 546, doi: 10.1093/pasj/psad018

De La Torre Luque, P., Gaggero, D., Grasso, D., et al. 2023, A&A, 672, A58, doi: 10.1051/0004-6361/202243714

Di Mauro, M., Manconi, S., & Donato, F. 2019, PhRvD, 100, 123015, doi: 10.1103/PhysRevD.100.123015

Donato, F., Maurin, D., Salati, P., et al. 2001, ApJ, 563, 172, doi: 10.1086/323684

Dorman, L. I. 2004, Cosmic Rays in the Earth's Atmosphere and Underground, Vol. 303, doi: 10.1007/978-1-4020-2113-8

Dubus, G. 2006, A&A, 456, 801, doi: 10.1051/0004-6361:20054779

Engel, R., Heck, D., & Pierog, T. 2011, ARNPS, 61, 467, doi: 10.1146/annurev.nucl.012809.104544

Event Horizon Telescope Collaboration, et al. 2021, ApJL, 910, L13, doi: 10.3847/2041-8213/abe4de

Evoli, C., Gaggero, D., Grasso, D., & Maccione, L. 2008, JCAP, 2008, 018, doi: 10.1088/1475-7516/2008/10/018

Evoli, C., Gaggero, D., Vittino, A., et al. 2017, JCAP, 2017, 015, doi: 10.1088/1475-7516/2017/02/015

Fan, Y.-Z., & Piran, T. 2008, FrPhC, 3, 306, doi: 10.1007/s11467-008-0033-z

Feng, J. L., Fisher, P., Wilczek, F., & Yu, T. M. 2002, PhRvL, 88, 161102, doi: 10.1103/PhysRevLett.88.161102

Ferrand, G., & Safi-Harb, S. 2012, AdSpR, 49, 1313, doi: https://doi.org/10.1016/j.asr.2012.02.004

Ferrari, A., Sala, P. R., Fasso, A., & Ranft, J. 2005, FLUKA: A Multi-Particle Transport Code, Tech. rep., SLAC National Accelerator Lab., Menlo Park, CA (United States). https://www.osti.gov/biblio/877507

Fleischhack, H. 2019a, in International Cosmic Ray Conference, Vol. 36, 36th International Cosmic Ray Conference (ICRC2019), 674, doi: 10.22323/1.358.0674

Fleischhack, H. 2019b, in International Cosmic Ray Conference, Vol. 36, 36th International Cosmic Ray Conference (ICRC2019), 675, doi: 10.22323/1.358.0675

Forro, M. 1947, PhRvD, 72, 868, doi: 10.1103/PhysRev.72.868

Fujita, Y., Bamba, A., Nobukawa, K. K., & Matsumoto, H. 2021, ApJ, 912, 133, doi: 10.3847/1538-4357/abf14a

Gaensler, B. M., & Slane, P. O. 2006, ARA&A, 44, 17, doi: https://doi.org/10.1146/annurev.astro.44.051905.092528

Gaisser, T. K., & Schaefer, R. K. 1992, ApJ, 394, 174, doi: 10.1086/171568

Ge, C., Liu, R.-Y., Niu, S., Chen, Y., & Wang, X.-Y. 2021, Innov, 2, 100118, doi: 10.1016/j.xinn.2021.100118

Gelfand, J. D., Slane, P. O., & Zhang, W. 2009, ApJ, 703, 2051, doi: 10.1088/0004-637X/703/2/2051

Giacinti, G., Mitchell, A. M. W., López-Coto, R., et al. 2020, A&A, 636, A113, doi: 10.1051/0004-6361/201936505

González, M. M., Avila Rojas, D., Pratts, A., et al. 2023, ApJ, 944, 178, doi: 10.3847/1538-4357/acb700

Grebenyuk, V., Karmanov, D., Kovalev, I., et al. 2019, AdSpR, 64, 2546, doi: 10.1016/j.asr.2019.10.004

Guillian, G., Hosaka, J., Ishihara, K., et al. 2007, PhRvD, 75, 062003, doi: 10.1103/PhysRevD.75.062003

H. E. S. S. Collaboration, Abdalla, H., Abramowski, A., & Aharonian, F. 2018, A&A, 612, A9, doi: 10.1051/0004-6361/201730824

H. E. S. S. Collaboration, Aharonian, F., Ait Benkhali, F., et al. 2024, A&A, 685, A96, doi: 10.1051/0004-6361/202348913

Hall, D. L., Munakata, K., Yasue, S., et al. 1999, JGR, 104, 6737, doi: 10.1029/1998JA900107

Halpern, J. P., Camilo, F., Gotthelf, E. V., et al. 2001, ApJL, 552, L125, doi: 10.1086/320347

HAWC Collaboration, Albert, A., Alfaro, R. J., et al. 2024, in 38th International Cosmic Ray Conference, 1422. https://pos.sissa.it/cgi-bin/reader/conf.cgi?confid=444

H.E.S.S. Collaboration, Abramowski, A., Aharonian, F., et al. 2016, Natur, 531, 476, doi: 10.1038/nature17147

H.E.S.S. Collaboration, Abdalla, H., Abramowski, A., et al. 2018, A&A, 612, A2, doi: 10.1051/0004-6361/201629377

H.E.S.S. Collaboration, Abdalla, H., Aharonian, F., et al. 2021, Sci, 372, 1081, doi: 10.1126/science.abe8560

Holder, J., Atkins, R., Badran, H., et al. 2006, APh, 25, 391, doi: 10.1016/j.astropartphys.2006.04.002

Hooper, D., & Linden, T. 2022, PhRvD, 105, 103013, doi: 10.1103/PhysRevD.105.103013

Huang, Y., Hu, S., Chen, S., et al. 2022, GRB Coordinates Network, 32677, 1. https://ui.adsabs.harvard.edu/abs/2022GCN.32677....1H/abstract

Hüntemeyer, P., & Ayala Solares, H. A. 2013, in International Cosmic Ray Conference, Vol. 33, International Cosmic Ray Conference, 1158

Jin, H.-B., Wu, Y.-L., & Zhou, Y.-F. 2015, PhRvD, 92, 055027, doi: 10.1103/PhysRevD.92.055027

Joncas, G., & Higgs, L. A. 1990, A&AS, 82, 113

Kennea, J. A., Williams, M., & Team, S. 2022, GRB Coordinates Network, 32635, 1. https://ui.adsabs.harvard.edu/abs/2022GCN.32635....1K/abstract

Kolb, C., Blondin, J., Slane, P., & Temim, T. 2017, ApJ, 844, 1, doi: 10.3847/1538-4357/aa75ce

Kostelecky, V. A., & Russell, N. 2011, RvMP, 83, 11, doi: 10.1103/RevModPhys.83.11

Kostelecky, V. A., & Samuel, S. 1989, PhRvD, D39, 683, doi: 10.1103/PhysRevD.39.683

Kouveliotou, C., Meegan, C. A., Fishman, G. J., et al. 1993, ApJL, 413, L101, doi: 10.1086/186969

Lara, A., Borgazzi, A., Guennam, E., Niembro, T., & Arunbabu, K. P. 2024, Journal of Geophysical Research (Space Physics), 129, e2024JA032478, doi: 10.1029/2024JA032478

León Vargas, H., Sandoval, A., Belmont, E., & Alfaro, R. 2017, AdAst, 2017, 1932413, doi: 10.1155/2017/1932413

Lesage, S., Veres, P., Roberts, O. J., et al. 2022, GRB Coordinates Network, 32642, 1. https://ui.adsabs.harvard.edu/abs/2022GCN.32642....1L/abstract

LHAASO-Collaboration. 2024, arXiv e-prints, arXiv:2410.08988, doi: 10.48550/arXiv.2410.08988

Li, H., & Ma, B.-Q. 2024, MPLA, 39, 2350201, doi: 10.1142/S0217732323502012

Li, J.-T., Beacom, J. F., Griffith, S., & Peter, A. H. G. 2024, ApJ, 961, 167, doi: 10.3847/1538-4357/ad158f

Liang, X.-H., Li, C.-M., Wu, Q.-Z., Pan, J.-S., & Liu, R.-Y. 2022, Universe, 8, 547, doi: 10.3390/universe8100547

Linden, T., Beacom, J. F., Peter, A. H. G., et al. 2022, PhRvD, 105, 063013, doi: 10.1103/PhysRevD.105.063013

Linden, T., Zhou, B., Beacom, J. F., et al. 2018, PhRvL, 121, 131103, doi: 10.1103/PhysRevLett.121.131103

MAGIC Collaboration, Acciari, V. A., Ansoldi, S., et al. 2019, Natur, 575, 455, doi: 10.1038/s41586-019-1750-x

MAGIC Collaboration, Acciari, V. A., Ansoldi, S., et al. 2020, MNRAS, 492, 5354, doi: 10.1093/mnras/staa014

Manchester, R. N., Hobbs, G. B., Teoh, A., & Hobbs, M. 2005, AJ, 129, 1993, doi: 10.1086/428488

Martínez-Huerta, H., Lang, R. G., & de Souza, V. 2020, Symmetry, 12, 1232, doi: 10.3390/sym12081232

Martínez-Huerta, H., & Pérez-Lorenzana, A. 2017, PhRvD, D95, 063001, doi: 10.1103/PhysRevD.95.063001

Mei, S., Blakeslee, J. P., Côté, P., et al. 2007, ApJ, 655, 144, doi: 10.1086/509598

Mirabel, I. F., & Rodríguez, L. F. 1999, ARA&A, 37, 409, doi: 10.1146/annurev.astro.37.1.409

Miroshnichenko, L. 2015, Solar Cosmic Rays, Fundamentals and applications, Vol. 405 (Heidelberg: Springer), doi: 10.1007/978-3-319-09429-8

Mitchell, J. W., Abe, K., Anraku, K., et al. 2005, AdSpR, 35, 135, doi: 10.1016/j.asr.2003.08.044

Mollerach, S., & Roulet, E. 2018, PrPNP, 98, 85, doi: 10.1016/j.ppnp.2017.10.002

Montini, P. 2016, NPPP, 279, 7, doi: 10.1016/j.nuclphysbps.2016.10.003

Moskalenko, I. 2023, Direct measurements of cosmic rays and their possible interpretations. https://arxiv.org/abs/2310.15442

Munakata, K., Mizoguchi, Y., Kato, C., et al. 2010, ApJ, 712, 1100, doi: 10.1088/0004-637X/712/2/1100

Myssowsky, L., & Tuwim, L. 1926, ZPhy, 39, 146, doi: 10.1007/BF01321981

Nagashima, K., Fujimoto, K., & Jacklyn, R. M. 1998, JGR, 103, 17429, doi: https://doi.org/10.1029/98JA01105

Nayerhoda, A., Salesa Greus, F., Casanova, S., et al. 2019, PoS, ICRC2019, 750, doi: 10.22323/1.358.0750

Nellen, L. 2016, PoS, ICRC2015, 850, doi: 10.22323/1.236.0850

Ng, C. Y., & Romani, R. W. 2004, ApJ, 601, 479, doi: 10.1086/380486

Ng, K. C. Y., Beacom, J. F., Peter, A. H. G., & Rott, C. 2016, PhRvD, 94, 023004, doi: 10.1103/PhysRevD.94.023004

Oakes, L., Armand, C., Charles, E., et al. 2019, Combined Dark Matter searches towards dwarf spheroidal galaxies with Fermi-LAT, HAWC, HESS, MAGIC and VERITAS. https://arxiv.org/abs/1909.06310

Ohira, Y., Kisaka, S., & Yamazaki, R. 2018, MNRAS, 478, 926, doi: 10.1093/mnras/sty1159

Orlando, E., & Strong, W. 2008, A&A, 480, 847, doi: 10.1051/0004-6361:20078817

Ostapchenko, S. 2006a, PhRvD, 74, 014026, doi: 10.1103/PhysRevD.74.014026

—. 2006b, NuPhS, 151, 143, doi: 10.1016/j.nuclphysbps.2005.07.026

—. 2011, PhRvD, 83, 014018, doi: 10.1103/PhysRevD.83.014018

Oughton, S., & Engelbrecht, N. E. 2021, NewA, 83, 101507, doi: 10.1016/j.newast.2020.101507

Padovani, P., Alexander, D. M., Assef, R. J., et al. 2017, A&ARv, 25, 2, doi: 10.1007/s00159-017-0102-9

Page, D. N., & Hawking, S. W. 1976, ApJ, 206, 1, doi: 10.1086/154350

Panov, A., Atkin, E., Gorbunov, N., et al. 2017, PoS, ICRC2017, doi: 10.22323/1.301.1094

Panov, A. D., Adams, J. H., Ahn, H. S., et al. 2009, BRASP, 73, 564, doi: 10.3103/S1062873809050098

Pierog, T., Karpenko, I., Katzy, J. M., Yatsenko, E., & Werner, K. 2015, PhRvC, 92, 034906, doi: 10.1103/PhysRevC.92.034906

Pineault, S., & Joncas, G. 2000, AJ, 120, 3218, doi: 10.1086/316863

Pope, I., Mori, K., Abdelmaguid, M., et al. 2024, ApJ, 960, 75, doi: 10.3847/1538-4357/ad0120

Prosin, V. V., Astapov, I. I., Bezyazeekov, P. A., et al. 2019, BRASP, 83, 1016, doi: 10.3103/S1062873819080343

Prosin, V. V., Berezhnev, S. F., Budnev, N. M., et al. 2014, NIMA, 756, 94, doi: 10.1016/j.nima.2013.09.018

Raue, M., Stawarz, L., Mazin, D., et al. 2012, IJMPS, 8, 184, doi: 10.1142/S2010194512004588

Rubtsov, G., Satunin, P., & Sibiryakov, S. 2017, JCAP, 1705, 049, doi: 10.1088/1475-7516/2017/05/049

Salazar-Gallegos, D., Hawc Collaboration, Fermi-Lat Collaboration, et al. 2021, APS, 2021, H10.004

Sari, R., & Esin, A. A. 2001, ApJ, 548, 787, doi: 10.1086/319003

Seckel, D., Stanev, T., & Gaisser, T. K. 1991, ApJ, 382, 652, doi: 10.1086/170753

Spiering, C. 2012, EPJH, 37, 515, doi: 10.1140/epjh/e2012-30014-2

Steinke, E. 1930, ZPhy, 64, 48, doi: 10.1007/BF01397427

Suzuki, H., Tsuji, N., Kanemaru, Y., et al. 2025, ApJL, 978, L20, doi: 10.3847/2041-8213/ad9d11

Takahashi, Y. 1998, NuPhS, 60, 83, doi: 10.1016/S0920-5632(97)00503-3

Tanabashi, M., Hagiwara, K., Hikasa, K., et al. 2018, PhRvD, 98, 030001, doi: 10.1103/PhysRevD.98.030001

Tang, Q.-W., Ng, K. C. Y., Linden, T., et al. 2018, PhRvD, 98, 063019, doi: 10.1103/PhysRevD.98.063019

Taricco, C., Arnone, E., Rubinetti, S., et al. 2022, PhRvR, 4, 023226, doi: 10.1103/PhysRevResearch.4.023226

Tavani, M., Bulgarelli, A., Vittorini, V., et al. 2011, Sci, 331, 736, doi: 10.1126/science.1200083

The MAGIC Collaboration, Acciari, V. A., Ansoldi, S., et al. 2022, PoS, ICRC2021, 796, doi: 10.22323/1.395.0796

Tibet Asγ Collaboration, Ayabe, S., Bi, X. J., et al. 2007, APh, 28, 137, doi: 10.1016/j.astropartphys.2007.05.002

Tibet ASγ Collaboration, Amenomori, M., Bao, Y. W., et al. 2021, NatAs, 5, 460, doi: 10.1038/s41550-020-01294-9

Tomassetti, N. 2023, PoS(ECRS), 007, doi: 10.22323/1.423.0007

Tsuboi, M., Handa, T., & Ukita, N. 1999, ApJS, 120, 1, doi: 10.1086/313165

Ureña-Mena, F., Carramiñana, A., Rosa-González, D., et al. 2023, arXiv:2310.02484, doi: 10.48550/arXiv.2310.02484

Valdes-Galicia, J., & Gonzalez, L. 2016, AdSpR, 57, 1294, doi: 10.1016/j.asr.2015.11.009

Vincent, P. 2005, ICRC, 5, 163. https://cds.cern.ch/record/963464

Vink, J. 2022, arXiv:2212.10677, doi: 10.48550/arXiv.2212.10677

Woo, J., An, H., Gelfand, J. D., et al. 2023, ApJ, 954, 9, doi: 10.3847/1538-4357/acdd5e

Woosley, S. E., & Bloom, J. S. 2006, ARA&A, 44, 507, doi: 10.1146/annurev.astro.43.072103.150558

Wright, C. J., Hindley, N. P., Alexander, M. J., et al. 2022, Natur, 609, 741, doi: 10.1038/s41586-022-05012-5

Xin, Y., Zeng, H., Liu, S., Fan, Y., & Wei, D. 2019, ApJ, 885, 162, doi: 10.3847/1538-4357/ab48ee

Yoon, Y. S., Anderson, T., Barrau, A., et al. 2017, ApJ, 839, 5, doi: doi:10.3847/1538-4357/aa68e4

Zdziarski, A. A., Malyshev, D., Chernyakova, M., & Pooley, G. G. 2017, MNRAS, 471, 3657, doi: 10.1093/mnras/stx1846

Zhang, B. 2011, CRPhy, 12, 206, doi: 10.1016/j.crhy.2011.03.004

Zirnstein, E. J., Heerikhuisen, J., Funsten, H. O., et al. 2016, ApJL, 818, L18, doi: 10.3847/2041-8205/8218/1/L18

**Academic Affiliations**

[2] *Universidad Autónoma de Chiapas, Tuxtla Gutiérrez, Chiapas, México.*

[3] *Instituto de Astronomía, Universidad Nacional Autónoma de México, Ciudad de México, México.*

[4] *Universidad de Costa Rica, San José 2060, Costa Rica.*

[5] *Universidad Michoacana de San Nicolas de Hidalgo, Morelia, Mexico .*

[6] *Department of Physics, Pennsylvania State University, University Park, PA, USA .*

[7] *Department of Physics and Astronomy, Michigan State University, East Lansing, MI, USA .*

[8] *Temple University, Department of Physics, 1925 N. 12th Street, Philadelphia, PA 19122, USA.*

[9] *Instituto Nacional de Astrofísica, Óptica y Electrónica, Puebla, Mexico .*

[10] *Institute of Nuclear Physics Polish Academy of Sciences, PL-31342 IFJ-PAN, Krakow, Poland .*

[11] *Facultad de Ciencias Físico Matemáticas, Benemérita Universidad Autonoma de Puebla, Puebla, Mexico .*

[12] *Dept. of Physics and Wisconsin IceCube Particle Astrophysics Center, University of Wisconsin—Madison, Madison, WI, USA.*

[13] *Departamento de Física, Centro Universitario de Ciencias Exactas e Ingenierias, Universidad de Guadalajara, Guadalajara, México .*

[14] *Max-Planck Institute for Nuclear Physics, 69117 Heidelberg, Germany.*

[15] *Department of Physics, Stanford University: Stanford, CA 94305-4060, USA.*

[16] *Department of Physics, University of Maryland, College Park, MD, USA .*

[17] *Tecnologico de Monterrey, Escuela de Ingeniería y Ciencias, Ave. Eugenio Garza Sada 2501, Monterrey, N.L., Mexico, 64849.*

[18] *Department of Physics, Michigan Technological University, Houghton, MI, USA .*

[19] *Los Alamos National Laboratory, Los Alamos, NM, USA .*

[20] *Universidad Politecnica de Pachuca, Pachuca, Hgo, Mexico .*

[21] *Department of Physics and Astronomy, University of Utah, Salt Lake City, UT, USA .*

[22] *Instituto de Geofísica, Universidad Nacional Autónoma de México, Ciudad de México, México .*

[23] *University of Seoul, Seoul, Rep. of Korea.*

[24] *NASA Goddard Space Flight Center, Greenbelt, MD 20771, USA .*

[25] *Centro de Investigación en Computación, Instituto Politécnico Nacional, México City, México.*

[26] *Departamento de Física y Matemáticas, Universidad de Monterrey, Av. Morones Prieto 4500, San Pedro Garza García 66238, Nuevo León, México.*

[27] *Dept of Physics and Astronomy, University of New Mexico, Albuquerque, NM, USA .*

[28] *Universidad Autónoma del Estado de Hidalgo, Pachuca, Mexico .*

[29] *Instituto de Ciencias Nucleares, Universidad Nacional Autónoma de Mexico, Ciudad de Mexico, Mexico .*

[30] *Department of Physics, Sungkyunkwan University, Suwon 16419, South Korea.*

[31] *Tsung-Dao Lee Institute & School of Physics and Astronomy, Shanghai Jiao Tong University, 800 Dongchuan Rd, Shanghai, SH 200240, China.*

[32] *NASA Marshall Space Flight Center, Astrophysics Office, Huntsville, AL 35812, USA.*

[33] *Physikalisch-Technische Bundesanstalt (PTB), Bundesallee 100, 38116 Braunschweig, Germany.*

[34] *School of Environmental Studies, Cochin University of Science and Technology, Cochin 602022, Kerala, India.*

[35] *Instituto de Física Corpuscular, CSIC, Universitat de València, E-46980, Paterna, Valencia, Spain.*

[36] *W. W. Hansen Experimental Physics Laboratory, Kavli Institute for Particle Astrophysics and Cosmology, Department of Physics and SLAC National Accelerator Laboratory, Stanford University, Stanford, CA 94305, USA.*

[37] *School of Physics and Center for Relativistic Astrophysics. Georgia Institute of Technology. Atlanta, GA 30332, USA.*

[38] *Tecnológico Nacional de México / ITS de Tantoyuca, Desviación Lindero Tametate S/N, La Morita, Tantoyuca, Veracruz, México.*

[40] *Institute for Nuclear Research of the Russian Academy of Sciences, Moscow, 117312, Russia.*

[41] *Department of Physics and Astronomy, University of Rochester, Rochester, NY 14627, USA.*

[42] *Département de Physique Nucléaire et Corpusculaire, Faculté de Science, Université de Genève, 1205, Geneva, Switzerland.*

[43] *Dipartimento di Fisica "Ettore Pancini", Università degli studi di Napoli "Federico II", Complesso Univ. Monte S. Angelo, I-80126 Napoli, Italy.*

[44] *INFN- Sezione di Napoli, Complesso Univ. Monte S. Angelo, I-80126 Napoli, Italy.*

[45] *INAF- Osservatorio Astronomico di Capodimonte,Salita Moiariello 16, I-80131, Napoli, Italy.*

[46] *Facultad de Ciencias, Universidad Nacional Autónoma de México Circuito interior s/n, Coyoacán, C.P. 0451.*

[47] *Department of Physics, Missouri University of Science and Technology, Rolla, MO, USA.*

[48] *Colegio de Ciencias y Humanidades Plantel Sur, Universidad Nacional Autónoma de México, 04500 Ciudad de México, México.*
[49] *Space Science Center University of New Hampshire Durham, New Hampshire, USA.*